\documentclass[12pt,a4paper]{article}

\usepackage{amsthm}

\usepackage[margin=1in]{geometry}
\usepackage{setspace}
\usepackage{titlesec}
\usepackage{fancyhdr}
\usepackage{tocloft}
\usepackage{hyperref}
\usepackage{times}
\usepackage{tabularx}
\titleformat{\section}{\bfseries\large}{\thesection}{1em}{}
\titleformat{\subsection}{\bfseries\normalsize}{\thesubsection}{1em}{}
\titleformat{\subsubsection}{\itshape\normalsize}{\thesubsubsection}{1em}{}

\usepackage{amsmath}
\usepackage{amssymb}
\usepackage{verbatim}
\usepackage{xcolor}
\usepackage{relsize}

\usepackage{tikz}

\usepackage{thmtools,thm-restate}  

\usepackage{paracol}

\usepackage[round]{natbib}

\usepackage{txfonts}  
  \DeclareSymbolFont{symbolsC}{U}{txsyc}{m}{n}
  \DeclareMathSymbol{\strictif}{\mathrel}{symbolsC}{74}
  \DeclareMathSymbol{\boxright}{\mathrel}{symbolsC}{128}

\newtheorem{df}{Definition}[section]  
\newtheorem{definition}{Definition}[section]  
\newtheorem{theorem}[df]{Theorem}
\newtheorem{corollary}[df]{Corollary}
\newtheorem{lemma}[df]{Lemma}
\newtheorem{proposition}[df]{Proposition}
\newtheorem{example}[df]{Example}

\newcommand{\cneg}[1]{{\sim}#1}
\newcommand{\with}{\mathbin{\& }}
\newcommand{\bigwith}{\text{\large $\&$}}

\newcommand{\cf}{\boxright}
\newcommand{\mcf}{\hspace{2pt}\Diamond\hspace{-4pt}\rightarrow}

\newcommand{\Hal}{\mathcal{H}}
\newcommand{\Halbox}{\mathcal{H}^{\Box}}
\newcommand{\HO}{\mathcal{HO}}

\newcommand{\A}{\mathbb{A}}

\newcommand{\F}{\mathcal{F}}
\newcommand{\G}{\mathcal{G}}
\newcommand{\K}{\mathcal{K}}

\newcommand{\Fs}{\mathbb{F}_\sigma}

\newcommand{\T}{\mathbb{T}}

\newcommand{\M}{\mathbb{M}}

\newcommand{\Ax}{\textsf{A}}
\newcommand{\Bx}{\textsf{B}}

\newcommand{\dom}{\mathrm{Dom}}
\newcommand{\ran}{\mathrm{Ran}}

\newcommand{\inte}{\mathrm{Int}}
\newcommand{\ext}{\mathrm{Ext}}

\newcommand{\ven}{\mathrm{VEnd}}
\newcommand{\vexo}{\mathrm{VExo}}

\newcommand{\Atom}{\mathrm{Atom}}

\newcommand{\SET}[1]{\mathbf{#1}}

\newcommand{\commf}[1]{
\textcolor{blue}{\textbf{+++ FB: #1 +++}}}

\newcommand{\corrf}[1]{#1}

\title{Axiomatizations of causal reasoning under indeterministic causal laws}

\author{Fausto Barbero\thanks{University of Helsinki, Department of Philosophy, History and Art Studies, Unioninkatu 40, 00170 Helsinki,  Finland}}

\date{}

\begin{document}





\maketitle


\begin{abstract}
We investigate the generalization of causal models to the case of indeterministic causal laws that was suggested in Halpern (2000). We give an overview of what differences in modeling are enforced by this more general perspective, and propose an implementation of generalized models in the style of the causal team semantics of Barbero \& Sandu (2020). In these models, the laws are not represented by functions (as in the deterministic case), but more generally by relations. We point out significant differences in the indeterministic vs. the deterministic case, both for what concerns the axiomatization of counterfactuals and the definability of causal notions in terms of them. We notice, for instance, that the notions of parenthood and direct cause, differently from the deterministic case, come apart.

We provide two main strongly complete axiomatizations, one for the class of indeterministic causal models and one for their team generalizations, which can also represent uncertainty about the state of the system. 
Then, we also identify axioms that allow specializing the axiomatizations to a number of significant subclasses, characterized by properties such as acyclicity, totality and determinism of the causal laws. 

We also see that a minor change in the notion of signature of the models has a dramatic impact; in particular, it makes so that the deterministic subclass of models become undefinable in the usual Halpern-style counterfactual language. We see that the definability of determinism can be recovered either by a restriction of the class of models, by allowing right-nested counterfactuals, or by introducing an operator for information update.

\vspace{5pt}

\end{abstract}


\section{Introduction}

When the existence of causal laws is considered at all, the discussion of causality in the philosophy of science tends to focus on \emph{deterministic} causal mechanisms, which uniquely determine an effect given full knowledge of the causes. 
The restriction to deterministic laws is in some case a deliberate choice to avoid technical complications \citep[see e.g.][sec. 4.3]{PauHal2013}, as many important and sometimes puzzling aspects of causation already emerge at this level. However, often the laws considered in applied sciences are not of this kind. A physical law may allow us to predict, for example, that
\begin{center}
    A cannonball shot at a certain angle will fall \emph{within a certain range}.
\end{center}
In contrast to more idealized laws of mechanics, this kind of law tells us that, \emph{even without taking into account some factor that is not mentioned by the law} (such as the direction or intensity of the wind), we can predict that the effect will (quite literally) fall within a certain range of possible values. These kinds of causal dependencies are discussed at length e.g. in  \citet[][chapter I.7]{Boh1957}, under the name of \emph{one-to-many} causal relationships.



Note that in the cannonball example the law is indeterministic due to the impossibility of taking into account all factors, i.e. it accounts for \emph{epistemic} limitations; the indeterministic law might e.g. just be a rough approximation of an underlying deterministic law. On the other hand, many physicists would think that the following statement expresses an irreducibly indeterministic law 
about the behaviour of a particle: 
\begin{center}
    If an atom of silver passes through a Stern-Gerlach apparatus, \corrf{it will be detected in either of two specific areas of the screen.} 
\end{center}
In other words, a physical law may tell us that the specifics of the outcome of a certain  experiment are unpredictable in principle -- even if we had a complete knowledge of the initial conditions of a physical system. In this case, an indeterministic law tells us that the particle will be detected somewhere within two areas, but nothing more. 
Indeterministic laws do not just arise within scientific theories, but also justify everyday statements such as 
\begin{center}
    If I toss the coin, it will land on heads or tails. 
\end{center}
Far from being a platitude, such a statement implicitly assumes an indeterministic causal law, which allows two values for the future state of the coin, while excluding other alternatives (such as the coin staying in my pocket or  breaking into pieces). 
Notice also that all these examples do not \emph{require} probabilities for their formulation and justification. We might not know what  probabilistic distributions are involved, or in some context it might even be mathematically impossible to associate a probability distribution to a given phenomenon \citep[see e.g.][for some examples of this kind]{Wys2023}. Thus, probabilistic laws are only a special case of indeterministic laws.

In the last decades, causal reasoning has been mathematically formalized in the fields of statistics and computer science known as \emph{causal discovery} \citep{SpiGlySch1993} and \emph{causal inference} \citep{Pea2000}, which have found applications e.g. in  machine learning \citep{PetJanSch2017,Sch2022},  epidemiology \citep{HerRob2023}, econometrics \citep{HecVyt2007} and social sciences \citep{MorWin2015}. Do indeterministic laws feature in these approaches? There are two main types of models considered in this context. The first are the \emph{Bayesian networks}; these consist of probabilistic distributions paired with graphs, and are intrinsically indeterministic models; but laws or mechanisms are not specified in any way by such models. Instead, \emph{structural equation models} (or \emph{causal models})  
%
use systems of equations 
 to encode the causal laws that link together the relevant variables of the scenario under examination. The equations take the form:
\[
Y:= f(X_1,\dots,X_n)
\]
where the term $f(X_1,\dots,X_n)$ stands for a \emph{function} of the variables $X_1,\dots, X_n$. The equations, together with some 
information about the state of the variables, allow one to formulate and often answer queries about the deterministic and probabilistic behaviour of the variables. A significant example of such a query is establishing the truth or falsity of an \emph{interventionist counterfactual} \citep{GalPea1998}:
\begin{center}
If variables $Z_1, \dots Z_m$ were set to values $z_1, \dots z_m$, then condition $\psi$ would hold. 
\end{center} 
These kinds of expressions can be studied with the methods of logic. The classic work of Halpern \citep{Hal2000,Hal2016} provided a wealth of complete deduction systems for classes of causal models, and the idea has since been extended in various directions \citep[e.g.][]{Bri2012,Zha2013,IbeIca2020,HalPet2022,BarSchVelXie2021,BarYan2022,BecHalHit2023,BarVir2023b,FanZha2023,KawSatKoh2023,DinManTziWanWan2023}. 
Now, the very fact that causal laws are represented by functions 
amounts to a restriction to deterministic laws. 
The natural way to extend these kinds of models to the indeterministic case is to replace these functional constraints with \emph{relational} constraints. 
If we write $R_Y$ for the causal law describing the behaviour of variable $Y$, we shall interpret the statement that $(x_1,\dots,x_n,y)\in R_Y$ as expressing the fact that, if variables $X_1,\dots,X_n$ are set to values $x_1,\dots,x_n$, then $Y$ \emph{might} take value $y$ (but it might also take any other value $y'$ such that $(x_1,\dots,x_n,y')\in R_Y$). Equivalently, we can recover the equational form by encoding the causal law as a \emph{multivalued function} $\F_Y$ that associates to each tuple of values for $X_1,\dots X_n$ a \emph{set} of possible values for $Y$.

The possibility of such an extension  is hinted at by Halpern in the conclusions of his paper on the axiomatization of interventionist counterfactuals:
\begin{quote}
    (...) a more general approach to modeling causality would allow there to be more than one value of $X$ once we have set all other variables. This would be appropriate if we model things at a somewhat coarser level of granularity, where the values of all the variables other than $X$ do not suffice to completely determine the value of $X$. I believe the results of this paper can be extended in a straightforward way to deal with this generalization, although I have not checked the details. \citep{Hal2000} 
\end{quote}
Perhaps stating that this generalization is ``straightforward'' discouraged researchers from pursuing this direction: or was it the discovery that, after all, the details are not so straightforward? To the best of our knowledge, the idea has been taken up again only in 2021, when Peters and Halpern \citep{PetHal2021,HalPet2022} introduced the more complex framework of \emph{generalized structural equation models} (GSEMs). In that context, causal laws are not mentioned at all, but it is evident that indeterministic causal laws can be modeled in a GSEM as a special case. More recently, a simpler framework, closer to Halpern's original idea for extending causal models was considered by Wysocki \citep{Wys2023}, who provided a cursory examination of differences between deterministic and indeterministic counterfactuals.\footnote{There is also recent new work \citep{Bec2025,Bec2025b} inspired by the conference version of this paper. We will discuss it in section \ref{sec: comparisons}.} The purpose of the present paper is to sharpen the understanding of the logic of interventionist counterfactuals in the indeterministic (but not probabilistic) context and ultimately provide complete axiomatizations for significant classes of indeterministic causal models. The models we use (\emph{relational causal teams}) are a generalization of the \emph{causal teams} proposed in \citet{BarSan2020} and are somewhat more general than Wysocki's (one key difference being that our models -- like structural equation models -- do not necessarily obey the Markov condition\footnote{Which, in the context of non-probabilistic models, amounts to the fact that there may be data dependencies among exogenous variables.}). In section \ref{sec: differences} we will describe, in general terms, some key differences between deterministic and indeterministic frameworks, and how they will affect our modeling choices. 
We then use the insights just gained to provide (section \ref{subs: relational causal teams}) a precise definition of 
indeterministic models. We introduce (section \ref{subs: H language}) 
a language $\Hal$ similar to those considered in \citet{Hal2000}, and its extension $\Hal^+$ that also allows right-nesting of counterfactuals. In section \ref{sec: H language} we provide a strongly complete axiomatization of language $\Hal$ over the class of all \emph{singleton} models (of a given signature); this allows us to make a direct comparison with tthe usual case of deterministic causal models. Later, section \ref{sec: general axiomatization} provides an axiomatization of $\Hal^+$ over the class of \emph{all} models; section \ref{sec: law components} lays down some necessary technical tools for the proof, by clarifying which models are indistinguishable using $\Hal$ or $\Hal^+$, and explaining to what extent the causal laws can be described in these languages. In section \ref{sec: PA, direct causes, dummy arguments} we analyze notions such as parenthood, direct cause and (non-)dummy arguments, we see that (diffently from the deterministic case) they come apart, and investigate their mutual relationships.  
 Some of these insights help us produce complete axiomatizations for 
 models where cyclic causal relationships are forbidden (section \ref{subs: causes and recursivity}); more precisely, we axiomatize two classes of models that we call, respectively, \emph{strictly recursive} and \emph{visibly recursive}. 
In section \ref{sec: further axiomatizations} we axiomatize two other interesting subclasses, that of \emph{deterministic} models (those in which the causal laws associate at most one output to each input value for the parents) and that of \emph{total} models (where the causal laws associate at least one possible effect to each input value for the parents). 
We also see that the \emph{recursive}  class of models, which forbids cycles of direct causation, can be axiomatized relatively to the class of total models. Combining all these results, we obtain an alternative axiomatization for the (usual, deterministic) recursive causal models. 
In section \ref{sec: simplified signatures}, we consider a slight variation of the framework, which adopts a notion of signature closer to that which is often used in the philosophical literature. Such simplified signatures only fix a variable domain and the allowed value ranges for the variables; they do not decide which variables are going to be associated with a causal law. While the use of such signatures is tempting, as it can potentially lead to more general completeness results, we see (section \ref{sec: undefinability}) that, in the indeterministic context, it leads to a barrage of undefinability results (which, among other things, prevent us from extending the axiomatization strategies used in the rest of the paper). Some of these results -- e.g. the undefinability of direct cause -- extend to the full signatures. Others, like the undefinability of the deterministic class of models, affect $\Hal$ only if  simplified signatures are adopted. We show three different strategies for recovering the definability of determinism while keeping the simplified signatures. First, we see it is again definable if we restrict attention to singleton models (section \ref{subs: determinism in singletons}). Secondly, we see definability is regained if one allows for right-nested counterfactuals (section \ref{sec: nested conditionals}). Thirdly, the same is achieved by extending $\Hal$ with a second conditional operator which describes learning or information update (section \ref{subs: determinism via learning}); we analyze potential axioms and rules for this language extension. 
Lastly, in section \ref{sec: comparisons} we compare our approach and results to the recent literature on indeterministic counterfactuals.

The reader who is interested in the conceptual aspects of indeterministic causation and counterfactuals but not in the technicalities of proving completeness results can pick the following shorter route: reading sections \ref{sec: differences}, \ref{sec: formal framework}, the first two pages of \ref{sec: H language}, sections \ref{sec: PA, direct causes, dummy arguments}, \ref{subs: causes and recursivity}, \ref{sec: simplified signatures} and \ref{sec: comparisons}. 



\section{Differences with the deterministic case}\label{sec: differences}


We describe here a few differences and challenges that arise when trying to describe models that are not limited to deterministic causal laws.

We need to introduce some notational conventions, for which we follow usage from the field of causal inference. We adopt capital letters $X, Y, \dots$ to denote \textbf{variables}, which are used e.g. to represent magnitudes such as pressure, temperature, etc.; binary variables may also be used to represent whether an event occurs or not. The \textbf{values} that a variable $X$ may take will be denoted by small letters such as $x,x',x''$. We use atomic formulas of the form $X = x$ to assert that (in a given context) $X$ takes value $x$; the set of atomic formulas of signature $\sigma$ will be denoted by $\Atom_\sigma$.
We assume that the variables come from \corrf{two disjoint finite sets $\mathcal E$ (\textbf{external} variables) and $ \mathcal I$ (\textbf{internal} variables); we write $\dom$ for $\mathcal E \cup \mathcal I$. Intuitively, the internal variables are those whose causal laws will be described by the models, while the external variables are those whose causes are not further analyzed.\footnote{\corrf{The more common terminology,  distinguishing between \emph{exogenous} and \emph{endogenous} variables, will be (re)defined later.}} Furthermore, we assume the existence of a function} 
$\ran$ that associates to each variable $X$ a finite set $\ran(X)$ of possible values. The \corrf{triple $(\mathcal E, \mathcal I,\ran)$} 
is called a \textbf{signature}. An \textbf{assignment} of signature $\sigma = \corrf{(\mathcal E, \mathcal I,\ran)}$  
is a function that assigns to each variable an allowed value (i.e., a function $f: \dom \rightarrow \bigcup_{V\in\dom} \ran(V)$ such that, for all $V\in\dom$, $s(V)\in\ran(V)$). We call $\A_\sigma$ the set of all such assignments.  

Boldface letters such as $\SET X$, $\SET x$ denote (depending on the context) either finite sets or finite tuples of variables, resp. of values. If $\SET X = (X_1,\dots,X_n)$, then  $\ran(\SET X)$ abbreviates $\ran(X_1) \times \dots \times \ran(X_n)$. If furthermore $\SET x = (x_1,\dots,x_n)$, we abbreviate as $\SET X = \SET x$ either a multiset of atomic formulas $X_1 = x_1,\dots, X_n = x_n$ or their conjunction. When $\SET X$ is a nonempty tuple of variables, and $s$ is an assignment, $s(\SET X)$ denotes the tuple of values that $s$ associates to the variables in $\SET X$.

We use $\SET W$, resp. $\SET W_X,\SET W_{XY}$ to denote tuples listing without repetitions the variables in $\dom$, resp. $\dom\setminus\{X\}, \dom\setminus\{X,Y\}$.

\subsection{Uncertainty}\label{subs:  uncertainty} 

While working with deterministic causal laws, it has been customary to describe a scenario by means of an assignment of values to the variables of the system.\footnote{At least, this has been most common in the philosophical literature; but also e.g. \cite{Hal2016}, section 2.7 follows this convention.} This may be appropriate, in some case, also in the presence of indeterministic laws. Consider  the scenario ``Alice tosses a coin, and it comes out heads''; we would model it by one causal law (saying that the coin tossing will lead either to a ``heads'' or a ``tails'' outcome) together with the following assignment:
\begin{center}
\begin{tabular}{|c|c|}
\hline
 \multicolumn{2}{|l|}{ \ \ $A$ \ \ $\strictif$ \ \ \ $C$ } \\
\hline
 \phantom{a}\textit{1}\phantom{a} & \phantom{a}$heads$\phantom{a} \\
\hline
\end{tabular}
\end{center}
where the Boolean variable $A$ tells us whether Alice tossed the coin ($A=1$) or not ($A=0$), and $C$ records the outcome of the toss; we drew the symbol $\strictif$ to emphasize that $A$ is an indeterministic cause of $C$.  However, one might also want to model the scenario ``Alice tossed a coin''.  The indeterministic law does not allow us to infer the outcome of the toss; thus, to represent this kind of scenario, we need \emph{two} assignments, describing the two alternative situations that are not excluded by the description of the scenario:  
\begin{center}
\begin{tabular}{|c|c|}
\hline
 \multicolumn{2}{|l|}{ \ \ $A$  \ \ $\strictif$ \ \ $C$ } \\
\hline
 \phantom{a}\textit{1}\phantom{a} & \phantom{a}$heads$\phantom{a} \\
\hline
\textit{1} & $tails$ \\
\hline
\end{tabular}
\end{center}
These kinds of considerations lead us to shift attention from causal models to the more general \emph{causal teams}  \citep{BarSan2020}, models which allow \emph{sets} (``teams'') 
of variable assignments 
compatible with the causal laws. This perspective is not alien to the previous literature on causal inference: it is implicit e.g. in the treatment of interventions in the presence of cyclic causal laws. Even when considering a full description of a scenario (i.e., a single assignment), such an intervention may produce a multiplicity of possible new scenarios \citep{GalPea1998,Hal2000}. 

In the presence of indeterministic causal laws, the need for teams also emerges when asking counterfactual queries about \corrf{a fully described scenario, e.g.: ``Alice did not toss and the coin is still in her pocket''. If we want to know what would have happened if Alice had tossed the coin}, we will need to consider an intervention that sets $A$ to $1$ ($do(A=1)$), which produces the same 2-assignment model that we considered in the second scenario.\footnote{\corrf{We will assume that the same team is produced by the intervention $do(A=1)$ when applied to our initial scenario ``Alice tossed the coin and it came heads'', i.e., when considering a subjunctive conditional with a \emph{true} antecedent we will also use non-actual worlds in the evaluation of the statement. This move seems justified because, in any case, the ``actual world'' cannot really be obtained via the intervention $do(A=1)$; rather, we obtain (one of the) worlds in which Alice's toss is dictated by an external intervention and not by its natural causes. As observed in \cite{Bri2012}, causal models do not even satisfy \emph{weak centering}.}}
We thus see that the class of causal models is not closed under interventions: intervening on an (indeterministic) causal \emph{model} produces an (indeterministic) causal \emph{team}. Our definitions will ensure that the class of indeterministic causal teams is closed under interventions.

\subsection{Specifying the causal laws}

In the deterministic case, the causal laws can be specified in at least two different ways, which are, for most purposes, equivalent:

\begin{enumerate}
    \item First pick out some variables, which will be considered \emph{endogenous}. For each endogenous variable $V$, we specify which other variables are \emph{direct causes} or \emph{parents} of $V$; call this set $PA_V$. We then specify the law for $V$ as a function $\F_V:\ran(PA_V) \rightarrow \ran(V)$.
    
    \item Assign to each variable $V$ 
    a function 
    $\F_V:\ran(\SET W_V)\rightarrow \ran(V)$. 
    Observe that some of the variables of 
    $\SET W_V$ are dummy arguments of $\F_V$; \emph{define} $PA_V$ as the set of variables of 
    $\SET W_V$ that are \emph{not} dummy for $\F_V$. \emph{Define} the set of endogenous variables as those whose parent set is nonempty.   
\end{enumerate}
The first approach is more natural and direct, but it has technical disadvantages, prominently the fact that it allows a proliferation of essentially equivalent\footnote{In the sense that two such models are not distinguishable by any reasonable formal language of interventionist counterfactuals.} models (for example, we might have two models that differ only in that, in the former, variable $Z$ is generated by the law $\F_Z(X,Y) = X+Y$, while in the latter the law is $\F_Z(X,Y,U,V,W) = X+Y$, i.e., the same function with three dummy arguments). For this reason, in technical papers the second approach is usually preferred.

Analogously, in the indeterministic case we might want to encode the laws as relations in $\ran(PA_V)\times \ran(V)$ or instead in $\ran(\SET W_V)\times \ran(V)$. Unfortunately, in the indeterministic case approach 2. seems not to be viable. We consider two examples that raise problems for this approach. 

\begin{example}[Jumpin' Alice]\label{example: jumpin Alice}
In this scenario we have Boolean variables $A$ (whether Alice jumps), $B$ (whether Bob tosses a coin) and $C$ representing three possible states of the coin (whether it is on heads, tails, or stays in Bob's pocket). Suppose we represent the causal law determining the state of the coin by the relation $\F_C \subseteq \ran(A)\times\ran(B)\times \ran(C)$, $\F_C = \{(0,0,in\text{-}pocket),(1,0,in\text{-}pocket),(0,1,heads),(0,1,tails),$
$(1,1,heads),(1,1,tails)\}$. Suppose also that we know that Alice did not jump, B tossed the coin and it came heads:
\begin{center}
\begin{tabular}{|c|c|c|}
\hline
 \multicolumn{3}{|l|}{ \ \ A \ \ \ \  \ \ B \ \ \ \ \ \ \ \ \ C } \\
\hline
 \phantom{a}0\phantom{a} & \phantom{a}1\phantom{a} & \phantom{a}heads\phantom{a} \\
\hline
\end{tabular}
\end{center}
Suppose we intervene on the system by forcing Alice to jump. According to Halpern's  definition of intervention \citep[][sec. 2.7]{Hal2016}, 
the possibile scenarios after such an intervention are those that agree 1) with the law, 2) with the condition $A=1$, 3) with the current state of the exogenous variables\footnote{\corrf{Roughly speaking, a variable is exogenous if causally unaffected by any other variable in the model. Later we will have a formal definition.}}\footnote{Applying Halpern's rule here is an extrapolation, since Halpern usually only considers interventions on endogenous variables. The reader who is troubled by this may extend the example with an additional variable $U$ representing the factors that determine Alice's decision, so that $A$ becomes an internal variable.} (different from $A$), i.e. $B=1$. There are two assignments consistent with these conditions, namely:
\begin{center}
\begin{tabular}{|c|c|c|}
\hline
 \multicolumn{3}{|l|}{ \ \ A \ \  \ \ \ \ B \ \ \ \ \ \ \ \ \ \ C } \\
\hline
 \phantom{a}1\phantom{a} & \phantom{a}1\phantom{a} & \phantom{a}heads\phantom{a} \\
\hline
 1 & 1 & tails \\
\hline
\end{tabular}
\end{center}
In other words, forcing Alice to jump \corrf{-- or imagining her to counterfactually jump --} makes us lose information about the outcome of a coin toss. This \corrf{may be unwanted, for example, if Alice's and Bob's feats take place, simultaneously, on distant planets}.
\end{example}
There seems to be a straightforward way to address this problem. \corrf{If Alice's and Bob's deeds are obviously unrelated in the scenario we wish to model, then $A$ should not be treated as a direct cause of $B$ (nor, more generally, there should be a path of direct causes from $A$ to $B$). This can be guaranteed using the approach 1. outlined at the beginning of the section.} 
Now, it seems reasonable that an intervention on a variable $A$ should only affect variables that are (directly or indirectly) causally dependent on $A$. 
Then, we should reject the second assignment in the table, which modifies the value of $C$, non-descendant of $A$. To this end, we will adopt a definition of intervention in the style of \citet{BarGal2022}, which, differently from Halpern's, does not violate this constraint on cyclic models.

There is more. As the following example shows, and contrarily to what happens in the deterministic case, identifying the dummy arguments of the relational laws is not sufficient for identifying the direct causes and parents of a variable.

\begin{example}[Two-coin Bob]\label{example: dummy parent}
    In this scenario Bob has two coins, say coin 1 and coin 2, and may toss one of the two ($B=1$ or $B=2$). Variable $O$ represents the outcome of the toss (heads or tails), and its behaviour is described by the relational law 
    \[
    \F_O = \{ (1,heads),(1,tails),(2,heads),(2,tails) \}.
    \]
    Furthermore, we know that Bob has tossed coin 1 and got heads:
\begin{center}
\begin{tabular}{|c|c|}
\hline
 \multicolumn{2}{|l|}{ \ \ B \ \ \ \ \ \ \ \ \ O } \\
\hline
 \phantom{a}1\phantom{a} & \phantom{a}heads\phantom{a} \\
\hline
\end{tabular}
\end{center}
Now, $B$ is a dummy argument of the law $\F_O$: changing the value of $B$ does not change the range of values that $O$ may attain ($\F_O(1) = \F_O(2) = \{heads, tails\}$). Seen as a multivalued function, $\F_O$ is a constant-valued law. But our intuition about the real world seems to disagree with the idea that $B$ is not a cause of $O$, and that $O$ should be considered external (uncaused) in this context. \corrf{Indeed, we presume that, if we force Bob to make another toss, the outcome may change. And we usually take the \emph{future} behaviour of a mechanism to be a reliable indicator of its \emph{counterfactual} behaviour. Thus,} $B$ should be considered a direct cause of $O$, even if it is a dummy argument of $\F_O$. \corrf{Our definition of intervention will ensure that} 
the intervention $do(B=1)$, forcing Bob to repeat the toss, will produce two alternative scenarios  
\begin{center}
\begin{tabular}{|c|c|}
\hline
 \multicolumn{2}{|l|}{ \ \ B \ \ \ \ \ \ \ \ \ O } \\
\hline
 \phantom{a}1\phantom{a} & \phantom{a}heads\phantom{a} \\
\hline
 1 & tails \\
\hline
\end{tabular}
\end{center}
where the range of possible values for $O$ has changed from $\{heads\}$ to $\{heads,tails\}$.
\end{example}

\noindent This second example should clarify that a relational causal law $\F_V$, by itself, determines the set of its own non-dummy arguments, but does not determine the (possibly larger) set of direct causes of $V$. Learning the set of direct causes is possible once we have a notion of intervention -- we may then check whether intervening on a given argument $Z$ may change the range of allowed values for $V$ in some model \corrf{(when everything else is held fixed)}. But on the other hand, to know whether intervening on $Z$ may affect $V$ we need to know whether $Z$ is a 
causal descendant of $V$. We can escape this vicious circle by declaring explicitly the set of parents of a variable as part of a specification of model, i.e. by following the approach 1. delineated at the beginning of this section (the set of direct causes of $Y$ will then be a subset of $PA_Y$). 

We remark that the problems raised by the two examples in this section are not artificially induced by switching from causal models to causal teams. The issue from example \ref{example: jumpin Alice} only concerns how an indeterministic causal law is modeled -- this only concerns the function component of a model. Example \ref{example: dummy parent} concerns a single-assignment model, which is the straightforward generalization to the indeterministic case of the causal models from the literature 
\citep[e.g.][]{Hal2016,Bri2012}. It will not arise by following the perspective of \cite{Hal2000}, in which all statements essentially concern the special case of teams that feature a single tuple of value for the exogenous variables and \emph{all} compatible tuples of values for the endogenous ones. We just take this to signal a limit to the expressiveness of said frameworks.\footnote{In any case, Halpern asserts (personal communication) that his intended semantics is that given in \cite{Hal2016}, not the earlier one from \cite{Hal2000}.}


\section{Formal framework}\label{sec: formal framework}

\subsection{Models: relational causal teams}\label{subs: relational causal teams}


The considerations from the previous section lead us to the following definition of a model. A \textbf{team} of signature $\sigma$ is a set of assignments of signature $\sigma$.


\begin{definition}
A 
\textbf{relational causal team} (of signature \corrf{$\sigma= (\mathcal E,\mathcal I,\ran)$)} 
is a pair $T = (T^-,\F)$, where:
\begin{itemize}
\item $T^-$ is a team of signature $\sigma$ (\emph{team component})
\item $\F$ (\emph{law component}) is a function that associates to each $V\in \corrf{\mathcal{I}}$: 
	\begin{itemize}
	\item a 
            set of variables $PA_V \subseteq \dom \setminus\{V\}$ (\emph{parents of $V$})
	\item a relation $\F_V \subseteq \ran(PA_V)\times\ran(V)$ ($V$-\emph{generating law})
	\end{itemize}
\item For all $V\in \corrf{\mathcal{I}}$, 
for all $s\in T^-$, $(s(PA_V),s(V))\in \F_V$ (\emph{compatibility constraint}).
\end{itemize}
\end{definition}
For brevity, we will often simply call it a \textbf{model}.
The last condition (the \textbf{compatibility constraint}) can be thought of as admitting in the team only ``solutions'' of the system of causal laws. 
\corrf{When dealing with a model $T$, we will often bypass mention of the signature by denoting its set of 
internal variables as $\inte(T)$.}



We remark that the causal law for an internal variable $V$ can also be, equivalently, represented by showing what set of values is associated by $\F_V$ to each tuple $pa$ in $PA_V$; we may denote such set as $\F_V(pa)$, thus treating the relation $\F_V$ as a multivalued function. \corrf{Note that we make exception to our earlier conventions and denote a tuple of values for the parents of a given variable by $pa,pa'$, and so on.}

We will say that a relational causal team is \textbf{total} if all causal laws are total multivalued functions, i.e.,  for each $V\in \inte(T)$ and for each $pa\in \ran(PA_V)$, $\F_V(pa)\neq\emptyset$. We will say it is \textbf{deterministic} if, for each $V\in \inte(T)$ and for each $pa\in \ran(PA_V)$, $|\F_V(pa)|\leq 1$.
The total deterministic models essentially coincide with the causal teams introduced in \citet{BarSan2020}. If we add the constraint that the team component must contain exactly one assignment, what we obtain are essentially the usual causal models.
\footnote{\corrf{As we shall see, however, our official definition of intervention will diverge from that used in \cite{BarSan2020} and \cite{Hal2016} when causal cycles are allowed -- even in the deterministic case.}} In general, a model with exactly one assingment will be called a \textbf{singleton} model; singleton models are the most straightforward generalization of causal models to the indeterministic case (we might also call them \textbf{indeterministic causal models}), and for some time they will be our main concern, in section \ref{sec: H language}. 
We will also say that a model is \textbf{empty} if such is its team component. We remark that empty models are not uninteresting: as we shall see, interventions on empty models may produce nonempty models.

We define the \textbf{\corrf{parenthood} graph} of $T$ to be the graph whose vertices are the variables of the domain, and where an arrow (directed edge) connects $X$ to $Y$ if and only if $X$ is a parent of $Y$. 
The parenthood graph allows us to apply some graph theoretic terminology to the variables; for example, we will say that $Y$ is a \emph{descendant} of $X$ if either $Y$ and $X$ are the same variable, or there is a path $X \rightarrow Z  \rightarrow \dots \rightarrow Z_n \rightarrow Y$ in the parenthood graph; otherwise, $Y$ is a \emph{nondescendant} of $X$.  \corrf{If the parenthood graph is acyclic we will say that the model is \textbf{strictly recursive}.} We will see later (section \ref{sec: PA, direct causes, dummy arguments}) how to define a notion of \emph{causal graph} closer to that used for the deterministic case (and related notions of endogeneity, exogeneity, recursivity).




Interventions can be defined similarly as in the deterministic case, using the parenthood graph as a guide. It was however argued in \citet{BarGal2022} that Halpern's general defining clause  \cite[from][]{Hal2000} leads to paradoxical outcomes: if the causal graph has cycles, intervening on a variable $X$ may affect a variable $Y$ that is not its descendant.  We then adopt the modifications suggested in \citet{BarGal2022}.\footnote{Judea Pearl seems to have been aware of this problem, as he adopts in \citet{Pea2017} a definition of intervention close to ours.} 
We will need some additional notational conventions. If $T^-$ is a team, and $\SET Y$ a nonempty set of variables in its domain, we let $T^-(\SET Y):= \{\SET y \in \ran(\SET Y) \mid \text{for some $s\in T^-$, } s(\SET Y) = \SET y \}$ be the set of (tuples of) values for $\SET Y$ that occur in $T^-$. 
Given a finite multiset of atomic formulas, say $X_1 =  x_1, \dots X_n = x_n$, we say it is \textbf{consistent} if, whenever $X_i$ and $X_j$ are the same variable, then $x_i$ and $x_j$ also coincide.  Given a consistent $\SET X =\SET x$, and a  relational causal team
$T = (T^-,\F)$ \corrf{of signature $(\mathcal E, \mathcal I,\ran)$}, the effect of an intervention $do(\SET X =\SET x)$ on $T$ is to produce a new model 
 $T_{\SET X =\SET x} = ((T_{\SET X = \SET x})^-, \F_{\SET X = \SET x})$ \corrf{of signature $(\mathcal E\cup \SET X, \mathcal I\setminus \SET X, \ran)$,} with components: 
\begin{itemize}
\item $\F_{\SET X = \SET x} := \F_{\upharpoonright \inte(T)\setminus \SET X}$ \hspace{20pt} (the restriction of $\F$ to $\inte(T)\setminus \SET X$)
\item $(T_{\SET X = \SET x})^- := \{s \in \A_\sigma \text{ compatible with } \F_{\SET X = \SET x}   \mid   s(\SET X) = \SET x \text{ and } s(\SET N_{\SET X})\in T^-(\SET N_\SET X)\}$
\end{itemize}
where $\SET N_{\SET X}$ is the set of nondescendants of $\SET X$ in the parenthood graph (in the special case where $\SET N_{\SET X} = \emptyset$, we conventionally take the condition $s(\emptyset) \in T^-(\emptyset)$ to be ``trivially true''). \corrf{Note also that the parenthood graph $G$ of $T$ is replaced with its subgraph $G_\SET X$ that omits all arrows going into $\SET X$ (i.e., $(V,Z)$ is an arrow of $G_\SET X$ iff it is an arrow of $G$ and $Z\notin \SET X$).} The definition above includes as a special case the empty intervention $do()$ corresponding to the empty multiset of atomic formulas. Such an intervention leaves both components of the causal team unchanged.\footnote{This is in contrast with Halpern's semantics, according to which new assignments may appear after an empty intervention (see \cite{BarGal2022}). In our case, this does not happen because our interventions leave unchanged the values of all nondescendants of the intervened variables, and in the case of empty interventions this set of variables is the whole variable domain.} 


\corrf{It is easy to prove from the definitions that}
the team component of intervened \emph{nonempty} teams can be presented more concretely as the union $(T_{\SET X = \SET x})^- = \bigcup_{s\in T^-} s_{\SET X = \SET x}^\F$, where each $s_{\SET X = \SET x}^\F$ is the outcome of the intervention applied to the single assignment $s$,\footnote{To see that nonemptyness is a necessary requirement, consider an arbitrary empty model $T$, say with variables $\SET W$. Now, $T_{\SET W = \SET w}$ is a singleton model (it contains the assignment $s(\SET W)=\SET w$), while $\bigcup_{s\in T^-} s_{\SET W = \SET w}^\F$, being the union of an empty family of sets, is empty.} which can be described 
\corrf{as follows:}
\begin{align*}
s_{\SET X = \SET x}^\F := \{ t\in\A_\sigma   \mid &  \forall X\in \SET X: t(X)= x \\
				 & \corrf{\forall V \in 
                \SET N_{\SET X}: t(V)=s(V)} \\
				& \forall V \in \inte(T)\setminus (\SET X \cup \SET N_{\SET X}): (t(PA_V),t(V)) \in \F_V \}. \\
\end{align*}
In the case of strictly recursive models, this presentation can be used to generate intervened models via a recursive procedure (analogously to the case of the usual recursive causal models).
Indeed, one can define the ``distance'' of a variable $Y$ from a tuple $\SET X$ as follows:
 \[
    \left\{
    \begin{array}{ll}
        d(\SET X, Y) = 0 & \text{if $Y\in\SET X \cup \SET N_\SET X$ } \\
         d(\SET X, Y) = \operatorname{max}\{d(\SET X, Z) \mid Z\in PA_Y\} +1 & \text{otherwise.} 
    \end{array}
    \right.     
\]
If $T=(T^-,\F)$ is strictly recursive (and the set of variables is finite), then clearly each variable has a finite distance from $\SET X$, and there is a maximum distance attained; furthermore, if $Z\in PA_Y$, then  $d(\SET X, Z) < d(\SET X, Y)$.  The team $T_{\SET X = \SET x}^-$ can then be produced by starting (step $0$) from the set $T_0$ of all assignments $t$ (over $\SET X \cup \SET N_\SET X$) such that $t(\SET X)=\SET x$ and $t(\SET N_\SET X)\in T^-(\SET N_\SET X)$. Supposing a team $T_n$ has been defined (over all variables $Y$ with $d(\SET X,Y)\leq n$) one then produces $T_{n+1}$ by adding to each assignment of $T_n$ all values of the $n+1$-distance variables that are compatible with $\F$. That is, if we write $\SET Z= (Z_1,\dots,Z_k)$ for a list of the variables at distance $n+1$ from $\SET X$, we may let $T_{n+1}:= \{t(\SET z/\SET Z) \mid t\in T_n, \SET z \in \F_{Z_{1}}(t(PA_{Z_1})) \times \dots \F_{Z_{k}}(t(PA_{Z_k})) \}$. Since our assumptions entail the existence of a maximum distance $\hat n$, this procedure ends, and we may let $T_{\SET X = \SET x}^- := T_{\hat n}$.\footnote{If some of the causal laws are not total, the procedure might also end before this stage, producing an empty model.}  We illustrate this construction in a simple case.

\begin{example}
Consider the following game. I can toss, or not,  a coin with my left hand ($L=0$ or $1$); the outcome is recorded by a variable $C_L$ with 3 possible values: heads (h), tails (t) or none (n) in case I do not toss. If I toss and get heads, then I will toss the coin that I hold in my right hand ($R=1$) whose outcome is recorded analogously in $C_R$. 
The parent sets and causal laws are then as follows: $PA_{C_L} = \{L\}$, $PA_{R} = \{C_L\}$, $PA_{C_R} = \{R\}$; $\F_{C_L} = \F_{C_R} =\{(0,n),(1,h),(1,t)\}$; $\F_R = \{(n,0),(h,1),(t,0)\}$. 
 The current situation is that neither coin has been tossed, which can be represented by a single assignment $s$:
\begin{center}
\begin{tabular}{|c|c|c|c|}
\hline
 \multicolumn{4}{|l|}{ \,  L \ $\strictif$ \ C$_L$ \ $\rightarrow$ R \ $\strictif$  C$_R$ } \\
\hline
 $\phantom{a}$0$\phantom{a}$ & $\phantom{a}$n$\phantom{a}$ & $\phantom{a}$0$\phantom{a}$ & $\phantom{a}$n$\phantom{a}$ \\
\hline
\end{tabular}
\end{center}

\noindent (the arrow $\rightarrow$ emphasizes that the law for $R$ is deterministic). Let $T = (\{s\}, \F)$. 
 To see what would have happened had I decided to toss with my left hand 
 (intervention $do(L=1)$), we need to update all columns (since all variables are descendants of $L$):
\begin{center}
\begin{tabular}{|c|c|c|c|}
\hline
\multicolumn{4}{|l|}{ \ \ L \ $\strictif\,$ \   C$_L$ \  $\rightarrow$ \  R \ \ $\strictif\,$  C$_R$} \\
\hline
 $\phantom{a}$\textbf{1}$\phantom{a}$ & \phantom{i}$\dots$\phantom{i} & $\phantom{i}\dots\phantom{i}$ & $\phantom{i}\dots\phantom{i}$ \\
\hline
\end{tabular}
\hspace{15pt} $\leadsto$ \hspace{15pt} 
\begin{tabular}{|c|c|c|c|}
\hline
 \multicolumn{4}{|l|}{  \ \ L \ $\strictif$ \ C$_L$ \ $\rightarrow$ \ R \  $\strictif$ \ C$_R$ } \\
\hline
 $\phantom{a}$1$\phantom{a}$ & $\phantom{a}$\textbf{h}$\phantom{a}$ & $\phantom{i}\dots\phantom{i}$ & $\phantom{i}\dots\phantom{i}$ \\
\hline
$\phantom{i}$1$\phantom{i}$ & $\phantom{i}$\textbf{t}$\phantom{i}$ & $\phantom{i}\dots\phantom{i}$ & $\dots$ \\
\hline
\end{tabular}
\end{center}

\begin{center}
    $\leadsto$ \hspace{15pt} 
\begin{tabular}{|c|c|c|c|}
\hline
 \multicolumn{4}{|l|}{  \ \ L \ $\strictif$ \ C$_L$  $\rightarrow$  R \  $\strictif$ \ C$_R$ } \\
\hline
 $\phantom{a}$1$\phantom{a}$ & $\phantom{a}$h$\phantom{a}$ & $\phantom{a}$\textbf{1}$\phantom{a}$ & $\phantom{i}\dots\phantom{i}$ \\
\hline
$\phantom{i}$1$\phantom{i}$ & $\phantom{i}$t$\phantom{i}$ & $\phantom{i}$\textbf{0}$\phantom{i}$ & $\dots$ \\
\hline
\end{tabular}
 \hspace{15pt} $\leadsto$ \hspace{15pt} $T_{L=1}$:
\begin{tabular}{|c|c|c|c|}
\hline
 \multicolumn{4}{|l|}{  \ \ L \ $\strictif$ \ C$_L$  $\rightarrow$  R \  $\strictif$ \  C$_R$ }\\
\hline
 $\phantom{a}$1$\phantom{a}$ & $\phantom{a}$h$\phantom{a}$ & $\phantom{a}$1$\phantom{a}$ & $\phantom{a}$\textbf{h}$\phantom{a}$ \\
\hline
$\phantom{i}$1$\phantom{i}$ & $\phantom{i}$h$\phantom{i}$ & $\phantom{i}$1$\phantom{i}$ & \textbf{t} \\
\hline
$\phantom{i}$1$\phantom{i}$ & $\phantom{i}$t$\phantom{i}$ & $\phantom{i}$0$\phantom{i}$ & \textbf{n} \\
\hline
\end{tabular}
\end{center}
\corrf{From the modified model we see, among other things, that if I had tossed the coin in my left hand, then all outcomes for the right-hand coin would have been possible, while outcome $n$ would have been impossible for the left-hand coin.}
\end{example}

It must be remarked that, while in \citet{BarGal2022} it could be proved that our definition of intervention and Halpern's coincide in the recursive deterministic case, the same does not hold in the (strictly) recursive \emph{in}deterministic case.\footnote{We thank Sander Beckers for this insight.} 

\begin{example}
    \corrf{
    Consider a model $T$ with four binary variables $X,Y, A,B$, $PA_Y = X$, $PA_B = A$, laws $\F_Y(x)=\{x\}$ (for all $x\in \ran(X)$) and $\F_B(a)=\{0,1\}$ (for all $a \in \ran(A)$), and the following team component, with a single assignment $s$:
    \begin{center}
\begin{tabular}{|c|c|c|c|}
\hline
 \multicolumn{4}{|l|}{ \,  X \ $\rightarrow$ \ Y \ \  \ \ \ \ \ A \ $\strictif$ \ B } \\
\hline
 $\phantom{a}$0$\phantom{a}$ & $\phantom{a}$0$\phantom{a}$ & $\phantom{a}$0$\phantom{a}$ & $\phantom{a}$0$\phantom{a}$ \\
\hline
\end{tabular}
\end{center}
Note that the parenthood graph, sketched above the table, is acyclic.
It is easy to see that, with our definitions, $T_{X=0} = T$. This is due to our requirement that only assignments that agree with $s$ on nondescendants of $X$ be included. If we had just required \citep[in the style of][]{Hal2000} agreement over the non-intervened exogenous variable ($A$), we would have obtained a larger team:
 \begin{center}
\begin{tabular}{|c|c|c|c|}
\hline
 \multicolumn{4}{|l|}{ \,  X \ $\rightarrow$ \ Y \ \ \ \  \ \ \ A \ $\strictif$ \ B } \\
\hline
 $\phantom{a}$0$\phantom{a}$ & $\phantom{a}$0$\phantom{a}$ & $\phantom{a}$0$\phantom{a}$ & $\phantom{a}$0$\phantom{a}$ \\
\hline
 $\phantom{a}$0$\phantom{a}$ & $\phantom{a}$0$\phantom{a}$ & $\phantom{a}$0$\phantom{a}$ & $\phantom{a}$1$\phantom{a}$ \\
\hline
\end{tabular}
\end{center}
}
\end{example}



\subsection{Formal languages}\label{subs: H language}

We consider a family of languages that is close to the language in \citet{Hal2000}. Given a consistent multiset  $\SET X = \SET x$, we call the expression $[\SET X = \SET x]$ a \emph{modal operator}.\footnote{Inside modal operators, we identify strings of symbols that represent the same multiset. This saves us the somewhat trivial issue of axiomatizing this form of equivalence. See \citet{BarSan2020} for a complete list of axioms for the antecedents of counterfactuals.} We also allow the empty multiset, in which case the operator will be written as $\Box$. For any given signature $\sigma = (\mathcal E, \mathcal I,\ran)$, we define a corresponding language $\mathcal H_\sigma$. The suffix $\sigma$ will be omitted when the signature is clear. 
\[
\text{Language } \mathcal H_\sigma: \hspace{20pt} 
X=x  \mid  \cneg\psi   \mid   \psi\with \chi    \mid   [\SET X=\SET x] \eta    
\] 
where $\SET X\cup\{X\} \subseteq \dom$, $x\in \ran(X)$, $\SET x\in \ran(\SET X)$, and $\eta$ has no occurrences of modal operators \corrf{(i.e., it is a Boolean combination of atoms like $X=x$)}. \corrf{For simplicity, and without loss of generality, we assume that $\SET X = \SET x$ is nonredundant (each variable appears in it at most once); it is then, \emph{a fortiori}, consistent.} 

Language $\Hal$ differs from Halpern's in some minor respects: 1) 
we allow for atomic formulas not prefixed by modal operators\footnote{In \citet{Hal2000} atomic formulas $Y = y$ are treated as abbreviations for $\Box Y=y$.} 
 2) 
we allow arbitrary variables in atomic formulas and inside the modal operators.\footnote{In \cite{Hal2000}, only \emph{endogenous} variables play an active role in the formal languages.} 
This alternative set of conventions is often followed in the philosophical literature \citep[e.g.][]{Bri2012}.

The restriction on $\eta$, which forbids the right-nesting of counterfactuals, comes from the original paper \cite{GalPea1998}, and has been preserved in large parts of the literature on causal inference. The reasons behind this self-imposed limitation can only be guessed; on one hand, it is interesting to note that, in the recursive deterministic case, it is easy to show that any nested counterfactual is equivalent to an unnested one \cite[see][]{Bri2012,BarSan2020}; thus, nested counterfactuals may appear as superfluous. On the other hand, this reduction is not straightforward if the recursivity constraint is removed; and in that case the axiomatization of right-nested counterfactuals is a considerably more difficult task than the unnested case.\footnote{This is illustrated by sections \ref{sec: law components} and \ref{sec: general axiomatization} below.} We will tackle this issue and consider, starting from section \ref{sec: law components}, the language $\Hal^+_\sigma$ that is obtained by allowing right-nesting of the modal operators. In other words, $\Hal^+_\sigma$ has the same formal definition as $\Hal_\sigma$, with the exception that no restriction is put on $\eta$. Thus, for example, formulas $[\SET X = \SET x][\SET Y= \SET y]W=w$ and  $[\SET X = \SET x](Z=z \with [\SET Y= \SET y]W=w)$ are $\Hal^+_\sigma$ formulas but not $\Hal_\sigma$ formulas.  This type of language is frequently used in the philosophical literature \citep[e.g.][]{Bri2012}.


 The semantics for languages $\Hal_\sigma$  and $\Hal^+_\sigma$ is given by the following clauses:
\begin{itemize}
\item $T\models X=x$ iff $s(X)=x$ for each $s\in T^-$.
\item $T\models \cneg \psi$ iff $T\not\models \psi$.
\item $T\models \psi \with  \chi$ iff $T\models \psi$ and $T\models \chi$.
\item $T\models [\SET X = \SET x]\psi$ iff for all $s\in T_{\SET X = \SET x}^-$, $(\{s\},\F)\models \psi$.
\end{itemize}
Formulas of the form $[\SET X = \SET x]\psi$ will be called \emph{counterfactuals}; formulas without occurrences of modal operators are said to be \emph{modal-free}.  
We can also define the \emph{might-counterfactual} $\langle\SET X = \SET x\rangle \psi$ 
as an abbreviation for $\cneg[\SET X = \SET x]\cneg\psi$; it is easy then to see that the semantics of such formulas is:
\begin{itemize}
\item $T\models \langle\SET X = \SET x\rangle \psi$ iff there is an $s\in T_{\SET X = \SET x}^-$ such that  $(\{s\},\F)\models \psi$.
\end{itemize}
As a special case, we write $\Diamond$ for $\cneg \Box \cneg$. Our definition of intervention yields the following derived semantic clauses:
\begin{itemize}
    \item $T\models \Box\psi$ iff for all $s\in T^-$, $(\{s\},\F)\models \psi$. 
    \item $T\models \Diamond\psi$ iff for some $s\in T^-$, $(\{s\},\F)\models \psi$. 
\end{itemize}

\noindent We can then define a few  additional operators:
\columnratio{0.607}
\setlength\columnsep{0pt}
\footnotelayout{m}
\begin{paracol}{2}
\begin{leftcolumn}
\begin{itemize}
    \item $\psi \sqcup \chi$ as $\cneg(\cneg\psi \with \cneg \chi)$
    \item $\psi\rightarrow \chi$ as $\cneg \psi\sqcup\chi$
    \item $\psi\leftrightarrow \chi$ as $(\psi\rightarrow \chi) \with  (\chi\rightarrow \psi)$
\end{itemize}
\end{leftcolumn}
\begin{rightcolumn}
 \begin{itemize}
    \item $\bot$ as $X=x \with \cneg X=x$\footnote{The reader used to team semantics must be warned that, with the present definition, $\bot$ is not satisfied by empty causal teams (it is satisfied by no model at all).}
    \item $\top$ as $\cneg \bot$
    \item $X\neq x$ as $\cneg X = x$
\end{itemize}   
\end{rightcolumn}
\end{paracol}
Note that  $X\neq x$ is just the classical negation of $X=x$, asserting that \emph{at least one assignment} does not assign $x$ to $X$.\footnote{This is in contrast with other papers involving causal team semantics.}

We will use index sets for iterated conjunctions and disjunctions; conventionally, if $I= \emptyset$, $\bigwith_{i\in I}\psi_i$ stands for $\top$ and $\bigsqcup_{i\in I}\psi_i$ stands for $\bot$. 



\section{An axiomatization for singleton models}\label{sec: H language}





\noindent

We will see that the general class of singleton models (which also allow for cyclic causation) obeys a set of axioms that is quite similar to that given by Halpern for the general class of \emph{deterministic} causal models \citep{Hal2000}. There is a significant omission: Halpern had an axiom stating that, if you intervene on all (endogenous) variables except one, say $Y$, then $Y$ will take a single value in the resulting model. This is false in our context, even restricting attention to singleton models, because if $Y$ is internal, an indeterministic law for $Y$ may produce multiple or no values upon intervention.\footnote{If we look at causal teams in general, also external variables disagree with Halpern's axiom. If $Y$ is external, a causal team may record more than one value for it, in distinct assignments; and many interventions do not make such values disappear.} On the other hand, we have a principle that describes the fact that the \emph{external} variables  
are not affected by interventions on \emph{all} other variables.  
It does so by saying that 
the available values for $Y$ are the same either before or after such an intervention:
\[
\langle \SET W_Y = \SET w\rangle Y=y \leftrightarrow \Diamond Y=y.
\]
This axiom scheme needs to be restricted to instances in which $Y$ is external according to the signature. This is the only axiom that requires having the distinction between $\mathcal E$ and $\mathcal I$ as part of the signature.\footnote{As shown e.g. in \citet{Bri2012,BarYan2022,BarVir2023b}, in the recursive deterministic case it is possible to give axiomatizations for simpler signatures that only describe the variable domain and ranges (we describe such signatures in section \ref{sec: simplified signatures}). We leave it as an open question whether this is possible in the indeterministic context, or even in the deterministic with cycles.} Halpern's system does not need such an axiom because the external (exogenous) variables do not appear in its syntax. 

%
A second new axiom (\emph{Flatness}) needs to be added to account for the fact that in $\mathcal H$ we also have non-modal formulas; 
it tells us that 
a formula $\SET Y = \SET y$ (which, remember, is a conjunction of aatoms) can be converted into modal statements $\Box \SET Y = \SET y$;
i.e., formula $\SET Y = \SET y$ just states that $\SET Y$ take values $\SET y$ \emph{in all assignments of the model}. A third new axiom (\emph{Full intervention}) states that intervening on all variables produces (at least) an assignment; we believe that such an axiom should have already been included in the axiomatizations for general deterministic models in \cite{Hal2000}.\footnote{It is difficult to establish unequivocally the correctness of this claim, not only because it is difficult to ascertain whether \emph{Full intervention} is derivable from Halpern's system, but also because of ambiguities in the definition of intervention in \cite{Hal2000} that make it difficult to decide whether this axiom is sound or not. However, in a personal communication, Halpern stated that his intended semantics for general deterministic models is that which is presented later in \cite{Hal2016}, section 2.7; and according to this semantics, the \emph{Full intervention} axiom is clearly sound. Halpern agreed that the axiom should be added, and he himself did so in later publications such as \cite{HalPet2022,BecHalHit2023}.} 
The fourth new axiom (\emph{Nonemptyness}) is specific for singleton models and states that they contain at least an assignment; it is simply $\Diamond\top$. We remark that Halpern's framework allowed evaluating formulas over assignments that are \emph{not} compatible with the causal laws; $\Diamond\top$ is unsound over such ``impossible models'', which here are excluded. 

For each signature $\sigma$ we define a Hilbert-style axiom system $\Ax_\sigma$ consisting of rules and axiom schemes, as follows; the symbol $\Ax_\sigma$ will also often be used just to denote the set of axioms. 

\begin{center}
\hspace{15pt} Rule MP. \Large$\frac{\psi \hspace{20pt}\psi \rightarrow \chi}{\chi}$\normalsize  
\hspace{80pt}
Rule NEC. \Large$\frac{\vdash\psi}{\vdash[\SET X = \SET x]\psi}$\footnotesize  (if $\psi$ modal-free) \normalsize
\end{center}

\noindent I0. $\Hal_\sigma$ instances of classical tautologies in $\cneg,\with$.


\noindent I1. 
$[\SET X = \SET x] Y=y \rightarrow [\SET X = \SET x] Y \neq y'$   \hspace{10pt}  (when $y\neq y'$)  \hspace{47pt} [Uniqueness]

 I1b. $Y=y \rightarrow  Y \neq y'$  \hspace{179pt} [Uniqueness] 

\noindent I2. 
$[\SET X = \SET x] \bigsqcup_{y\in\ran(Y)} Y=y$\hspace{171pt}   [Definiteness]

 I2b. $\bigsqcup_{y\in\ran(Y)} Y=y$\hspace{187pt}   [Definiteness]

\noindent I3. 
$\langle\SET X = \SET x\rangle  (Z = z \with  \SET Y = \SET y) \rightarrow \langle\SET X = \SET x,  Z =z \rangle \SET Y = \SET y$ \hspace{46pt} [Weak composition]

\noindent I4. 
$[\SET X = \SET x,  Y=y] Y=y$ \hspace{181pt} [Effectiveness]

\noindent I5. 
 $([\SET X = \SET x]\psi \with  [\SET X = \SET x](\psi\rightarrow \chi)) \rightarrow [\SET X = \SET x]\chi$ \hspace{71pt} [K-axiom]

\noindent I6. 
$(\langle \SET X = \SET x, V=v \rangle (Y=y \with \SET Z = \SET z) \with \langle \SET X = \SET x,Y=y \rangle(V=v \with \SET Z = \SET z)) \rightarrow$ 
$\langle \SET X = \SET x\rangle(V=v \with Y=y \with  \SET Z = \SET z)$ 

\hspace{40pt} (for $V\neq Y$, and $\SET Z = \dom \setminus (\SET X \cup \{V,Y\})$)    \hspace{38pt}  [Weak reversibility]



\noindent I7. 
$\SET Y = \SET y \leftrightarrow \Box \SET Y = \SET y$. \hspace{189pt} [Flatness] 


\noindent I8. 
$\langle \SET W_Y = \SET w\rangle Y=y \leftrightarrow \Diamond Y=y$ \hspace{10pt} (if $Y\in \mathcal E$) \hspace{84pt}  [External variables]

\noindent I9. $\langle \SET W = \SET w\rangle\top$ \hspace{225pt} [Full intervention]

\noindent I10. $\Diamond \top$ 
\hspace{254pt}  [Nonemptyness]

\noindent In the axioms, $\SET X$ and $\SET Y$ are allowed to be empty tuples; in such case the modal $[\SET X = \SET x]$ reduces to $\Box$, $\langle\SET X = \SET x\rangle$ to $\Diamond$, and e.g. the conjunction $\SET Y = \SET y$ is taken to stand for the formula $\top$. 

\vspace{5pt}

 Note that rule NEC is restricted to non-modal $\psi$; this is to guarantee that $[\SET X = \SET x]\psi$ will be an $\Hal$ formula. 
Axioms I0-I6 essentially coincide with the part of Halpern's axiomatization (for the general class of causal models) that is meaningful for our language and sound on indeterministic singleton models (we will see that I6 fails in causal teams with multiple assignments).\footnote{Axiom I3 is also analogous to the principle of \emph{Cautious monotonocity} in the field of nonmonotonic logic \citep{StrAnt2024}.} 
We can also see that the schemes I1b and I2b (which can be thought as no-intervention cases of I1 and I2) characterize the fact that the team contains at least one assignment, resp. at most one assignment.

\begin{lemma}\label{lemma: I1+I2 implies singleton}
$T$ is a singleton model iff for all $Y\in \mathcal E \cup \mathcal I$, 1) $T\models \bigsqcup_{y\in\ran(Y)} Y=y$ and 2) for all $y,y'\in \ran(Y)$ with $y\neq y'$, $T\models Y=y \rightarrow Y\neq y'$. 

Furthermore, condition 2) may be replaced by 2') $T\models \Diamond\top$.
\end{lemma}

\begin{proof}
  $\Leftarrow$) Suppose $T$ is empty; then, $T\models Y=y \with Y=y'$, contradicting 2).

  Since $T$ satisfies 1), its team component is constant over all variables in the domain; thus, it can contain at most one assignment.

  $\Rightarrow$) Straightforward.\footnote{It is a special case of the forthcoming soundness theorem \ref{thm: soundness}.} 

  Finally, it is easy to see that $\Diamond \top \equiv Y=y \rightarrow Y\neq y'$. 
\end{proof}

\noindent It seems to us, however, that these two non-modal formula schemes are insufficient for deriving formally all the modal consequences of the nonemptyness of the model; that is why we add axiom scheme I10.

 We shall write 
$\vdash_{\Ax_\sigma}$ 
for derivability in axiom system $\Ax_\sigma$. 
We shall just write $\Ax, \vdash_{\Ax}$ when the signature is irrelevant (or even just $\vdash$ if it is clear that we are discussing derivability in $\Ax$).
We will see that $\Ax_\sigma$ is sound and complete over the class of all singleton models of signature $\sigma$.

\begin{theorem}\label{thm: soundness}
    Axiom system $\Ax$ is sound over singleton models.

    Furthermore, all axioms except for I6 and I10 are sound over the class of all models.    
\end{theorem}


\begin{proof}
    We consider the least intuitive axioms, namely I3 (weak composition), I6 (weak reversibility), I8 (external variables) and I9 (full intervention); the rest is left to the reader.


Axiom I3 (weak composition): $T =(T^-,\F)\models \langle\SET X = \SET x\rangle  (Z=z \with  \SET Y = \SET y)$ means that there is an $s\in T_{\SET X = \SET x}^-$ such that $s(Z\SET Y) = z\SET y$. 
Now, 
$s$ is compatible with $\F_{\SET X = \SET x,Z=z}\subseteq \F_{\SET X = \SET x}$. Furthermore, $s(\SET N_{\SET X})\in T^-(\SET N_{\SET X})$, so since $\SET N_{\SET XZ}$ is a subtuple of $\SET N_{\SET X}$, $s(\SET N_{\SET XZ})\in T^-(\SET N_{\SET XZ})$. Then, by definition of intervention, $s\in T_{\SET X = \SET x, Z=z}^-$; thus, since $s(\SET Y)=\SET y$, $T\models \langle\SET X = \SET x,  Z =z \rangle \SET Y = \SET y$.


Axiom I6 (weak reversibility): Suppose that $T = (\{t\},\F)$ safisfies $\psi: \langle \SET X = \SET x, V=v \rangle (Y=y \with \SET Z = \SET z)$ and $\chi: \langle \SET X = \SET x,Y=y \rangle(V=v \with \SET Z = \SET z)$. By $\psi$, there is an $s \in T_{\SET X = \SET x, V=v}^-$ such that $s(Y)=y$ and $s(\SET Z)=\SET z$. 
But (since $\SET X \cup \SET Z \cup \{Y\}\cup \{V\} = \dom $) $\chi$ tells us that $s$ is also in $T_{\SET X = \SET x,Y=y}^-$. In particular, if $V$ is 
\corrf{internal}, $s$ is compatible also with the law $\F_V$; if we prove that $s(\SET N_{\SET X}) \in T^-(\SET N_{\SET X})$ (i.e., $s(\SET N_{\SET X}) = t(\SET N_{\SET X})$), then we can conclude that $s\in T_{\SET X = \SET x}^-$. But this follows from the fact that $\SET N_{\SET X} = \SET N_{\SET XY} \cup \SET N_{\SET XV}$, $s(\SET N_{\SET XY}) = t(\SET N_{\SET XY})$ (since $s\in T_{\SET X = \SET x,Y=y}^-$ and $t$ is the unique assignment in $T^-$) 
and $s(\SET N_{\SET XV}) = t(\SET N_{\SET XV})$ (for similar reasons). 
Since $s\in T_{\SET X = \SET x}^-$, then, we have $T\models \langle \SET X = \SET x\rangle(V=v \with Y=y \with  \SET Z = \SET z)$. 


Axiom I8 (\corrf{external} variables): 
Let $Y$ be an external variable.

Assume furthermore that $T\models \Diamond Y=y$. Then there is an $s\in T^-$ such that $s(Y)=y$. 
\corrf{Since $Y$ is external, by the definition of intervention} the assignment $t(\SET W_Y) = \SET w, t(Y)=y$ is in $T_{\SET W_Y = \SET w}^-$. Thus, $T\models \langle \SET W_Y = \SET w\rangle Y=y$.

Vice versa, assume $T\models\langle \SET W_Y = \SET w\rangle Y=y$. Then, there is a $t\in T_{\SET W_Y = \SET w}^-$ with $t(Y)=y$. But then, \corrf{since $Y$ is external, by definition of intervention $y\in T^-(Y)$.}
Thus, $T\models \Diamond Y=y$.

Axiom I9 (full intervention): Let $T =(T^-,\F)$ be a model. Since $\F_{\SET W = \SET w} = \emptyset$, the assignment $s$ with $s(\SET W)=\SET w$ is compatible with $\F_{\SET W = \SET w}$. Furthermore, in the parenthood graph of $\F_{\SET W = \SET w}$ the set $\SET N_\SET W$ of nondescendants of $\SET W$ is empty; thus, the condition $s(\SET N_\SET W)\in T^-(\SET N_\SET W)$ is trivially satisfied by $s$. Thus, by the definition of intervention, $s \in (T_{\SET W=\SET w})^-$. Thus, $T_{\SET W=\SET w}\models \Diamond \top$, so $T\models \langle\SET W=\SET w\rangle \top$.  
\end{proof}

\noindent In general, models can have empty teams, so that explains why I10 does not hold for all models. The following is a minimal example of a model that falsifies weak reversibility (axiom I6).\footnote{We thank an anonymous reviewer for noticing the failure of weak reversibility and illustrating it with a more complex counterexample.} 

\begin{example}
    Consider a model $T$ with two binary external variables, $L$ and $R$, say representing whether a switch on the left (resp. a switch on the right) has been pressed. We know that exactly one switch has been pressed. We thus have the following team:
    \begin{center}
    \begin{tabular}{|c|c|}
    \hline
    \multicolumn{2}{|l|}{ \ \  L \ \ \ \ \ \   R } \\
    \hline
    \phantom{a}1\phantom{a} & \phantom{a}0\phantom{a} \\
    \hline
    0 & 1 \\
    \hline
    \end{tabular}
    \end{center}
    Intervening to press either the left of the right switch will produce the following teams:
    \begin{center}
    $T_{L=1}$: \begin{tabular}{|c|c|}
    \hline
    \multicolumn{2}{|l|}{ \ \ L \ \ \ \ \ \   R } \\
    \hline
    \phantom{a}\textbf{1}\phantom{a} & \phantom{a}0\phantom{a} \\
    \hline
    \textbf{1} & 1 \\
    \hline
    \end{tabular}
    \hspace{40pt} $T_{R=1}$: \begin{tabular}{|c|c|}
    \hline
    \multicolumn{2}{|l|}{ \ \  L \ \ \ \ \ \   R } \\
    \hline
    \phantom{a}1\phantom{a} & \phantom{a}\textbf{1}\phantom{a} \\
    \hline
    0 & \textbf{1} \\
    \hline
    \end{tabular}
    \end{center}
    From the tables, it is immediate to see that $T\models \langle L=1 \rangle R=1$, $T\models \langle R=1\rangle L=1$ but $T\not\models \Diamond (L=1 \with R=1)$. Thus, the simplest instance of axiom I6 is falsified.
\end{example}

The following lemma lists basic properties of $\Ax$ (proofs can be found in the appendix). We note that these are essentially results in the modal logic K, as their proofs only use MP, NEC and axioms I0 (instances of tautologies) and I5 (K-axiom). 

\begin{restatable}{lemma}{BasicProperties}\label{lemma: basic properties of A}
The following hold in $\Ax$.
    \begin{enumerate}
        \item (Deduction theorem) If $\Gamma,\psi \vdash \chi$, then $\Gamma \vdash \psi\rightarrow \chi$.

        \item ($\Box$-Monotonicity) If $\Gamma\vdash[\SET X = \SET x]\psi$, $\vdash \psi \rightarrow \psi'$ and $[\SET X = \SET x]\psi'\in\Hal$, then $\Gamma\vdash [\SET X = \SET x]\psi'$.
        
        \item ($\Diamond$-Monotonicity) If $\Gamma\vdash \langle\SET X = \SET x\rangle\psi$, $\vdash \psi \rightarrow \psi'$ and $\langle\SET X = \SET x \rangle\psi' \in\Hal$, then $\Gamma\vdash \langle\SET X = \SET x \rangle\psi'$.

        \item  $\vdash([\SET X = \SET x]\psi \with  [\SET X = \SET x]\chi) \leftrightarrow [\SET X = \SET x](\psi \with \chi)$.

        \item (Replacement) Suppose $\Gamma\vdash \theta \leftrightarrow \theta'$. Then $\Gamma\vdash \varphi \leftrightarrow \varphi[\theta'/\theta]$, provided this is an $\Hal$ formula.
        
        \item $\vdash\cneg[\SET X = \SET x]\psi \leftrightarrow \langle\SET X = \SET x\rangle\cneg\psi$
       
        \item $\vdash\cneg\langle\SET X = \SET x\rangle\psi \leftrightarrow [\SET X = \SET x]\cneg\psi$

        \item $\vdash(\langle\SET X = \SET x\rangle \psi \sqcup \langle\SET X = \SET x\rangle \chi) \leftrightarrow \langle\SET X = \SET x\rangle(\psi\sqcup\chi)$.

         \item  $\vdash([\SET X = \SET x]\psi \with \langle\SET X = \SET x\rangle \top)\rightarrow \langle\SET X = \SET x\rangle\psi$.
    


        \item $\vdash([\SET X = \SET x]\psi \with \langle \SET X = \SET x \rangle \chi) \rightarrow \langle \SET X = \SET x \rangle (\psi \with \chi)$.

        \item $\vdash\langle\SET X = \SET x\rangle(\psi \with \chi) \rightarrow (\langle\SET X = \SET x\rangle\psi \with \langle\SET X = \SET x\rangle\chi)$.
    \end{enumerate}
\end{restatable}

We say that a set of $\mathcal H_\sigma$ formulas is ($\Ax_\sigma$)-\textbf{consistent} if $\Gamma \not\vdash_{\Ax_\sigma} \bot$. 
$\Gamma$ is \textbf{maximally consistent} if it is consistent and, furthermore, if $\Gamma'\supseteq \Gamma$ is consistent, then $\Gamma' = \Gamma$. 
We will later consider also a proof system $\Bx$ for the $\Hal^+$ languages. For both systems the following classical results hold:




\begin{restatable}{lemma}{MCSProperties}\label{lemma: properties of MCS}
Let $\Gamma$ be a maximally consistent set of $\mathcal H_\sigma$ formulas (or $\Hal^+_\sigma$ formulas). Then:
\begin{enumerate}
\item Closure under $\vdash$: if $\Gamma\vdash\psi$, then $\psi\in\Gamma$.
\item Completeness: for every formula $\psi$, either $\psi$ or $\cneg\psi$ is in $\Gamma$.
\item Closure under $\with $: if $\psi,\chi\in\Gamma$, then $\psi\with \chi\in\Gamma$. 
\item Primality: if $\psi \sqcup \chi\in\Gamma$, then $\psi\in\Gamma$ or $\chi\in\Gamma$.
\item Impossibility of contradiction: $\Diamond \bot \notin\Gamma$.
\end{enumerate} 
\end{restatable}

\begin{restatable}[Lindenbaum]{lemma}{Lindenbaum}\label{lemma: Lindenbaum}
Any consistent set $\Delta$ of $\mathcal H_\sigma$ formulas (resp. $\Hal^+_\sigma$ formulas) can be extended to a maximal consistent set.
\end{restatable}

The canonical relational causal team associated to a maximal consistent set $\Gamma$ (of formulas of a fixed signature $\sigma = (\mathcal E, \mathcal I, \ran)$), which we shall denote as $\T^\Gamma=((\T^\Gamma)^-,\F^\Gamma)$, is defined as follows:

\begin{itemize}
\item  We let $(\T^\Gamma)^- := \{ s\in \A_\sigma \mid \Diamond \SET W = s(\SET W)\in \Gamma\}$.
\item For each pair of variables $X,Y$ \corrf{with $Y\in \mathcal I$} we let $X\in PA_Y$ if and only if 
there are $\SET w,y,x \in \ran(\SET W_{XY},Y,X)$ such that either of the following holds:
  \begin{enumerate}
  \item $\langle\SET W_{XY} = \SET w\rangle Y=y \with \cneg\langle\SET W_{XY} = \SET w,X=x\rangle Y= y\in\Gamma$
  \item $\langle\SET W_{XY} = \SET w,X=x\rangle Y= y \with \cneg\langle\SET W_{XY} = \SET w\rangle Y=y  \in \Gamma$
  \end{enumerate}
  
\item In case $Y\in \mathcal I$, 
we let $(pa,y)\in \F^\Gamma_Y$  if and only if, for any $\SET w\in \ran(\SET W_{Y})$ such that $\SET w_{\upharpoonright PA_Y} =pa$, $\langle \SET W_Y=\SET w \rangle Y=y \in \Gamma$.



\end{itemize}

\begin{restatable}[Canonical team]{lemma}{CanonicalTeam}\label{lemma: canonical team}
Let $\Gamma$ be a maximal 
consistent set of $\Hal_\sigma$ formulas that contains all axioms of $\Ax_\sigma$. Then $\T^\Gamma$ is a singleton model, i.e.:
\begin{enumerate}
\item For all $Y\in \inte(\T^\Gamma)$, $\F_Y^\Gamma$ is well-defined.
\item For all $Y\in \inte(\T^\Gamma)$ and $s\in (\T^\Gamma)^-$, $(s(PA_Y),s(Y))\in \F^\Gamma_Y$.
\item $(\T^\Gamma)^-$ is a singleton.
\end{enumerate}
\end{restatable}

\begin{proof}

1) We need to show that the specific choice of a $\SET w$ extending $pa$ does not matter for the definition of $\F_Y^\Gamma$. In other words, we need to prove that, if $\SET w,\SET w' \in \ran(\SET W_Y)$ are such that $\SET w_{\upharpoonright PA_Y} = \SET w'_{\upharpoonright PA_Y} = pa$, then 
\[
\langle \SET W_Y = \SET w\rangle Y=y \in \Gamma \text{ if and only if } \langle \SET W_Y = \SET w'\rangle Y=y \in \Gamma.
\]
We prove this by induction on the number $n$ of variables on which $\SET w,\SET w'$ differ. If $n=0$, the statement is trivial. If $n=1$, $\SET w,\SET w'$ differ over a single variable $X\notin PA_Y$. Write $\SET w^*$ for $\SET w_{\upharpoonright W_{XY}} = \SET w'_{\upharpoonright W_{XY}}$. Assume $\langle \SET W_Y = \SET w\rangle Y=y \in \Gamma$. Since $X\notin PA_Y$, by the definition of $PA_Y$ in the canonical model, clause 2.,  we have then that $\cneg\langle \SET W_{XY} = \SET w^*\rangle Y=y \notin \Gamma$. But then, by lemma \ref{lemma: properties of MCS}, 2., 
$\langle \SET W_{XY} = \SET w^*\rangle Y=y \in \Gamma$. Again, since $X\notin PA_Y$, by clause 1., $\cneg\langle \SET W_Y = \SET w'\rangle Y=y \notin \Gamma$; so, similarly as before, we conclude $\langle \SET W_Y = \SET w'\rangle Y=y \in \Gamma$. The converse is analogous.



Now 
assume that the statement holds for some $n\geq 1$, and suppose $\SET w, \SET w'$ differ on $n+1$ variables $X_1,\dots, X_{n+1}$ (which take values  $x_1,\dots, x_{n+1}$ in $\SET w$ and values $x'_1,\dots, x'_{n+1}$ in $\SET w'$). Write $\SET Z$ for $\SET W_Y \setminus \{X_1, \dots, X_{n+1}\}$ and $\SET z$ for $\SET w_{\upharpoonright \SET Z}$. 
By the inductive hypothesis (for case $n$) we have:
\[
\langle \SET W_Y = \SET w\rangle Y=y \in \Gamma \text{ if and only if } \langle \SET Z X_1\dots X_n X_{n+1} = \SET z x'_1\dots x'_n x_{n+1}\rangle Y=y \in \Gamma 
\]
Since $\SET z x'_1\dots x'_n x_{n+1}$ and $\SET w'$ differ only on one variable ($X_{n+1}$), 
by the base case proved above we obtain that $ \langle \SET Z X_1\dots X_n X_{n+1} = \SET z x'_1\dots x'_n x_{n+1}\rangle Y=y\in\Gamma$ if and only if  $\langle \SET W_Y = \SET w'\rangle Y=y \in \Gamma$.

2) Let $s\in (\T^\Gamma)^-$. By definition of $(\T^\Gamma)^-$, $\Diamond \SET W = s(\SET W)\in\Gamma$. Thus, by axiom I3, MP, and lemma \ref{lemma: properties of MCS}, 1.,  $\langle \SET W_Y = s(\SET W_Y)\rangle Y = s(Y)\in \Gamma$. Since the restriction of $s(\SET W_Y)$ to $PA_Y$ is $s(PA_Y)$, then, by definition of $\F^\Gamma$ we have $(s(PA_Y),s(Y))\in \F^\Gamma_Y$. 

3) Let us first prove that $(\T^\Gamma)^-$ is nonempty. By axiom I2b, $\bigsqcup_{\SET w \in \ran(\SET W)} \SET W = \SET w \in \Gamma$. By the primality of $\Gamma$ (lemma \ref{lemma: properties of MCS}, 4.), there is a $\SET w^*\in \ran(\SET W)$ such that $\SET W = \SET w^* \in \Gamma$. By I7, $\Box\SET W = \SET w^* \in \Gamma$. Since, by I10, $\Diamond\top \in \Gamma$, by lemma \ref{lemma: basic properties of A}, 9., we obtain  $\Diamond\SET W = \SET w^* \in \Gamma$. But then, by definition, the assignment $s$ with $s(\SET W) = \SET w^*$ is in $(\T^\Gamma)^-$.

Suppose for the sake of contradiction that there are two distinct assignments, $s_1$ and $s_2$, both contained in $(\T^\Gamma)^-$. Then, by definition of $(\T^\Gamma)^-$ both $\Diamond \SET W = s_1(\SET W)\in\Gamma$ and $\Diamond \SET W = s_2(\SET W)\in\Gamma$. 
As proved above, $\Box\SET W = \SET w^* \in \Gamma$ for some $\SET w^*\in \ran(\SET W)$. Now, $s_1(\SET W)$ and $s_2(\SET W)$ cannot be both equal to $\SET w^*$; wlog, suppose $s_2(\SET W) \neq \SET w^*$; in particular, there is a variable $Y\in \SET W$ such that $s_2(Y)\neq y^* = \SET w^*_{\upharpoonright Y}$. By axiom I1b  
$\vdash Y=y^* \rightarrow \cneg Y=s_2(Y)$, and by I0, 
 $\vdash\SET W = \SET w^* \rightarrow Y=y^*$. Thus, by two applications of monotonicity (lemma \ref{lemma: basic properties of A}, 2.) we obtain (*): $\Box \cneg Y=s_2(Y)\in\Gamma$. On the other hand, since $\vdash \SET W = s_2(\SET W) \rightarrow Y = s_2(Y)$, by  monotonicity (lemma \ref{lemma: basic properties of A}, 3.) we have $\Diamond  Y = s_2(Y)\in\Gamma$. Thus, by lemma \ref{lemma: basic properties of A}, 10., $\Diamond(Y = s_2(Y) \with \cneg Y = s_2(Y))\in\Gamma$. By monotonicity again, we obtain $\Diamond \bot\in\Gamma$, contradicting lemma \ref{lemma: properties of MCS}, 5. 
\end{proof}

In order to prove a truth lemma, we need first to show that $\mathcal H$ formulas can be put, up to equivalence, in a simple normal form, similarly as in \citet{Hal2000}.




\begin{restatable}{lemma}{RemoveRedundancies}\label{lemma: remove redundancies in diamonds}
Let $w,w'\in \ran(W)$, $w\neq w'$. Then,
\begin{enumerate}
 \item $\vdash  \langle \SET X = \SET x, W=w \rangle (W=w \with \chi) \leftrightarrow \langle \SET X = \SET x, W=w\rangle \chi$.
\item $\vdash \langle \SET X = \SET x, W=w \rangle (W=w' \with \chi) \leftrightarrow \langle \SET X = \SET x, W=w \rangle \bot.$
\end{enumerate} 
\end{restatable}


\begin{proof}
1., $\Rightarrow$) By I0, $\vdash (W=w \with \chi) \rightarrow \chi$. Thus, by monotonicity (lemma \ref{lemma: basic properties of A}, 3.), $\vdash  \langle \SET X = \SET x, W=w \rangle (W=w \with \chi) \rightarrow \langle \SET X = \SET x, W=w\rangle \chi$.

1., $\Leftarrow$) Assume $\cneg \langle \SET X = \SET x, W=w \rangle (W=w \with \chi)$, i.e. $\cneg\cneg [\SET X = \SET x, W=w] \cneg (W=w \with \chi)$. By I0 we get $[\SET X = \SET x, W=w] \cneg (W=w \with \chi)$. On the other hand, by I4, $[\SET X = \SET x, W=w]W=w$. Thus, by lemma \ref{lemma: basic properties of A}, 4.,  $[\SET X = \SET x, W=w] (W=w \with \cneg (W=w \with \chi))$. Then, by I0 and monotonicity, $[\SET X = \SET x, W=w]\cneg \chi$. By I0, $\cneg \cneg [\SET X = \SET x, W=w]\cneg \chi$, which is the same as $\cneg \langle \SET X = \SET x, W=w \rangle \chi$.

2., $\Leftarrow$) By I0 and monotonicity.

2.,  $\Rightarrow$) By axiom I4, $\vdash [\SET X = \SET x, W=w] W=w$. Then, by I1, $\vdash [\SET X = \SET x, W=w] \cneg W=w'$. Together with $\langle \SET X = \SET x, W=w \rangle (W=w' \with \chi)$, by lemma \ref{lemma: basic properties of A}, 10., 
we obtain $\langle \SET X = \SET x, W=w \rangle (\cneg W=w' \with (W=w' \with \chi))$. Then, by I0 and replacement,  $\langle \SET X = \SET x, W=w \rangle \bot$.
\end{proof}

\begin{lemma}\label{lemma: eliminating negations} \

\corrf{$\vdash [\SET X = \SET x]\varphi \leftrightarrow [\SET X = \SET x]\varphi[\bigsqcup_{y'\in \ran(Y)\setminus\{y\}} Y=y'/\cneg Y=y]$.}
\end{lemma}

\begin{proof}
By replacement, it suffices to prove $\vdash (\bigsqcup_{y'\in \ran(Y)\setminus\{y\}} Y=y')\leftrightarrow \cneg Y=y$.

For the left-to-right direction, using I0 we have $\bigwith_{y'\in \ran(Y)\setminus\{y\}}(Y=y'\rightarrow \cneg Y=y) \rightarrow [(\bigsqcup_{y'\in \ran(Y)\setminus\{y\}}Y=y')\rightarrow \cneg Y=y]$. But, by axiom I1b, for each  $y'\neq y$ we have $\vdash Y=y' \rightarrow \cneg Y=y$; thus, $\vdash \bigwith_{y'\in \ran(Y)\setminus\{y\}}(Y=y'\rightarrow \cneg Y=y)$ by I0. Finally we obtain $(\bigsqcup_{y'\in \ran(Y)\setminus\{y\}} Y=y')\rightarrow \cneg Y=y$ by MP.

From right to left, assume $\cneg Y=y$. By axiom I2b, we obtain $\bigsqcup_{y'\in\ran(Y)}Y=y'$. By I0, then, we have $\cneg Y=y \with \bigsqcup_{y'\in\ran(Y)}Y=y'$; but then, again by I0 and MP, we obtain $\bigsqcup_{y'\in \ran(Y)\setminus\{y\}} Y=y'$.
\end{proof}

\begin{lemma}\label{lemma: adding trivial conjuncts}
    \corrf{Consider a formula $\varphi: [\SET X = \SET x]\bigsqcup_{i\in I}\bigwith_{j\in J_i\ } \chi_i^j$ and let $\varphi'$ be obtained by replacing a disjunct $\bigwith_{j\in J_{i^*}\ } \chi_{i^*}^j$ with $(\bigsqcup_{z\in \ran(Z)} Z = z) \with \bigwith_{j\in J_{i^*}\ } \chi_{i^*}^j$. Then, $\vdash \varphi \leftrightarrow \varphi'$.}
\end{lemma}

\begin{proof}
By I2b, $\vdash \bigsqcup_{z\in \ran(Z)} Z = z$. By I0, $\vdash (\bigsqcup_{z\in \ran(Z)} Z = z) \rightarrow (\bigwith_{j\in J_{i^*}\ } \chi_{i^*}^j \leftrightarrow  [(\bigsqcup_{z\in \ran(Z)} Z = z) \with \bigwith_{j\in J_{i^*}\ } \chi_{i^*}^j])$.  Thus, $\vdash \bigwith_{j\in J_{i^*}\ } \chi_{i^*}^j \leftrightarrow  [(\bigsqcup_{z\in \ran(Z)} Z = z) \with \bigwith_{j\in J_{i^*}\ } \chi_{i^*}^j]$ by MP. Then, we obtain the statement by replacement.  
\end{proof}


\begin{restatable}[Normal form]{lemma}{NormalForm}\label{lemma: normal form}
Every $\mathcal H_\sigma$ formula $\varphi$ is provably equivalent to a Boolean combination of formulas 
 $\langle\SET X = \SET x\rangle \SET Y= \SET y$, where $\SET X \cap \SET Y = \emptyset$ and $\SET X \cup \SET Y = \dom $.
\end{restatable}


\begin{proof}
First observe that all subformulas of the form $\SET Z = \SET z$ that do not occur in the scope of a modal operator can be replaced, by axiom I7 and replacement, with 
$\Box\SET Z = \SET z$. 
Secondly, if $\varphi$ has subformulas of the form $[\SET X =\SET x]\psi$, by I0 and replacement they can be replaced with $\cneg\cneg[\SET X =\SET x]\cneg\cneg\psi$, i.e. $\cneg\langle\SET X =\SET x\rangle\cneg\psi$. So, $\varphi$ is equivalent to a Boolean combination of formulas of the form $\theta:\langle\SET X =\SET x\rangle \psi$ (where $\SET X$ might also be an empty tuple).

Next, by I0 and replacement  such $\psi$ can be rewritten in disjunctive normal form, i.e. $\bigsqcup_{i\in I}  \bigwith_{j\in J_i}\ \chi_i^j$, where each $\chi_i^j$ is either of the form $Y=y$ or $\cneg Y=y$. 

Now, by lemma \ref{lemma: eliminating negations}, each conjunct of the form $\cneg Y=y$ can be rewritten as $\bigsqcup_{y'\in \ran(Y)\setminus\{y\}}Y=y'$. Furthermore, if a disjunct $\bigwith_{j\in J_i}\ \chi_i^j$ does not mention some variable $Z \in \dom \setminus \SET X$, by lemma \ref{lemma: adding trivial conjuncts} 
we can add to it a conjunct of the form $\bigsqcup_{z\in \ran(Z)}Z=z$.

Applying distributivity of $\with $ over $\sqcup$ (i.e. I0 plus replacement), we transform $\theta$ into a formula of the form $\langle \SET X = \SET x \rangle \bigsqcup_{i\in I'} \SET Y = \SET y_i$, where $\SET X\cup\SET Y = \dom $. Then, by lemma \ref{lemma: basic properties of A}, 8., we rewrite it as $\bigsqcup_{i\in I'}\langle \SET X = \SET x \rangle  \SET Y = \SET y_i$. We can, finally, eliminate all the conjuncts (within the subformulas $ \SET Y = \SET y_i$) that contain variables of $\SET X$ by using lemma \ref{lemma: remove redundancies in diamonds} 
(plus replacement).
\end{proof}




    

\begin{restatable}[Truth lemma]{lemma}{TruthLemma}\label{lemma: truth lemma}
Let $\Gamma\supseteq \Ax$ be a maximally consistent set of $\mathcal H_\sigma$ formulas, and $\varphi$ an $\mathcal H_\sigma$ formula. Then, $\varphi \in \Gamma \iff \T^\Gamma \models \varphi$.
\end{restatable}

\begin{proof}
By lemma \ref{lemma: normal form}, lemma \ref{lemma: properties of MCS}, 1. and soundness, we can assume that $\varphi$ is a Boolean combination of formulas of the form $\langle\SET X = \SET x\rangle \SET Y = \SET y$, with $\SET X \cap \SET Y = \emptyset$ and $\SET X \cup \SET Y = \dom $.

We proceed by induction on $\varphi$.

Case $\varphi$ is $\psi \with \chi$. Then, $\T^\Gamma \models \psi \with \chi$ iff $\T^\Gamma\models\psi$ and $\T^\Gamma \models \chi$ iff (i.h.) $\psi\in\Gamma$ and $\chi\in\Gamma$, iff (by lemma \ref{lemma: properties of MCS}, 3.) $\psi\with\chi\in\Gamma$.

Case $\varphi$ is $\cneg\psi$. Then, $\T^\Gamma \models \cneg\psi$ iff $\T^\Gamma \not\models \psi$ iff (i.h.) $\psi\notin \Gamma$ iff (lemma \ref{lemma: properties of MCS}, 2.) $\cneg\psi \in \Gamma$.

Case $\varphi$ is $\langle\SET X = \SET x\rangle \SET Y = \SET y$. We proceed by a subinduction on $n = |\dom  \setminus \SET X| =|\SET Y|$.
\begin{itemize}
\item Case $n=0$. This is straightforward: we both have $\T^\Gamma \models \langle\SET X = \SET x\rangle \top$ by the definition of intervention, and $\langle\SET X = \SET x\rangle \top \in \Gamma$ by axiom I9.



\item Case $n=1$. Suppose $\langle\SET X = \SET x\rangle  Y =  y \in \Gamma$.  If $Y$ is \corrf{internal} 
in $\T^\Gamma$, by the definition of $\F_Y^\Gamma$, 
we have $(\SET x_{\upharpoonright PA_Y},y)\in \F_Y^\Gamma$. Thus, by definition of intervention, $\T^\Gamma \models \langle\SET X = \SET x\rangle  Y =  y$.

If instead $Y$ is external 
in $\T^\Gamma$, 
axiom I8 yields $\Diamond Y=y \in \Gamma$. By I2 
 and lemma \ref{lemma: basic properties of A}, 10., 
 we obtain $\Diamond ( (\bigsqcup_{\SET x'\in \ran(\SET X)}\SET X = \SET x') \with Y=y) \in \Gamma$. By I0 and replacement, $\Diamond  \bigsqcup_{\SET x'\in \ran(\SET X)}(\SET X = \SET x' \with Y=y) \in \Gamma$. By lemma \ref{lemma: basic properties of A}, 8., $\bigsqcup_{\SET x'\in \ran(\SET X)}\Diamond(\SET X = \SET x' \with Y=y) \in \Gamma$. By lemma \ref{lemma: properties of MCS}, 4., there is an $\SET x^*\in\ran(\SET X)$ such that $\Diamond(\SET X = \SET x^* \with Y=y) \in \Gamma$. By definition of $\T^\Gamma$, there is an $s\in (\T^\Gamma)^-$ (namely $s(\SET XY)=\SET x^*y$). Since $Y$ is 
 \corrf{external}, $\F_{\SET X=\SET x}$ is empty; thus, by definition of intervention, 
 there is a $t\in s^\F_{\SET X = \SET x}$ with $t(Y)=y$. Thus, $\T^\Gamma \models \langle \SET X = \SET x\rangle Y=y$.

Vice versa, suppose $\T^\Gamma \models \langle \SET X = \SET x\rangle Y=y$. Then, again, we have two cases: $Y$ is internal or external. 
In the former case, by definition of intervention, $(\SET x_{\upharpoonright PA_Y},y)\in \F_Y^\Gamma$. But then, by definition of $\F_Y^\Gamma$, and lemma \ref{lemma: canonical team}, 1., this means that $ \langle \SET X = \SET x\rangle Y=y \in \Gamma$.

Suppose $Y$ is \corrf{external}; 
then $\T^\Gamma \models \langle \SET X = \SET x\rangle Y=y$ entails $\T^\Gamma \models \Diamond Y=y$. There is then an  $s\in(\T^\Gamma)^-$ such that $s(Y) =y$; write $\SET x^*$ for $s(\SET X)$. Then $\T^\Gamma \models \Diamond(\SET X = \SET x^* \with Y=y)$. By definition of $(\T^\Gamma)^-$, $\Diamond(\SET X = \SET x^* \with Y=y)\in \Gamma$.  By I0 and monotonicity, then, $\Diamond Y=y \in \Gamma$. 
Thus, by axiom I8,  $\langle \SET X = \SET x\rangle Y=y\in\Gamma$.

\item Case $n>1$. Suppose $\langle\SET X = \SET x\rangle \SET Y = \SET y \in\Gamma$. Let $Y_1,Y_2\in\SET Y$, $Y_1\neq Y_2$. Let $y_1 = \SET y_{\upharpoonright Y_1}$ and $y_2 = \SET y_{\upharpoonright Y_2}$, and define:
\[
\SET Y' = \SET Y \setminus \{Y_1\}    \hspace{30pt} \SET Y'' = \SET Y \setminus \{Y_2\}  \hspace{30pt}  \SET Y^\circ = \SET Y' \cap \SET Y''
\]

\vspace{-15pt}

\[
  \SET y' = \SET y \setminus \{y_1\}  \hspace{35pt} \SET y'' = \SET y \setminus \{y_2\}  \hspace{35pt} y^\circ = \SET y' \cap \SET y''. 
\]
From $\langle\SET X = \SET x\rangle \SET Y = \SET y \in \Gamma$, by I3 (composition) we obtain $\langle\SET X = \SET x, Y_1 = y_1\rangle \SET Y' = \SET y' \in \Gamma$ and $\langle\SET X = \SET x, Y_2 = y_2\rangle \SET Y'' = \SET y'' \in \Gamma$; by the inductive hypothesis, then, $\T^\Gamma \models\langle\SET X = \SET x, Y_1 = y_1\rangle \SET Y' = \SET y'$ and $\T^\Gamma \models\langle\SET X = \SET x, Y_2 = y_2\rangle \SET Y'' = \SET y''$. By the soundness of I6, then, $\T^\Gamma \models\langle\SET X = \SET x\rangle (Y_1 = y_1 \with Y_2 = y_2 \with \SET Y^\circ = \SET y^\circ)$, i.e. $\T^\Gamma \models \langle\SET X = \SET x\rangle \SET Y = \SET y$.

Vice versa, assume $\T^\Gamma \models \langle\SET X = \SET x\rangle \SET Y = \SET y$. With the same notations as above, by the soundness of I3 we obtain $\T^\Gamma \models\langle\SET X = \SET x, Y_1 = y_1\rangle \SET Y' = \SET y'$ and $\T^\Gamma \models\langle\SET X = \SET x, Y_2 = y_2\rangle \SET Y'' = \SET y''$. By i.h., $\langle\SET X = \SET x, Y_1 = y_1\rangle \SET Y' = \SET y' \in \Gamma$ and $\langle\SET X = \SET x, Y_2 = y_2\rangle \SET Y'' = \SET y'' \in \Gamma$. Then, by axiom I6, $\langle\SET X = \SET x\rangle \SET Y = \SET y\in\Gamma$.
\end{itemize}
\end{proof}

We write $\Gamma\models^1_\sigma\varphi$ 
if every singleton model
of signature $\sigma$ that satisfies $\Gamma$ also satisfies $\varphi$.

\begin{restatable}[Strong completeness of $\Ax$]{theorem}{StrongCompleteness}\label{thm: completeness of Ax}
For $\Gamma \cup \{\varphi\} \subseteq \mathcal H_\sigma$, 
\[
\Gamma \models^1_\sigma \varphi \iff \Gamma \vdash_{\Ax_\sigma} \varphi.
\]
\end{restatable}

\begin{proof}
\corrf{The soundness direction is given by theorem \ref{thm: soundness}.} 
For completeness, suppose $\Gamma \not\vdash \varphi$. Then, by routine reasoning $\Gamma\cup\Ax\cup\{\cneg\varphi\}$ is consistent, so by lemma \ref{lemma: Lindenbaum} there is a maximally consistent $\Delta \supseteq \Gamma\cup\Ax\cup\{\cneg\varphi\}$. Now, $\T^\Delta$ is a singleton model (lemma \ref{lemma: canonical team}), and by lemma \ref{lemma: truth lemma}, $\T^\Delta \models \Gamma\cup\{\cneg\varphi\}$. 
Thus, $\Gamma \not\models^1_\sigma\varphi$. 
\end{proof}

\section{Definability of law components}\label{sec: law components} 

Towards the formulation of a complete proof system for $\Hal^+$ over the general class of models, we need some additional concepts and terminology, which we introduce here. 

Note that any fixed signature $\sigma$ allows finitely many possible law components $\F$; we will denote as $\Fs$ the set of these objects. The subtleties of formalization described in section \ref{sec: differences} suggest that it might be difficult to talk about the law components \emph{within language $\Hal^+$}; in particular, it might be difficult to tell apart law components that only disagree about the parent sets of some variables. We introduce a notion of equivalence of law components to take this obstacle into account. When discussing several function components, say $\F,\G$,  we will differentiate between their parent sets by using indexed notation such as $PA_V^\F$, resp. $PA_V^\G$. 

\begin{definition}
    Let $\F,\G$ two law components of signature $\sigma = (\mathcal E, \mathcal I, \ran)$. 
    We say that $\F$ and $\G$ are \textbf{equivalent}, $\F \equiv\G$, if, for all $V \in \mathcal I$ and all $pa \in \ran(PA^\F_V \cup PA^\G_V)$, $\F_V(pa_{\upharpoonright PA^\F_V}) = \G_V(pa_{\upharpoonright PA^\G_V})$.  

    Two models $S=(S^-,\F)$ and $T = (T^-, \G)$ of signature $\sigma$ are \textbf{equivalent}, $S\equiv T$, if $S^- = T^-$ and $\F \equiv \G$.
\end{definition}

The next theorem shows that this notion of equivalence does indeed characterize all that can be said about law components in language $\Hal^+$ (as well as in $\Hal$). 

\begin{theorem}\label{thm: equivalence implies HPLUS equivalence}
    Let $S=(S^-,\F),T =(T^-,\G)$ be two models of signature $\sigma = (\mathcal E,\mathcal I,\ran)$. Then, $S \equiv T$ iff for all $\varphi\in \Hal^+_\sigma$, $S\models \varphi \iff T \models \varphi$ iff for all $\varphi\in \Hal_\sigma$, $S\models \varphi \iff T \models \varphi$.
\end{theorem}

\begin{proof}
    1) First, we prove that $S \equiv T$ implies that $S$ and $T$ agree over all $\Hal^+_\sigma$ formulas. We proceed by induction on $\varphi$; we prove the statement simultaneously for all pairs of equivalent teams of arbitrary signatures. The atomic case ($\varphi$ is $Y=y$) immediately follows from the fact that $S$ and $T$ have the same team component. The cases for $\varphi$ of the forms $\cneg\psi$ and $\psi \with\chi$ are straightforward.

    If $S\models \varphi = [\SET X = \SET x]\psi$, then, for all $s\in S_{\SET X = \SET x}^-$, $(\{s\},\F_{\SET X = \SET x})\models \psi$. Since  $\F\equiv\G$, we have $\F_V(pa_{\upharpoonright PA^\F_V}) = \G_V(pa_{\upharpoonright PA^\G_V})$ for all $V\in \mathcal I\setminus \SET X$. Since $V\in \mathcal I\setminus \SET X$ is the set of internal variables of $\F_{\SET X = \SET x}$ and $\G_{\SET X = \SET x}$, and for all such variables  $(\F_{\SET X = \SET x})_V = \F_V$ and $(\G_{\SET X = \SET x})_V = \G_V$,  then, $\F_{\SET X = \SET x} \equiv \G_{\SET X = \SET x}$. Thus, for each $s\in S_{\SET X = \SET x}^-$, $(\{s\},\F_{\SET X = \SET x}) \equiv (\{s\},\G_{\SET X = \SET x})$. By the inductive hypothesis, then, for all $s\in (S_{\SET X = \SET x})^-$ we have $(\{s\},\G_{\SET X = \SET x})\models \psi$. If we  manage to prove that $S_{\SET X = \SET x}^- = T_{\SET X = \SET x}^-$, then, we are done.

    It suffices to show that $S_{\SET X = \SET x}^- \subseteq T_{\SET X = \SET x}^-$; the converse proof is symmetric. Suppose $s\in S_{\SET X = \SET x}^-$; by definition of intervention, this means that $s(\SET N_{\SET X})\in S^-(\SET N_{\SET X}), s(\SET X) = \SET x$ and $s$ satisfies the constraints given by $\F_{\SET X = \SET x}$. The latter means that, for all $V\in \mathcal{I}\setminus \SET X$, 
    $(s(PA^\F_V),s(V))\in \F_V$. In other words, $s(V) \in \F_V(s(PA^\F_V)) = \G_V(s(PA_V^\G))$, where the equality is given by $\F \equiv \G$. Thus $(s(PA^\G_V),s(V))\in \G_V$ for all $V\in \mathcal{I}\setminus \SET X$; together with $s(\SET N_{\SET X})\in S^-(\SET N_{\SET X}) = T^-(\SET N_{\SET X})$ and $s(\SET X) = \SET x$, this yields $s\in T_{\SET X = \SET x}^-$.

    2) We now deduce $S \equiv T$ from the assumption that $S$ and $T$ agree on the truth values of $\Hal_\sigma$ formulas. Since $S$ and $T$ satisfy the same formulas of the form $\Diamond \SET W = \SET w$, we immediately obtain $S^- = T^-$. Now let $pa \in \ran(PA^\F_Y \cup PA^\G_Y)$.  Suppose $y\in \F_Y(pa_{\upharpoonright PA^\F_Y})$. 
    By definition of intervention, for any $\SET w\in \ran(\SET W_Y)$ with $\SET w_{\upharpoonright PA^\F_Y \cup PA^\G_Y} = pa$ we have $S\models \langle \SET W_Y = \SET w\rangle Y=y$. But then, by the assumption, $T\models \langle \SET W_Y = \SET w\rangle Y=y$. Thus, by the same reasoning, $y\in \G_Y(pa_{\upharpoonright PA^\G_Y})$. The converse is analogous. Thus, we conclude that $\F \equiv \G$.   
\end{proof}

\noindent We can improve on this result by showing that each equivalence class of law components can be characterized by a single $\Hal$ formula. For each signature $\sigma = (\mathcal E,\mathcal I, \ran)$ and each $\F \in \Fs$, we define a formula
\[
\Phi^\F :=  \bigwedge_{\substack{Y\in\mathcal I \\ \SET w \in \ran(\SET W_Y) \\y\in\F_Y(\SET w_{\upharpoonright PA_Y^\F})  }}  \langle \SET W_Y=\SET w\rangle Y=y   \with          
\bigwedge_{\substack{Y\in\mathcal I \\ \SET w \in \ran(\SET W_Y) \\ y\notin\F_Y(\SET w_{\upharpoonright PA_Y^\F})  }}
\cneg \langle \SET W_Y=\SET w\rangle Y=y.
\]
The $\Phi^\F$ formulas will be crucial in the formulation of two axioms; the next result shows that they indeed characterize law components up to equivalence.

\begin{theorem}\label{thm: phiF}
    Let $T= (T^-,\G)$ be a model. Then, $T\models \Phi^\F$ $\iff$ $\G\equiv \F$.
\end{theorem}

\begin{proof}
    $\Rightarrow$) Suppose $T\models \Phi^\F$. Let $Y\in \mathcal I$, $pa\in\ran(PA_Y^\F \cup PA_Y^\G)$.  Pick a $\SET w \in \ran(\SET W_Y)$ such that $\SET w_{\upharpoonright PA_Y^\F \cup PA_Y^\G} = pa$.
    
    Suppose first that $y\in \F_Y(pa_{\upharpoonright PA_Y^\F})$. Thus, we may also write $y\in \F_Y(\SET w_{\upharpoonright PA_Y^\F})$. Since $T\models \Phi^\F$, then, in particular $T\models  \langle \SET W_Y=\SET w\rangle Y=y$. By definition of intervention, then, $y\in \G(\SET w_{\upharpoonright PA_Y^\G}) = \G(pa_{\upharpoonright PA_Y^\G})$, as needed.

    Vice versa, suppose $y\in \G(pa_{\upharpoonright PA_Y^\G})$. Then, by definition of intervention, $T\models  \langle \SET W_Y=\SET w\rangle Y=y$.  Suppose for the sake of contradiction that $y\notin \F_Y(pa_{\upharpoonright PA_Y^\F})$. Then, $y\notin \F_Y(\SET w_{\upharpoonright PA_Y^\F})$. So, since $T\models \Phi^\F$, we obtain that $T\models \cneg \langle \SET W_Y=\SET w\rangle Y=y$, contradicting our earlier statement.

    Thus, $\G \equiv \F$.

    $\Leftarrow$) Suppose $\G \equiv \F$.  Let $Y\in \mathcal I$, $pa\in \ran(PA_Y^\F \cup PA_Y^\G)$. Pick a $\SET w \in \ran(\SET W_Y)$ such that $\SET w_{\upharpoonright PA_Y^\F \cup PA_Y^\G} = pa$. 

    Assume first that $y\in \F_Y(\SET w_{\upharpoonright PA_Y^\F})$. Since $\G \equiv \F$, we have $y\in \G_Y(\SET w_{\upharpoonright PA_Y^\G})$. Thus, by definition of intervention, $T\models \langle \SET W_Y=\SET w\rangle Y=y$, as needed.

    If we assume instead that $y\notin \F_Y(\SET w_{\upharpoonright PA_Y^\F})$, since $\G \equiv \F$ we obtain $y\notin \G_Y(\SET w_{\upharpoonright PA_Y^\G})$. By definition of intervention, then, $T\not\models \langle \SET W_Y=\SET w\rangle Y=y$, and thus $T\models \cneg\langle \SET W_Y=\SET w\rangle Y=y$. 
\end{proof}

\section{A general axiomatization}\label{sec: general axiomatization} 

In the following we prove the soundness and completeness of the following system $\Bx_\sigma$ for language $\Hal^+$ over the class of all models (again, $\Bx_\sigma$ will often just denote the set of axioms of this system).


\begin{center}
\hspace{15pt} Rule MP. \Large$\frac{\psi \hspace{20pt}\psi \rightarrow \chi}{\chi}$\normalsize  
\hspace{80pt}
Rule NEC. \Large$\frac{\vdash\psi}{\vdash[\SET X = \SET x]\psi}$\normalsize 
\end{center}

\noindent J0. $\Hal^+_\sigma$ instances of a classical tautologies in $\cneg,\with$.


\noindent I1. 
$[\SET X = \SET x] Y=y \rightarrow [\SET X = \SET x] Y \neq y'$  \hspace{10pt}  (when $y\neq y'$)  \hspace{56pt} [Uniqueness]


\noindent I2. 
$[\SET X = \SET x] \bigsqcup_{y\in\ran(Y)} Y=y$\hspace{180pt}   [Definiteness]

\noindent I3. 
$\langle\SET X = \SET x\rangle  (Z = z \with  \SET Y = \SET y) \rightarrow \langle\SET X = \SET x,  Z =z \rangle \SET Y = \SET y$ \hspace{55pt} [Weak composition]

\noindent I4. 
$[\SET X = \SET x,  Y=y] Y=y$ \hspace{190pt} [Effectiveness]

\noindent I5. 
 $([\SET X = \SET x]\psi \with  [\SET X = \SET x](\psi\rightarrow \chi)) \rightarrow [\SET X = \SET x]\chi$ \hspace{80pt} [K-axiom]

\noindent J6. 
$\SET W = \SET w \rightarrow$ I6.  \hspace{219pt} [Reversibility for small teams]

\noindent I7. 
$\SET Y = \SET y \leftrightarrow \Box \SET Y = \SET y$. \hspace{198pt} [Flatness] 

\noindent I8. 
$\langle \SET W_Y = \SET w\rangle Y=y \leftrightarrow \Diamond Y=y$ \hspace{10pt} (if $Y\in \mathcal E$) \hspace{93pt}  [External variables]

\noindent I9. $\langle \SET W = \SET w\rangle\top$ \hspace{234pt} [Full intervention]


\noindent J10. 
$\langle \SET X = \SET x \rangle \langle \SET W_Y = \SET w\rangle Y=y \leftrightarrow \langle \SET W_Y = \SET w\rangle Y=y$     (if $Y\in \mathcal I$ and $Y\notin \SET X$) \hspace{7pt} [Diamond screen-off]

\noindent J11. 
$[\SET X = \SET x]\langle \SET W_Y = \SET w\rangle Y=y \leftrightarrow \langle \SET W_Y = \SET w\rangle Y=y$     (if $Y\in \mathcal I$ and $Y\notin \SET X$) \hspace{8pt} [Box screen-off]

\noindent J12. 
$[\SET X = \SET x]\Diamond\psi \leftrightarrow [\SET X = \SET x]\psi$ \hspace{164pt} [Conditional diamond]

\noindent J13. 
$\langle\SET X = \SET x \rangle\Diamond \psi \leftrightarrow \langle\SET X = \SET x \rangle\psi$ \hspace{163pt} [Mighty diamond]

\noindent J14. 
$\Diamond \langle\SET X = \SET x \rangle\psi \leftrightarrow \langle\SET X = \SET x \rangle\psi$ \hspace{163pt}  [Double diamond]


\noindent J15. 
$\langle\SET X = \SET x \rangle \psi \rightarrow \langle\SET X = \SET x \rangle(\psi \with \Diamond \top)$  \hspace{132pt} [Persistency of diamond] 

\noindent J16. $\bigsqcup_{\F\in \Fs} \Phi^\F$  \hspace{231pt} [Certainty of laws]

\noindent J17. 
$\Phi^\F \rightarrow ([\SET X = \SET x]\psi  \leftrightarrow  [\SET X = \SET x]\bigsqcup_{\substack{ t \text{ such that} \\ (\{t\},\F_{\SET X = \SET x})\models \psi}} \SET W=t(\SET W))$ \hspace{32pt} [Modal collapse]

\vspace{3pt}

\noindent As in system $\Ax$, we allow $\SET X$ and $\SET Y$ to be empty tuples, as a special case.

\vspace{5pt}

Eight of the axioms (those marked with a letter I) are shared with system $\Ax_\sigma$. Note the absence of I10 and I1b (which only hold in nonempty models) and I2b (which holds iff there is at most one assignment). Axiom J6 says that \emph{Weak reversibility} holds for singleton and empty models (under the same restrictions as for axiom I6, which here have not been repeated). 
The remaining axioms deal with the nesting of modal operators. Axioms J10-J11 say that an intervention on all variables except an internal $Y$ trumps any previous intervention on variables different from $Y$. Axioms J12-13-14-15 are various consequences of the fact that the semantics of the modal operators only involves looking at certain singleton submodels. We note that the ``empty intervention'' versions of these axioms imply that any sequence of $\Diamond$ and $\Box$ modalities can always be reduced to the \emph{leftmost} modality occurring in it (contrarily, e.g., to what happens in modal logic S5). J16 says that any model has a fixed set of causal laws. Together with J17, it will be used to show that, when reasoning with maximal consistent sets of formulas, the $\Hal^+$ formulas can always be ``reduced'' to $\Hal$ formulas (note that, differently from $[\SET X = \SET x]\psi$, the rightmost member of J17 is always an unnested formula). We have to admit that axiom scheme J17 is rather unsatisfying due to its complexity and semantic flavour; however, note that membership in this scheme is decidable due to the finiteness of the models. 
Furthermore, the problem of deciding which formulas are instances of J17 reduces to the proof-theoretic problem of finding proofs of consequences of the forms $\SET W = s(\SET W), \Phi^{\F_{\SET X = \SET x}} \vdash \psi$ and $\SET W = s(\SET W), \Phi^{\F_{\SET X = \SET x}} \vdash \cneg\psi$ in system $\Ax$, which we already know to be complete.

\begin{theorem}\label{thm: soundness of B}
    System $\Bx_\sigma$ is sound over the class of all models of signature $\sigma$.
\end{theorem}

\begin{proof}
See the proof of theorem \ref{thm: soundness} for the soundness of axioms J0, I1-I5, I7-I9.

Concerning axiom J6, suppose $T\models \SET W = \SET w$. Then, either $T$ is a singleton model or is empty. If it a singleton model, then it satisfies I6 (see theorem \ref{thm: soundness} for a proof). Suppose now that  $T$ is empty, and that it safisfies $\psi: \langle \SET X = \SET x, V=v \rangle (Y=y \with \SET Z = \SET z)$ and $\chi: \langle \SET X = \SET x,Y=y \rangle(V=v \with \SET Z = \SET z)$. By $\psi$, there is an $s \in T_{\SET X = \SET x, V=v}^-$ such that $s(Y)=y$ and $s(\SET Z)=\SET z$. 
But (since $\SET X \cup \SET Z \cup \{Y\}\cup \{V\} = \dom $) $\chi$ tells us that $s$ is also in $T_{\SET X = \SET x,Y=y}^-$. In particular, if $V$ is 
\corrf{internal}, $s$ is compatible also with the law $\F_V$; if we prove that (*): $s(\SET N_{\SET X}) \in T^-(\SET N_{\SET X})$, 
then we can conclude that $s\in T_{\SET X = \SET x}^-$. But, since $s \in T_{\SET X = \SET x, V=v}^-$ and $T^-$ is empty, by definition of intervention we must have $\SET N_{\SET XV} = \emptyset$; and similarly, we obtain $\SET N_{\SET XY} = \emptyset$. But then $\SET N_{\SET X} = \SET N_{\SET XY} \cup \SET N_{\SET XV} = \emptyset$, so that condition (*) is trivially  satisfied.  


For axiom J10, note that $T\models \langle \SET X = \SET x \rangle \langle \SET W_Y = \SET w\rangle Y=y$ iff, for some $s\in T_{\SET X = \SET x}^-$, $(\{s\},\F_{\SET X = \SET x})\models \langle \SET W_Y = \SET w\rangle Y=y$. The latter holds iff there is a $t\in (\{s\},\F_{\SET X = \SET x})_{ \SET W_Y = \SET w}^-$ such that $(\{t\}, (\F_{\SET X = \SET x})_{\SET W_Y = \SET w})\models Y=y$. But $(\F_{\SET X = \SET x})_{\SET W_Y = \SET w} = \F_{\SET W_Y = \SET w}$ (since $\SET X \subseteq \SET W_Y$). Let us show that $t\in T_{\SET W_Y = \SET w}^-$. Since $t\in (\{s\},\F_{\SET X = \SET x})_{ \SET W_Y = \SET w}^-$, by definition of intervention, $t(\SET W_Y) = \SET w$, and, since $Y\in \mathcal I$,  $t(Y)\in \F_Y(\SET w_{\upharpoonright PA_Y})$, as needed. 
Thus,  $t\in  T_{\SET W_Y = \SET w}^-$, and therefore $T\models \langle \SET W_Y = \SET w\rangle Y=y$. 

Vice versa, if $T\models \langle \SET W_Y = \SET w\rangle Y=y$, there is a $t\in  T_{\SET W_Y = \SET w}^-$ such that $(\{t\}, \F_{\SET W_Y = \SET w})\models Y=y$. As before $(\F_{\SET X = \SET x})_{\SET W_Y = \SET w} = \F_{\SET W_Y = \SET w}$. It suffices then to show that $t\in (T_{\SET X = \SET x})_{\SET W_Y = \SET w}^-$. Since $t\in  T_{\SET W_Y = \SET w}^-$, we have $t(\SET W_Y) = \SET w$, as needed. Since $Y$ is internal (in all the involved causal teams, since $Y\notin \SET X$), then $t(Y)=y \in (\F_{\SET W_Y = \SET w})_Y(\SET w_{\upharpoonright PA_Y}) = ((\F_{\SET X = \SET x})_{\SET W_Y = \SET w})_Y(\SET w_{\upharpoonright PA_Y})$; thus, $t\in (T_{\SET X = \SET x})_{\SET W_Y = \SET w}^-$, as needed. 
The soundness of axiom J11 is proved analogously. 

The soundness of axioms J12, J13 and J14 is a straightforward consequence of the semantic clause for the modal operators, which just requires checking whether the consequent is satisfied in singleton models.
E.g., for axiom J12, note that $T\models [\SET X = \SET x]\Diamond\psi$ iff all singleton submodels of $T$ satisfy $\Diamond\psi$ iff (being singleton) they satisfy $\psi$, iff $T\models  [\SET X = \SET x]\psi$.  

For axiom J15, suppose $T\models \langle \SET X = \SET x \rangle \psi$. Then, there is an $s\in T_{\SET X = \SET x}^-$ such that $(\{s\}, \F_{\SET X = \SET x})\models \psi$. Obviously 
$(\{s\}, \F_{\SET X = \SET x}) \models \Diamond\top$. 
Thus,  $(\{s\}, \F_{\SET X = \SET x})\models \psi \with \Diamond \top$. But $s\in T_{\SET X = \SET x}^-$; thus, $T\models \langle\SET X = \SET x \rangle(\psi \with \Diamond \top)$.

For axiom J16, note that $T = (T^-,\F)\models \Phi^\F$ by theorem \ref{thm: phiF}. Thus, $T\models \bigsqcup_{\F\in \Fs} \Phi^\F$.

For axiom J17, suppose $T\models \Phi^\F$.

$\Rightarrow$) Assume $T\models [\SET X = \SET x]\psi$. Then, for all $s\in (T_{\SET X = \SET x})^-$, (*): $(\{s\},\F_{\SET X = \SET x})\models \psi$. But we also have that $(\{s\},\F_{\SET X = \SET x})\models \SET W = s(\SET W)$; thus, by (*), $(\{s\},\F_{\SET X = \SET x})\models \bigsqcup_{\substack{ t \text{ such that} \\ (\{t\},\F_{\SET X = \SET x})\models \psi}} \SET W=t(\SET W)$. Since this holds for all $s\in T_{\SET X = \SET x}^-$, $T\models [\SET X = \SET x]\bigsqcup_{\substack{ t \text{ such that} \\ (\{t\},\F_{\SET X = \SET x})\models \psi}} \SET W=t(\SET W)$.

$\Leftarrow$) Assume $T\models [\SET X = \SET x]\bigsqcup_{\substack{ t \text{ such that} \\ (\{t\},\F_{\SET X = \SET x})\models \psi}} \SET W=t(\SET W)$. Then, for all $s\in (T_{\SET X = \SET x})^-$, $(\{s\},\F_{\SET X = \SET x})\models \bigsqcup_{\substack{ t \text{ such that} \\ (\{t\},\F_{\SET X = \SET x})\models \psi}} \SET W=t(\SET W)$. Thus, by the semantic clauses, $s=t$ for some $t$ such that $(\{t\},\F)\models \psi$. In other words, $(\{s\},\F)\models \psi$. Thus, $T\models [\SET X = \SET x]\psi$.
\end{proof}

Since this system includes modal logic K (MP, NEC, axioms J0 and I5), we automatically obtain the same theorems as in lemma \ref{lemma: basic properties of A}. But notice that points 2., 3. and 5. have more relaxed statements.

\begin{lemma}\label{lemma: basic properties of B}
The following hold in $\Bx$.
    \begin{enumerate}
        \item (Deduction theorem) If $\Gamma,\psi \vdash \chi$, then $\Gamma \vdash \psi\rightarrow \chi$.

        \item ($\Box$-Monotonicity) If $\Gamma\vdash[\SET X = \SET x]\psi$ and, $\vdash \psi \rightarrow \psi'$, then $\Gamma\vdash [\SET X = \SET x]\psi'$.
        
        \item ($\Diamond$-Monotonicity) If $\Gamma\vdash \langle\SET X = \SET x\rangle\psi$ and $\vdash \psi \rightarrow \psi'$, then $\Gamma\vdash \langle\SET X = \SET x \rangle\psi'$.

        \item  $\vdash([\SET X = \SET x]\psi \with  [\SET X = \SET x]\chi) \leftrightarrow [\SET X = \SET x](\psi \with \chi)$.

        \item (Replacement) Suppose $\Gamma\vdash \theta \leftrightarrow \theta'$. Then, $\Gamma\vdash \varphi \leftrightarrow \varphi[\theta'/\theta]$.
        
        \item $\vdash\cneg[\SET X = \SET x]\psi \leftrightarrow \langle\SET X = \SET x\rangle\cneg\psi$
       
        \item $\vdash\cneg\langle\SET X = \SET x\rangle\psi \leftrightarrow [\SET X = \SET x]\cneg\psi$

        \item $\vdash(\langle\SET X = \SET x\rangle \psi \sqcup \langle\SET X = \SET x\rangle \chi) \leftrightarrow \langle\SET X = \SET x\rangle(\psi\sqcup\chi)$.

         \item  $\vdash([\SET X = \SET x]\psi \with \langle\SET X = \SET x\rangle \top)\rightarrow \langle\SET X = \SET x\rangle\psi$.
    


        \item $\vdash([\SET X = \SET x]\psi \with \langle \SET X = \SET x \rangle \chi) \rightarrow \langle \SET X = \SET x \rangle (\psi \with \chi)$.

        \item $\vdash\langle\SET X = \SET x\rangle(\psi \with \chi) \rightarrow (\langle\SET X = \SET x\rangle\psi \with \langle\SET X = \SET x\rangle\chi)$.
    \end{enumerate}
\end{lemma}

Let us clarify the relationship between systems  $\Ax_\sigma$ and $\Bx_\sigma$. The intuitive picture is that system $\Bx_\sigma$ is conservative over $\Ax_\sigma$ when restricting attention to singleton models. Formally, this means:

\begin{theorem}\label{thm: B as conservative extension of A}
    Let $\Gamma\cup \{\varphi\}\subseteq \Hal_\sigma$. Then, $\Gamma\vdash_{\Ax_\sigma} \varphi$ iff $\Diamond \top, \bigsqcup_{\SET w \in \ran(\SET W)} \SET W = \SET w,\Gamma\vdash_{\Bx_\sigma} \varphi$.
\end{theorem}

\begin{proof}
For brevity, let us write $\Delta$ for the set of assumptions $\{\Diamond \top\} \cup \{\bigsqcup_{\SET w \in \ran(\SET W)} \SET W = \SET w\} \cup \Gamma$. 
Let us first observe that, if $\theta$ is an axiom of $\Ax_\sigma$, then $\Delta\vdash_{\Bx_\sigma} \theta$. Indeed, most of of the axioms of $\Ax_\sigma$ are included in $\Bx_\sigma$.
For axiom I10, obviously $\Delta\vdash_{\Bx_\sigma} \Diamond\top$.
For I2b, 
note first that $\Delta\vdash_{\Bx_\sigma} \bigsqcup_{\SET w \in \ran(\SET W)} \SET W = \SET w$ trivially. But then  $\Delta \vdash_{\Bx_\sigma} \bigsqcup_{y\in \ran(Y)} Y=y$ follows by J0 and MP.

For axiom I1b, 
write $\Xi$ for $\Delta \cup \{Y=y\} \cup\{Y=y'\}$. 
By axiom I7, $\Xi \vdash_{\Bx_\sigma} \Box Y=y$ and $\Xi \vdash_{\Bx_\sigma} \Box Y=y'$. On the other hand, by I1, $\Box Y=y \rightarrow \Box Y \neq y'$; thus, by MP, $\Xi \vdash_{\Bx_\sigma} \Box Y\neq y'$. By J0 and lemma \ref{lemma: basic properties of B}, 4.,  $\Xi \vdash_{\Bx_\sigma} \Box (Y=y' \with  Y\neq y')$. Thus, by J0 and replacement, $\Xi \vdash_{\Bx_\sigma} \Box\bot$. Since $\Diamond \top\in \Delta$, by lemma \ref{lemma: basic properties of B}, 9.,  $\Xi \vdash_{\Bx_\sigma} \Diamond\bot$, i.e. $\Xi \vdash_{\Bx_\sigma}\cneg\Box \top$. On the other hand, $\Xi \vdash_{\Bx_\sigma} \Box \top$ by J0 and NEC. So, by J0, $\Xi \vdash_{\Bx_\sigma}\bot$. By the deduction theorem, then, $\Delta \vdash_{\Bx_\sigma} (Y=y \with Y=y')\rightarrow\bot$. By J0, then, $\Delta \vdash_{\Bx_\sigma} Y=y \rightarrow Y \neq y'$.

Finally, for axiom I6, just note that, by J6, J0 and MP, $\Delta \vdash_{\Bx_\sigma} \bigsqcup_{\SET w\in \ran(\SET W)} $I6. Thus, by J0 again, $\Delta \vdash_{\Bx_\sigma}$I6. 

Since all axioms of $\Ax_\sigma$ are $\vdash_{\Bx_\sigma}$-derivable from $\Delta$, and $\Bx$ includes all the rule instances of $\Ax$, it is straightforward to transform any proof of the form $\Gamma\vdash_{\Ax_\sigma} \varphi$ into a proof of $\Delta\vdash_{\Bx_\sigma} \varphi$.

Vice versa, if $\Delta\vdash_{\Bx_\sigma} \varphi$, by the soundness of $\Bx$ (theorem \ref{thm: soundness of B}), $\Delta\models \varphi$. But this (by lemma \ref{lemma: I1+I2 implies singleton}) means that every singleton model that satisfies $\Gamma$ also satisfies $\varphi$; i.e., $\Gamma \models^1 \varphi$. But then $\Gamma\vdash_{\Ax_\sigma} \varphi$ by the completeness theorem for $\Ax_\sigma$ (theorem \ref{thm: completeness of Ax}). 
\end{proof}

We will now consider maximally $\Bx_\sigma$-consistent sets and their relationships to canonical teams. 
Note that, for sets $\Gamma$ of $\Hal^+_\sigma$ formulas that are maximally consistent according to $\Bx_\sigma$, we can define their associate canonical causal team $\T^\Gamma$ exactly as in section \ref{sec: H language}; we may write $\T^\Gamma_\Bx$ where there is the risk of confusion. As before, we can prove:

\begin{lemma}\label{lemma: canonical team for B}
Let $\Gamma\supseteq \Bx_\sigma$ be a maximal 
consistent set of $\Hal^+_\sigma$ formulas. Then $\T^\Gamma$ is a model, i.e.:
\begin{enumerate}
\item For all $Y\in \inte(\T^\Gamma)$, $\F_Y^\Gamma$ is well-defined.
\item For all $Y\in \inte(\T^\Gamma)$ and $s\in (\T^\Gamma)^-$, $(s(PA_Y),s(Y))\in \F^\Gamma_Y$.
\end{enumerate}
\end{lemma}
The proof is as for $\Ax$, since it only uses axiom I3 and classical logic.

The following corollary of theorem \ref{thm: B as conservative extension of A} shows that the notational abuse of writing  $\T^\Gamma$ for  $\T^\Gamma_\Bx$ is not too dangerous, at least when $\T^\Gamma_\Bx$ is a singleton model.

\begin{corollary}\label{cor: singleton TB is T}
    Let $\Gamma\supseteq \Bx_\sigma$ be a maximal $\Bx_\sigma$-consistent set of $\Hal^+_\sigma$-formulas such that $\T_\Bx^\Gamma$ is a singleton model, and let $\Gamma_0$ be the set of $\Hal_\sigma$ formulas in $\Gamma$. Then, $\Gamma_0$ is maximally $\Ax_\sigma$-consistent, and $\T^{\Gamma_0} = \T_\Bx^\Gamma$. 
\end{corollary}

\begin{proof}
First, we show that $\Diamond \top, \bigsqcup_{\SET w\in\ran(\SET W)} \SET W = \SET w \in \Gamma$.

For the former, note that since $\T^\Gamma_\Bx$ is a singleton model, by its definition there is a $\SET w\in\ran(\SET W)$ such that $\Diamond \SET W = \SET w \in \Gamma$. But $\vdash_{\Bx_\sigma} \SET W = \SET w \rightarrow \top$ by J0, so $\Diamond \top \in \Gamma$ by monotonicity. 

For the latter, note that, by the definition of $\T^\Gamma_\Bx$ and the fact that it is a singleton model, we also have that,  for all $\SET w'\neq \SET w$, $\Diamond \SET W = \SET w' \notin\Gamma$. Since $\Gamma$ is maximal, $\cneg\Diamond \SET W = \SET w' \in\Gamma$. By lemmas \ref{lemma: basic properties of B}, 7. and lemma \ref{lemma: properties of MCS}, 3., we obtain $\bigwedge_{\SET w' \neq \SET w} \Box\cneg \SET W = \SET w' \in\Gamma$. By lemma \ref{lemma: basic properties of B}, 4., $\Box\bigwedge_{\SET w' \neq \SET w} \cneg \SET W = \SET w' \in\Gamma$. Thus, since we have $\Box \bigsqcup_{\SET w\in\ran(\SET W)} \SET W = \SET w\in \Gamma$ by I2, by lemma \ref{lemma: properties of MCS}, 3., and lemma \ref{lemma: basic properties of B}, 4. we obtain $\Box(\bigwedge_{\SET w \neq \SET w} \cneg \SET W = \SET w' \with \bigsqcup_{\SET w\in\ran(\SET W)} \SET W = \SET w)\in\Gamma$. Thus, by J0 and replacement, $\Box\SET W = \SET w \in \Gamma$. By I7, $\SET W = \SET w \in \Gamma$. Thus, by I0, $\bigsqcup_{\SET w\in\ran(\SET W)} \SET W = \SET w \in \Gamma$.  

We use these facts to prove that $\Gamma_0$ is closed under $\Ax$-derivations. Suppose $\Gamma_0 \vdash_\Ax \varphi$. Then, by theorem \ref{thm: B as conservative extension of A}, $\Diamond \top, \bigsqcup_{\SET w\in\ran(\SET W)} \SET W = \SET w,\Gamma_0 \vdash_\Bx \varphi$; but then, by what we proved above, $\Gamma \vdash_\Bx \varphi$. Since $\Gamma$ is maximal, then, $\varphi\in\Gamma$; since $\varphi \in \Hal$, we conclude $\varphi\in\Gamma_0$.

Now suppose for the sake of contradiction that $\Gamma_0$ is $\Ax$-inconsistent; then $\Gamma_0\vdash_{\Ax_\sigma} \bot$, but since $\Gamma_0$ is closed under $\Ax$-derivations, $\bot \in\Gamma_0\subseteq\Gamma$, contradicting the $\Bx$-consistency of $\Gamma$.

Finally, suppose for the sake of contradiction that $\Gamma_1\supset \Gamma_0$ is an $\Ax_\sigma$-consistent set of $\Hal_\sigma$ formulas; say, $\varphi\in\Gamma_1\setminus \Gamma_0$. Then, $\varphi\notin\Gamma$, so, by the maximality of $\Gamma$, $\cneg\varphi \in \Gamma$. But $\cneg\varphi$ is an $\Hal$ formula, so $\cneg\varphi \in \Gamma_0 \subseteq \Gamma_1$. Thus, after all, $\Gamma_1$ is inconsistent. 
   
Now, since $\Gamma_0$ is maximally $\A_\sigma$-consistent, it defines the canonical team $\T^{\Gamma_0}$, which by definition is the same as $\T_\Bx^\Gamma$.
\end{proof}

Theorem \ref{thm: B as conservative extension of A} also ensures that the truth lemma will work correctly at least for the $\Hal$ formulas in $\Gamma$, provided $\T^\Gamma$ is a singleton model.

\begin{corollary}\label{cor: truth corollary}
    Let $\Gamma\supseteq \Bx_\sigma$ be a set of $\Hal^+_\sigma$ formulas that is maximally 
    consistent, and $\varphi$ be an $\Hal_\sigma$ formula.
    \begin{enumerate}
    \item If $\varphi$ is an atomic  formula, then $\varphi\in \Gamma \iff \T^\Gamma \models \varphi$.
    
    \item If $\T^\Gamma$ is a singleton model, then  $\varphi \in \Gamma \iff \T^\Gamma \models \varphi$.
    \end{enumerate}
\end{corollary}

\begin{proof}
    1) Suppose $Y=y \in \Gamma$. Then, $\Box Y=y \in \Gamma$ by I7. 
    Suppose for the sake of contradiction that there is a $t\in (\T^\Gamma)^-$ such that $t(Y)= y' \neq y$. Then, by definition of $\T^\Gamma$, $\Diamond\SET W = t(\SET W) \in \Gamma$. Thus, $\Diamond Y=y'\in\Gamma$ by lemma \ref{lemma: basic properties of B}, 11. Thus, by lemma \ref{lemma: basic properties of B}, 10., $\Diamond (Y=y \with Y=y')\in\Gamma$. On the other hand, by I1 and MP we have  $\Box Y\neq y' \in \Gamma$. Again by lemma \ref{lemma: basic properties of B}, 10., we obtain $\Diamond (Y=y \with Y=y' \with Y\neq y')\in \Gamma$, so, by I0 and monotonicity, $\Diamond \bot \in \Gamma$. This contradicts lemma \ref{lemma: properties of MCS}, 5. Thus, all assignments in $(\T^\Gamma)^-$ satisfy $Y=y$, and so $\T^\Gamma \models Y=y$. 

    Vice versa, suppose $\T^\Gamma \models Y=y$. Suppose for the sake of contradiction that $Y=y \notin \Gamma$. By axiom I7,  $\Box Y=y \notin \Gamma$. By maximality, $\cneg \Box Y=y\in \Gamma$, and by lemma \ref{lemma: basic properties of B}, 6., $\Diamond \cneg Y=y \in \Gamma$. on the other hand, by axioms I2, $\Box \bigsqcup_{\SET w \in \ran(\SET W)} \SET W = \SET w\in\Gamma$. Thus, by lemma \ref{lemma: basic properties of B}, 10., $\Diamond (\cneg Y=y \with\bigsqcup_{\SET w \in \ran(\SET W)} \SET W = \SET w)\in\Gamma$. By I0 and monotonicity, $\Diamond \bigsqcup_{\SET w \in \ran(\SET W), \SET w_{\upharpoonright Y}\neq y} \SET W = \SET w \in \Gamma$. By lemma \ref{lemma: basic properties of B}, 8., $ \bigsqcup_{\SET w \in \ran(\SET W), \SET w_{\upharpoonright Y}\neq y}\Diamond \SET W = \SET w \in \Gamma$. Thus, by primality of $\Gamma$, there is a $\SET w \in \ran(\SET W)$ such that $\SET w_{\upharpoonright Y}\neq y$ and $\Diamond \SET W = \SET w \in \Gamma$. Thus, the assignment $s(\SET W) = \SET w$ is in $(\T^\Gamma)^-$. But $s(Y)\neq y$; thus, $\T^\Gamma \not\models Y=y$, contradicting our initial assumption.

    2) Let $\Gamma_0$ be the set of all $\Hal$ formulas in $\Gamma$.
   From right to left, suppose $\T^\Gamma\models \varphi$; by corollary \ref{cor: singleton TB is T}, this gives $\T^{\Gamma_0} \models\varphi$. But then, since $\varphi\in\Hal$, by the truth lemma \ref{lemma: truth lemma} for $\Gamma_0$, $\varphi\in\Gamma_0 \subseteq \Gamma$. 
    
    From left to right, since $\T^{\Gamma_0}=\T^\Gamma$ is a singleton model, $\T^{\Gamma_0}\models \Diamond \top, \bigsqcup_{\SET w \in \ran(\SET W)} \SET W = \SET w$. 
    Now, if $\varphi\in \Gamma$, then $\varphi\in \Gamma_0$ (since it is an $\Hal$ formula). Then, by the maximality of $\Gamma_0$,  $\Gamma_0 \vdash_{\Ax_\sigma} \varphi$. 
    Thus, by theorem \ref{thm: B as conservative extension of A}, we have $\Diamond \top, \bigsqcup_{\SET w \in \ran(\SET W)} \SET W = \SET w,\Gamma_0\vdash_{\Bx_\sigma} \varphi$. Furthermore, $\T^{\Gamma_0} \models \Gamma_0$ by the truth lemma for $\T^{\Gamma_0}$. Thus, by the soundness of $\Bx_\sigma$, $\T^\Gamma = \T^{\Gamma_0}\models \varphi$. 
\end{proof}

This corollary gives us some control on the $\Hal_\sigma$-formulas that are $\Bx$-provable for singleton models. To extend our grasp to $\Hal^+$ formulas, we aim to show that the $\Hal_\sigma$ formulas that hold in a singleton model completely determine the truth or falsity of $\Hal^+_\sigma$ formulas. We already know that this holds semantically (even for non-singleton models) by theorem \ref{thm: equivalence implies HPLUS equivalence}; we seek here a proof-theoretic counterpart of this result. We need a couple of intermediate lemmas; the first one shows that, if $\Gamma\supseteq \Bx_\sigma$ is a maximal consistent set of formulas, then there is exactly one $\F$ (up to equivalence) such that $\Phi^\F \in \Gamma$.

\begin{lemma}\label{lemma: Gamma determines F}
    If $\Gamma\supseteq \Bx_\sigma$ is a maximal 
    consistent set of $\Hal^+$ formulas, then there is an $\F\in\Fs$ such that $\Phi^\F \in \Gamma$; if $\Phi^\F, \Phi^\G \in \Gamma$, then $\F\equiv\G$.
\end{lemma}

\begin{proof}
    Existence. Since $\Gamma\supseteq \Bx_\sigma$, by axiom J16 we have $\bigsqcup_{\F\in \Fs} \Phi^\F\in \Gamma$. Thus, by the primality of $\Gamma$, there is an $\F$ such that $\Phi^\F\in \Gamma$.

    Uniqueness. Suppose for the sake of contradiction that $\Phi^\F, \Phi^\G \in \Gamma$, where $\F\not\equiv\G$. The statement $\F\not\equiv\G$ implies that there is a $Y \in \mathcal I$ and a tuple $pa\in\ran(PA_Y^\F \cup PA_Y^\G)$ such that $\F_Y(pa_{\upharpoonright PA_Y^\F}) \neq \G_Y(pa_{\upharpoonright PA_Y^\G})$. Suppose wlog that there is a $y \in \F_Y(pa_{\upharpoonright PA_Y^\F}) \setminus \G_Y(pa_{\upharpoonright PA_Y^\G})$. Then,   by J0, $\Phi^\F \in \Gamma$ implies $\langle \SET W_Y = \SET w \rangle Y=y\in\Gamma$ and $\Phi^\G \in \Gamma$  implies $\cneg\langle \SET W_Y = \SET w \rangle Y=y\in\Gamma$ for some $\SET w$ such that $\SET w_{\upharpoonright PA_Y^\F \cup PA_Y^\G} = pa$, contradicting the consistency of $\Gamma$.  
\end{proof}

\begin{lemma}\label{lemma: modal collapse}
    Let $\Gamma\supseteq \Bx_\sigma$ be a maximal 
    consistent set of $\Hal^+_\sigma$ formulas. Then, for each $\varphi\in \Hal^+_\sigma$ there is a $\varphi^*_\Gamma\in\Hal_\sigma$ such that $\Gamma \vdash_{\Bx_\sigma} \varphi \leftrightarrow \varphi^*_\Gamma$.

    Furthermore, $\varphi^*_\Gamma$ is determined just by the causal laws encoded in $\Gamma$, i.e., if $\Phi^\F\in\Gamma$ and $\Phi^\F\in\Gamma'$, then $\varphi^*_\Gamma = \varphi^*_{\Gamma'}$. 
\end{lemma}

\begin{proof}
We define $\varphi^*_\Gamma$ and prove  $\Gamma \vdash \varphi \leftrightarrow \varphi^*_\Gamma$ by induction on $\varphi$.

If $\varphi$ is atomic, we let $\varphi^*_\Gamma:= \varphi$; then,  $\Gamma \vdash \varphi \leftrightarrow \varphi^*_\Gamma$ follows from J0 (since $\Bx_\sigma \subseteq \Gamma$).

If $\varphi$ is $\psi \with \chi$, then we let $\varphi^*_\Gamma := \psi^*_\Gamma \with \chi^*_\Gamma$. By the i.h., $\Gamma \vdash \psi \leftrightarrow \psi^*_\Gamma$ and $\Gamma \vdash \chi \leftrightarrow \chi^*_\Gamma$; thus, $\Gamma \vdash \varphi \leftrightarrow \varphi^*_\Gamma$ follows by I0 and MP. The case for $\varphi$ of the form $\cneg \psi$ is similar (letting $\varphi^*_\Gamma := \cneg\psi^*_\Gamma$).

If $\varphi$ is $[\SET X = \SET x]\psi$, then observe that 
 by lemma \ref{lemma: Gamma determines F}, there is one function component $\F$ (up to equivalence) such that $\Phi^\F\in \Gamma$. We let $\varphi^*_\Gamma:= [\SET X = \SET x] \bigsqcup_{\substack{ t \text{ such that} \\ (\{t\},\F_{\SET X = \SET x})\models \psi}} \SET W=t(\SET W)$. We note that, by theorem \ref{thm: equivalence implies HPLUS equivalence}, if $\F \equiv \G$ then this formula is the same as $[\SET X = \SET x]\bigsqcup_{\substack{ t \text{ such that} \\ (\{t\},\G_{\SET X = \SET x})\models \psi}} \SET W=t(\SET W)$ (as obviously $\F \equiv \G$ implies $\F_{\SET X = \SET x} \equiv \G_{\SET X = \SET x}$); thus, the definition of $\varphi^*_\Gamma$ is determined by the equivalence class of $\F$. Now, since $\Phi^\F\in \Gamma$, by axiom J17 $\Gamma\vdash \varphi \leftrightarrow \varphi^*_\Gamma$.
\end{proof}

\noindent Upon closer inspection, this lemma also proves that, if an $\Hal^+$ formula completely determines the causal laws (i.e., it entails $\Phi^\F$ for some $\F$) then it is equivalent to an $\Hal$ formula. We will not need this fact in the following.

\begin{theorem}\label{thm: rigidity theorem}
    For every $\Gamma_0\supseteq \Ax_\sigma$ maximal $\Ax_\sigma$-consistent set of $\Hal_\sigma$ formulas there is a unique maximal $\Bx_\sigma$-
    consistent set $\Gamma \supseteq \Gamma_0 \cup \Bx_\sigma$ of $\Hal^+$ formulas.
\end{theorem}

\begin{proof}
Existence. Since $\Gamma_0$ includes $\Ax_\sigma$ and is maximally $\Ax_\sigma$-consistent, by lemma \ref{lemma: truth lemma} $\Gamma_0$ has a (singleton) model $T$. Let $\Gamma:= \{\varphi\in\Hal^+_\sigma \mid T\models \varphi\}$. Since $T\models \Gamma$, by the soundness of $\Bx_\sigma$ (lemma \ref{thm: soundness of B}) we have $\Bx_\sigma \subseteq \Gamma$. And since it is the set of $\Hal^+_\sigma$ formulas satisfied by a model, $\Gamma$ is maximally 
consistent. 

Uniqueness. Suppose for the sake of contradiction that there are two distinct $\Gamma,\Gamma'$ satisfying the statement. Say, wlog, that there is a $\varphi\in \Gamma \setminus \Gamma'$. Thus, by maximality of $\Gamma'$, $\cneg\varphi\in\Gamma'$. By lemma \ref{lemma: modal collapse}, there is a $\varphi^*_\Gamma \in \Hal_\sigma$ such that $\Gamma\vdash \varphi \leftrightarrow \varphi^*_\Gamma$; thus, by MP, (*): $\varphi^*_\Gamma\in \Gamma_0\subseteq \Gamma$. By the same lemma, we also have $\Gamma'\vdash \cneg\varphi \leftrightarrow (\cneg\varphi)^*_{\Gamma'}$. But the proof of the lemma shows that $(\cneg\varphi)^*_{\Gamma'} = \cneg\varphi^*_{\Gamma'}$. Furthermore, note that, since $\Phi^\F \in \Hal$, we have $\Phi^\F\in\Gamma  \iff  \Phi^\F \in \Gamma_0  \iff  \Phi^\F \in \Gamma'$. Since the $\varphi^*_\Gamma$ are determined by the $\F$ such that $\Phi^\F \in\Gamma$, we conclude that $\varphi^*_\Gamma = \varphi^*_{\Gamma'}$. Therefore, $\Gamma'\vdash \cneg\varphi \leftrightarrow \cneg \varphi^*_\Gamma$. So, $\cneg \varphi^*_\Gamma \in\Gamma_0\subseteq\Gamma'$. This, together with (*), contradicts the consistency of $\Gamma_0$.       
\end{proof}

We now move towards a completeness proof. the failure of \emph{Weak reversibility} over general models (and the apparent lack of an adequate substitute axiom) prevents us from applying the typical proof strategy used for logics of causal reasoning; \emph{Weak reversibility} seems essential for the usual proof of the truth lemma (cp. lemma \ref{lemma: truth lemma}). Our strategy will be to introduce an (artificial) modal semantics for $\Hal^+$, so that we can use the standard techniques for proving completeness in modal logics \citep[see e.g. ][]{BlaRijVen2001}\footnote{The overall approach used here is inspired by \cite{WanCao2013}.}; this was also the reason for switching from $\Hal$ to $\Hal^+$. In particular, we will build a canonical modal model whose worlds can be identified (modulo equivalence) with our canonicl teams, and prove a truth lemma for it.

Let us fix some notation. We assume that there is a fixed alphabetical order of the variables. For a given signature $\sigma$, we write $I_\sigma$ for the set of all conjunctions of the form $\SET X = \SET x$, where the variables in $\SET X$ are listed alphabetically and without repetitions. 

\begin{definition}
    An unintended model (umodel) for $\sigma$ is a tuple $M=(W,R_{\top}, (R_{\SET X = \SET x})_{\SET X = \SET x \in I_\sigma},V)$, where $W$ is an arbitrary set (of ``worlds''), $R_{\top}$ and the $R_{\SET X = \SET x}$ are binary relations on $W$, 
    and $V$ is a function $W\times\Atom_\sigma\rightarrow \{0,1\}$.
\end{definition}

We can then define the notion of truth at a world. Given a umodel $M$, a world $u$ of $M$ and an $\Hal^+$ formula $\varphi$, the relation $M,u\Vdash \varphi$ is given by the following inductive clauses:
\begin{itemize}
    \item $M, u \Vdash Y=y$ iff $V(u, Y=y)=1$.
    \item $M, u \Vdash \psi \with \chi$ iff $M, u \Vdash \psi$ and $M, u \Vdash \chi$.
    \item $M, u \Vdash \cneg \psi$ iff $M, u \not\Vdash \psi$. 
    \item $M, u \Vdash \Box \psi$ iff for all $v$ such that $u R_{\top} v$, $M,v\Vdash \psi$.
    \item $M, u \Vdash [\SET X = \SET x]\psi$ iff for all $v$ such that $u R_{\SET X = \SET x} v$, $M,v\Vdash \psi$.
\end{itemize}

Umodel-world pairs $M,w$ might in general not be equivalent to any causal team (in the sense that they might not satisfy the same set of $\Hal^+$ formulas as a causal team). However, we introduce a canonical umodel whose worlds are maximal consistent sets\footnote{Throughout the rest of this section, the notions of consistency and maximal consistency are relative to system $\Bx$ unless otherwise specified.}, and they do behave like causal teams. We need to introduce a bit of extra notation. If $\sigma = (\mathcal E,\mathcal I,\ran)$ is a signature and $\SET X \subseteq \mathcal E \cup \mathcal I$, we denote as $\sigma_\SET X$ the signature $(\mathcal E\cup \SET X, \mathcal I\setminus \SET X,\ran)$ and call it a \textbf{derived signature} of $\sigma$. Note that, conveniently, $\Hal^+_{\sigma_\SET X} = \Hal^+_\sigma$ and $\Hal_{\sigma_\SET X} = \Hal_\sigma$; on the other hand, $\Bx_{\sigma_\SET X}$ may differ from $\Bx_\sigma$ (they may contain different instances of axioms I8, J10, J11, J16 and J17). Thus, in general, $\T^\Gamma_{\sigma_\SET X}$ might differ from $\T^\Gamma_\sigma$, even if $\Gamma$ is both a set of  $\Hal^+_{\sigma_\SET X}$ and $\Hal^+_\sigma$ formulas. We denote by $\hat\sigma$ the set of all derived signatures of $\sigma$; note that $\sigma \in \hat\sigma$.  

The \textbf{canonical umodel} $\M_\sigma$ has the following components: 
\begin{itemize}
    \item $W$ is the set of all maximal consistent sets $\Gamma \supseteq \Bx_\tau$ of $\Hal^+_\tau$ formulas, where $\tau\in \hat\sigma$.
    \item $\Gamma R_\top \Delta$ iff, for some $\tau\in\hat\sigma$,  $\Gamma,\Delta\supseteq \Bx_\tau$ and:
        \begin{enumerate}
            \item for some $\SET w\in \ran(\SET W)$,  $\Diamond\top, \Box \SET W = \SET w \in \Delta$
            \item for all $Y\in \mathcal I, y\in\ran(Y)$ and $\SET w \in \ran(\SET W_Y)$, if $\langle \SET W_Y = \SET w\rangle Y=y \in \Gamma$, then it is also in $\Delta$
            \item for all $\chi \in \Hal^+_\sigma$, if $\chi\in\Delta$, then $\Diamond \chi \in \Gamma$.
        \end{enumerate}
        
   \item $\Gamma R_{\SET X = \SET x} \Delta$ iff for some $\tau\in\hat\sigma$, we have $\Gamma\supseteq \Bx_\tau$, $\Delta\supseteq \Bx_{\tau_\SET X}$ and:
        \begin{enumerate}
            \item for some $\SET w\in \ran(\SET W_\SET X)$,  $\Diamond\top, \Box (\SET X\SET W_\SET X = \SET x \SET w) \in \Delta$
            \item for all $Y\in \mathcal I\setminus \SET X, y\in\ran(Y)$ and $\SET w \in \ran(\SET W_Y)$, if $\langle \SET W_Y = \SET w\rangle Y=y \in \Gamma$, then it is also in $\Delta$
            \item for all $\chi \in \Hal^+_\sigma$, if $\chi\in\Delta$, then $\langle \SET X = \SET x \rangle \chi \in \Gamma$.
        \end{enumerate}
    
    \item $V(\Gamma, Y=y)=1$ iff $Y=y \in \Gamma$
\end{itemize}
We will simply write $\M$ when the intended signature is not ambiguous. In the definition of the accessibility relations for $\M$, clause 3. is the standard one used in the definition of canonical models in modal logic. The purpose of the additional clause 1. is to guarantee  that, if a world $\Delta$ is accessible from some other world, then it behaves analogously as a singleton causal model. This is to mimick the fact that the evaluation of a formula $[\SET X = \SET x]\psi$ on a causal model amounts to checking whether $\psi$ holds on \emph{singleton} submodels of the intervened model. Finally, clause 2. (together with 3.) guarantees that such singleton models have the same causal laws as the initial causal model -- except for the variables $\SET X$ that have been intervened upon. Concerning these variables, the fact that $\Delta$ has signature $\tau_\SET X$ guarantees that they behave as external variables.


\begin{lemma}\label{lemma: Box version of canonical accessibility}
With notations as above:
\begin{enumerate}
    \item In the definition of $\Gamma R_\top \Delta$, clause 3. can be replaced with: for all $\varphi \in \Hal^+_\sigma$, $\Box \varphi \in \Gamma$ implies $\varphi \in \Delta$. 
    \item In the definition of $\Gamma R_{\SET X = \SET x} \Delta$, clause 3. can be replaced with: for all $\varphi \in \Hal^+_\sigma$, $[\SET X = \SET x] \varphi \in \Gamma$ implies $\varphi \in \Delta$.
\end{enumerate}
\end{lemma}

\begin{proof}
We only prove point 2., as the proof of point 1. is analogous. 



 From left to right, suppose $\Gamma R_{\SET X = \SET x} \Delta$ and $\varphi \notin \Delta$. By lemma \ref{lemma: properties of MCS}, 2., $\cneg \varphi \in \Delta$. By clause 3. of the definition of $\M$, then, $\langle\SET X = \SET x\rangle \cneg \varphi \in \Gamma$. By lemma \ref{lemma: basic properties of B}, 6. $\cneg [\SET X = \SET x]  \varphi \in \Gamma$. Thus, by the consistency of $\Gamma$, $[\SET X = \SET x]  \varphi \notin \Gamma$.  

 From right to left, 
 suppose  $[\SET X = \SET x]\varphi \in \Gamma$ implies $\varphi \in \Delta$ for all formulas $\varphi \in \Hal^+_\sigma$. By contraposition, $\varphi \notin \Delta$ (which is equivalent to $\cneg \varphi \in \Delta$ by lemma \ref{lemma: properties of MCS}, 2.) implies $[\SET X = \SET x] \varphi \notin \Gamma$, thus $\cneg[\SET X = \SET x] \varphi \in \Gamma$ by lemma \ref{lemma: properties of MCS}, 2., and finally $\langle \SET X = \SET x\rangle \cneg\varphi \in \Gamma$ by lemma \ref{lemma: basic properties of B}, 6. Since every formula $\chi$ is equivalent to a negated formula (say, $\chi \equiv \cneg\cneg\chi$), this is just a reformulation of clause 3. in the definition of $\M$.
\end{proof}

\begin{lemma}\label{lemma: possible that psi and Box W = w}
    Let $\psi$ be an $\Hal^+_\sigma$ formula and $\Gamma \supseteq \Bx_\sigma$ be a maximal consistent set of $\Hal^+_\sigma$ formulas.

    \begin{enumerate}
        \item Suppose $\Diamond \psi \in \Gamma$. Then, there is a $\SET w \in \ran(\SET W)$ such that $\Diamond( \psi \with \Box\SET W = \SET w)\in\Gamma$.

        \item Suppose $\langle\SET X = \SET x\rangle \psi \in \Gamma$. Then, there is a $\SET w \in \ran(\SET W_\SET X)$ such that $\langle\SET X = \SET x\rangle(\psi \with \Box \SET X\SET W_{\SET X} = \SET x \SET w)\in\Gamma$.
    \end{enumerate}
\end{lemma}

\begin{proof}
We prove only statement 2. as the proof of statement 1. is analogous.


 Suppose $\langle \SET X = \SET x \rangle \psi \in \Gamma$. By axiom J2, $[\SET X = \SET x]\bigsqcup_{\SET x\SET w \in \ran(\SET X\SET W_{\SET X})}\SET X\SET W_{\SET X} = \SET x\SET w \in \Gamma$. By axiom I7 and replacement, $[\SET X = \SET x]\bigsqcup_{\SET x\SET w \in \ran(\SET X\SET W_{\SET X})}\Box\SET X\SET W_{\SET X} = \SET x\SET w \in \Gamma$. Thus, by lemma \ref{lemma: basic properties of B}, 10., $\langle \SET X = \SET x \rangle (\psi \with \bigsqcup_{\SET x\SET w \in \ran(\SET X\SET W_{\SET X})}\Box\SET X\SET W_{\SET X} = \SET x\SET w)\in \Gamma$. By J0 and replacement, $\langle \SET X = \SET x \rangle \bigsqcup_{\SET x\SET w \in \ran(\SET X\SET W_{\SET X})}(\psi \with \Box\SET X\SET W_{\SET X} = \SET x\SET w) \in \Gamma$. By lemma \ref{lemma: basic properties of B}, 8., $\bigsqcup_{\SET x\SET w \in \ran(\SET X\SET W_{\SET X})}\langle \SET X = \SET x \rangle(\psi \with \Box\SET X\SET W_{\SET X} = \SET x\SET w) \in \Gamma$. By primality of $\Gamma$ (lemma \ref{lemma: properties of MCS}, 4.), there is an $\SET x'\SET w \in \ran(\SET X\SET W_{\SET X})$ such that $\langle \SET X = \SET x \rangle(\psi \with \Box\SET X\SET W_{\SET X} = \SET x'\SET w)\in \Gamma$.
%
 Now, suppose for the sake of contradiction that $\SET x' \neq \SET x$. Note that, from the previous statement, by lemma \ref{lemma: basic properties of B}, 11., and J0 we obtain $\langle\SET X = \SET x\rangle\Box\SET X\SET W_{\SET X} = \SET x'\SET w\in \Gamma$. By I7 and replacement, $\langle\SET X = \SET x\rangle\SET X\SET W_{\SET X} = \SET x'\SET w\in \Gamma$. Thus, $\langle\SET X = \SET x\rangle\SET X = \SET x'\in \Gamma$ by I0 and monotonicity.     But $[\SET X = \SET x]\SET X = \SET x\in \Gamma$ by axiom I4; thus, by lemma \ref{lemma: basic properties of B}, 10., $\langle\SET X = \SET x\rangle(\SET X = \SET x \with \SET X = \SET x')\in \Gamma$. By the same lemma, J1 and J0, we obtain $\langle\SET X = \SET x\rangle \bot \in \Gamma$, i.e. $\cneg[\SET X = \SET x]\cneg \bot \in \Gamma$. On the other hand, $[\SET X = \SET x]\cneg \bot \in \Gamma$ by J0 and NEC, contradicting the consistency of $\Gamma$. Thus, $\SET x' = \SET x$.    
\end{proof}

\begin{lemma}[Existence lemma]\label{lemma: existence lemma}
Let $\Gamma \supseteq \Bx_\sigma$ be a maximal 
consistent set of $\Hal^+_\sigma$ formulas, $\varphi$ an $\Hal^+_\sigma$ formula. Then:
\begin{enumerate}
    \item if $\Diamond \varphi \in \Gamma$, then there is a maximal $\Bx_\sigma$-consistent set $\Delta$ of $\Hal^+_\sigma$ formulas such that  $\Gamma R_\top \Delta$ and $\varphi \in \Delta$. 
    \item if $\langle \SET X = \SET x \rangle \varphi \in \Gamma$, then there is a maximal consistent $\Bx_{\sigma_\SET X}$ set $\Delta$ of $\Hal^+_\sigma$ formulas such that  $\Gamma R_{\SET X = \SET x} \Delta$ and $\varphi \in \Delta$.
\end{enumerate}
    
\end{lemma}

\begin{proof}
   We only prove point 2., as the proof of point 1. is analogous.

    Suppose $\langle \SET X = \SET x \rangle  \varphi \in \Gamma$. We first observe that, since $\langle \SET X = \SET x \rangle  \varphi\in\Gamma$, by lemma \ref{lemma: possible that psi and Box W = w} there is a $\SET w\in \ran(\SET W)$ such that $\langle \SET X = \SET x \rangle (\varphi \with \Box \SET W = \SET w)\in\Gamma$ and $\SET w_{\upharpoonright \SET X} = \SET x$. Let $\Delta_1 = \{ \psi \mid [\SET X = \SET x] \psi \in \Gamma\}$, $\Delta_2 = \{\langle \SET W_Y = \SET w\rangle Y=y \in \Gamma \mid Y\in \mathcal I \setminus \SET X, \SET wy \in \ran(\SET W_YY)\}$ and     
    $\Delta_0 = \Delta_1 \cup \Delta_2 \cup \{\varphi, \Diamond \top, \Box \SET W = \SET w\}$. Suppose for the sake of contradiction that $\Delta_0$ is not consistent in $\Bx_{\sigma_\SET X}$. Since $\Bx_{\sigma_\SET X}$ is finitary, this means that there are formulas $\psi_1,\dots,\psi_n\in \Delta_1$ and $\chi_1: \langle \SET W_{Y_1} = \SET w_1\rangle Y_1=y_1, \dots, \chi_m: \langle \SET W_{Y_m} = \SET w_m\rangle Y_m=y_m\in \Delta_2$ such that $\psi_1, \dots, \psi_n,\chi_1,\dots,\chi_m,\varphi, \Diamond \top, \Box \SET W = \SET w \vdash \bot$. By some applications of the deduction theorem and J0, we obtain $\vdash (\psi_1 \with \dots \with \psi_n \with \chi_1 \with \dots \with \chi_m \with \Diamond \top) \rightarrow \cneg(\varphi \with \Box \SET W = \SET w)$. By NEC, I5,  lemma \ref{lemma: basic properties of B}, 4. and replacement,  $\vdash ([\SET X = \SET x]\psi_1 \with \dots \with [\SET X = \SET x]\psi_n \with [\SET X = \SET x]\chi_1 \with \dots \with [\SET X = \SET x]\chi_m  \with [\SET X = \SET x]\Diamond \top) \rightarrow [\SET X = \SET x]\cneg(\varphi \with \Box \SET W = \SET w)$. We can remove the diamond using axiom J12 and replacement,  and then by $m$ applications of J11 we obtain $\vdash ([\SET X = \SET x]\psi_1 \with \dots \with [\SET X = \SET x]\psi_n \with \chi_1 \with \dots \with \chi_m \with  [\SET X = \SET x]\top) \rightarrow [\SET X = \SET x]\cneg(\varphi \with \Box \SET W = \SET w)$ (applying J11 is correct because $Y_1,\dots, Y_m$ are internal and not in $\SET X$). But now, $[\SET X = \SET x] \psi_1,\dots,[\SET X = \SET x]\psi_n\in \Gamma$ (by definition of $\Delta_1$), $\chi_1, \dots, \chi_m\in \Gamma$ (by definition of $\Delta_2$) and $[\SET X = \SET x]\top \in\Gamma$ (by J0 and NEC). Thus, by MP, $[\SET X = \SET x]\cneg(\varphi \with \Box \SET W = \SET w)\in\Gamma$. Then, by lemma \ref{lemma: basic properties of B}, 7.,  $\cneg\langle \SET X = \SET x \rangle (\varphi \with \Box \SET W = \SET w)\in\Gamma$. But we already know that $\langle \SET X = \SET x \rangle (\varphi \with \Box\SET W = \SET w)\in\Gamma$, so we contradict the consistency of $\Gamma$.
   
   Thus, $\Delta_0$ is consistent. By the Lindenbaum lemma there is a maximal $\Bx_{\sigma_\SET X}$-consistent set $\Delta \supseteq \Delta_0$. Note that $[\SET X = \SET x] \psi\in \Gamma$ implies $\psi \in \Delta$ for all $\psi$. By lemma \ref{lemma: Box version of canonical accessibility}, then, we obtain clause 3. of the definition of $R_{\SET X = \SET x}$. Furthermore, we have that 
   $\Diamond \top\in\Delta$ and $\Box \SET W = \SET w \in \Delta$ with $\SET w_{\upharpoonright \SET X} = \SET x$; 
   this is clause 1. of the definition of $R_{\SET X = \SET x}$. Finally, $\Delta_2 \subseteq\Delta$; this is clause 2. Thus, $\Gamma R_{\SET X = \SET x} \Delta$. Finally, we observe that $\varphi\in\Delta$, as needed.  
\end{proof}

\begin{lemma}[Truth lemma for $\M$]\label{lemma: truth lemma for M}
    Let $\Gamma \supseteq \Bx_\sigma$ be a maximally consistent set of $\Hal^+_\sigma$ formulas, $\varphi$ an $\Hal^+_\sigma$ formula. Then, $\M,\Gamma\Vdash \varphi$ iff $\varphi\in\Gamma$.
\end{lemma}

\begin{proof}
    By some applications of I0 and replacement, we can assume up to equivalence that all modal operators $[\SET X =\SET  x]$ occurring in $\varphi$ are replaced with  $\cneg\cneg[\SET X =\SET x]\cneg\cneg$, i.e. $\cneg \langle\SET X =\SET x \rangle\cneg$. We can then proceed by induction on the subformulas $\psi$ of $\varphi$, thought as combinations of atoms, $\cneg$, $\with$ and $\langle \SET X =\SET x\rangle$ operators; we prove the statement simultaneously for all $\Gamma$.
    
    If $\psi$ is an atom $Y=y$, then    $\M, \Gamma \Vdash Y=y$ iff (by definition of $\Vdash$) $V(\Gamma,Y=y)=1$ iff (by definition of $\M$) $Y=y\in\Gamma$. The cases for $\psi= \chi_1 \with\chi_2$ or $\psi = \cneg \chi$ are also straightforward.

    If $\psi$ is of the form $\langle\SET X =\SET x\rangle\chi$, then $\M, \Gamma \Vdash \langle\SET X =\SET x\rangle\chi$ iff (by definition of $\Vdash$) there is a $\Delta$ such that $\Gamma R_{\SET X = \SET x} \Delta$ and  $\M,\Delta\Vdash \chi$, iff (induction hypothesis) there is a $\Delta$ such that $\Gamma R_{\SET X = \SET x} \Delta$ and $\chi \in \Delta$. By clause 3. of the definition of $R_{\SET X = \SET x}$, then, the latter implies $\langle\SET X =\SET x\rangle\chi \in \Gamma$. Vice versa, if $\langle\SET X =\SET x\rangle\chi \in \Gamma$, by the existence lemma \ref{lemma: existence lemma} there is a $\Delta$ such that $\Gamma R_{\SET X = \SET x} \Delta$ and $\chi \in \Delta$.
\end{proof}

\begin{corollary}
   If $\Gamma$ is a world of $\M_\sigma$ of signature $\sigma_\SET X$, then it satisfies the axioms of $\Bx_{\sigma_\SET X}$.
\end{corollary}

\begin{lemma}[Canonical singletons]\label{lemma: canonical singletons}
    Let $\Delta \supseteq \Bx_\sigma$ be a maximally 
    consistent set of $\Hal^+_\sigma$ formulas, and let $\Delta_0$ be the set of $\Hal$ formulas in $\Delta$. If $\Diamond \top, \Box \SET W = s(\SET W) \in \Delta$, then $\Delta_0 = \{\chi \in \Hal_\sigma \mid (\{s\}, \F^\Delta) \models \chi \}$.\footnote{Remember that $\F^\Delta$ denotes the law component of the canonical causal team $\T^\Delta$.}
\end{lemma}

\begin{proof}
    By lemma \ref{lemma: basic properties of B}, 9.,  $\Diamond \top, \Box \SET W = s(\SET W) \in \Delta_0\subseteq\Delta$ implies that $\Diamond \SET W = s(\SET W) \in\Delta_0$. Furthermore, by axiom I1,  $\vdash\Box \SET W = s(\SET W) \rightarrow \Box \SET W \neq t(\SET W)$ if $s\neq t$; thus, $\Box \SET W \neq t(\SET W)\in\Delta$. 
    If we had $\Diamond \SET W = t(\SET W)\in \Delta_0$ for some $t\neq s$, then, by lemma \ref{lemma: basic properties of B}, 10., we would obtain $\Diamond (\SET W = t(\SET W) \with \SET W \neq t(\SET W))$; thus, by J0 and replacement, $\Diamond\bot\in \Delta_0\subseteq \Delta$; this contradicts lemma \ref{lemma: properties of MCS}, 5. Thus, $s$ is the unique assignment such that $\Diamond \SET W = s(\SET W) \in\Delta$. By definition of $(\T^\Delta_\Bx)^-$, then,  we have $(\T^\Delta_\Bx)^- = \{s\}$. Thus, $(\{s\}, \F^{\Delta_0}) = (\{s\}, \F^\Delta) = \T^\Delta$. 
    By corollary \ref{cor: truth corollary}, then, $\Delta =  \{\chi \in \Hal^+_\sigma \mid \T^\Delta \models \chi \}$; thus, $\Delta_0 = \{\chi \in \Hal_\sigma \mid \T^{\Delta}\models \chi \} = \{\chi \in \Hal_\sigma \mid (\{s\}, \F^\Delta) \models \chi \}$. 
\end{proof}

\noindent As we said early on, the canonical model $\M$ is designed so that each of its worlds $\Gamma$ would behave as the canonical causal team $\T^\Gamma$, and each world $\Delta$ with $\Gamma R_{\SET X = \SET x} \Delta$ would behave as one of the singleton models that are produced by the intervention $do(\SET X = \SET x)$ applied to $\T^\Gamma$. The second of these claims, unfortunately, will not in general be correct; the law component $\F^\Delta$ of $\T^\Delta$ might disagree with the law component $\F^\Gamma_{\SET X = \SET x}$ of $\T^\Gamma_{\SET X = \SET x}$ about the parent sets of some variables.\footnote{Intuitively, the reason is that, after fixing the value of $\SET X$ to $\SET x$, information is lost, and relationships of direct cause that were ``visible'' in $\Gamma$ are not detectable anymore in $\Delta$.} 
We will see that, nonetheless, $\T^\Delta$ is $\equiv$-equivalent to a corresponding singleton submodel of $\T^\Gamma_{\SET X = \SET x}$, and thus it satisfies the same $\Hal^+$ formulas.  


\begin{lemma}\label{lemma: post-intervention equivalence}
    Let $\tau\in \hat\sigma$. Let $\Gamma \supseteq \Bx_\tau$ and  $\Delta\supseteq \Bx_{\tau_\SET X}$ be maximal 
    consistent sets of $\Hal^+_\sigma$ formulas. If $\Gamma R_{\SET X = \SET x} \Delta$, then $\F^\Delta \equiv \F^\Gamma_{\SET X = \SET x}$.
\end{lemma}

\begin{proof}
    The causal laws for an internal variable $Y$ do not necessarily have the same parent set in $\F^\Delta$ as in $\F^\Gamma_{\SET X = \SET x}$; so, we shall write $PA^\Delta_Y$, resp. $PA^\Gamma_Y$ to differentiate them. Note that $\Gamma R_{\SET X = \SET x} \Delta$ entails that $\F^\Delta$ and $\F^\Gamma_{\SET X = \SET x}$ have the same signature (namely, $\tau_\SET X$), so that the definition of $\equiv$ may apply. Let us consider a tuple of values $pa \in \ran(PA^\Delta_Y \cup PA^\Gamma_Y)$; to conclude that $\F^\Delta \equiv \F^\Gamma_{\SET X = \SET x}$ we need to prove that $\F^\Delta_Y(pa_{\upharpoonright PA^\Delta_Y}) = \F^\Gamma_Y(pa_{\upharpoonright PA^\Gamma_Y})$ when $Y\in \mathcal I\setminus \SET X$.\footnote{Note that $\F^\Gamma_Y = (\F^\Gamma_{\SET X = \SET x})_Y$ when $Y\in \mathcal I\setminus \SET X$.}

    Suppose $y\in \F^\Delta_Y(pa_{\upharpoonright PA^\Delta_Y})$. By definition of $\F^\Delta$, this means that, for any $\SET w \in \ran(\SET W_Y)$ with $\SET w_{\upharpoonright PA^\Delta_Y} = pa_{\upharpoonright PA^\Delta_Y}$, $\langle \SET W_Y = \SET w\rangle Y = y \in \Delta$. In particular, this holds for any $\SET w \in \ran(\SET W_Y)$ with $\SET w_{\upharpoonright PA^\Delta_Y \cup PA^\Gamma_Y} =pa$; fix one such tuple $\SET w$. 
    By clause 3. of the definition of $R_{\SET X = \SET x}$, then, $\langle \SET X = \SET x \rangle\langle \SET W_Y = \SET w\rangle Y = y \in  \Gamma$. Since 
    $Y\notin \SET X$, by axiom J10 we have $\langle \SET W_Y = \SET w\rangle Y = y \in  \Gamma$. Thus, by definition of $\F^\Gamma_Y$, $y \in \F^\Gamma_Y(pa_{\upharpoonright PA^\Gamma_Y})$.

    Vice versa, suppose $y \in \F^\Gamma_Y(pa_{\upharpoonright PA^\Gamma_Y})$. Reasoning as before, there is a $\SET w \in \ran(\SET W_Y)$ with $\SET w_{\upharpoonright PA^\Delta_Y \cup PA^\Gamma_Y} =pa$ such that $\langle \SET W_Y = \SET w\rangle Y = y \in \Gamma$. But then, $\langle \SET W_Y = \SET w\rangle Y = y \in \Delta$ by clause 2. of the definition of $R_{\SET X = \SET x}$. 
    %
\end{proof}


The following two lemmas show that the singleton subteams of a canonical model $\T^\Gamma$, resp. those of $\T^\Gamma_{\SET X = \SET x}$, can be identified (up to $\equiv$-equivalence) with well-chosen canonical models.

\begin{lemma}\label{lemma: singleton subteams as canonical teams}
Let $\Gamma\supseteq \Bx_\sigma$ be a maximal 
consistent set of $\Hal^+_\sigma$ formulas. If $s\in (\T^\Gamma)^-$, then $(\{s\},\F^\Gamma) \equiv \T^\Delta$, where $\Delta = \{\varphi\in\Hal^+_\sigma  \mid  \Diamond(\SET W=s(\SET W) \with \varphi)\in\Gamma \}$.  Furthermore, $\Gamma R_\top \Delta$.   
\end{lemma}

\begin{proof}
First of all, note that it is easy to show that, for any formula $\varphi$, $(\{s\},\F^\Gamma)\models \varphi$ iff $\T^\Gamma \models  \Diamond(\SET W=s(\SET W) \with \varphi)$. Therefore, $(\{s\},\F^\Gamma)\models \Delta$ and it satisfies no $\Hal^+_\sigma$ formula outside of $\Delta$. Thus, $\Delta$ is a maximal $\Bx_\sigma$-consistent set of $\Hal^+_\sigma$ formulas.


Thus, $\T^\Delta$ is well-defined. Since $\Delta$ is the set of \emph{all} formulas satisfied by $(\{s\},\F^\Gamma)$, and $(\{s\},\F^\Gamma)$ is a model of signature $\sigma$, by the soundness of $\Bx_\sigma$ we have $\Delta \supseteq \Bx_\sigma$. Let $\Delta_0$ be the set of all $\Hal$ formulas in $\Delta$. If we can prove that $\T^\Delta$ is a singleton model, then by corollary \ref{cor: truth corollary} $\T^\Delta \models \Delta_0$, i.e., $\T^\Delta$ and $(\{s\},\F^\Gamma)$ satisfy the same $\Hal_\sigma$ formulas; thus, by theorem \ref{thm: equivalence implies HPLUS equivalence}, $\T^\Delta \equiv (\{s\},\F^\Gamma)$. 

Let us then prove that  $\T^\Delta$ is a singleton model. Since $s\in (\T^\Gamma)^-$, by definition of $\T^\Gamma$ we have $\Diamond \SET W = s(\SET W)\in \Gamma$. Thus, $\Diamond (\SET W = s(\SET W) \with \SET W = s(\SET W))\in \Gamma$ by I0 and replacement; so, by the definition of $\Delta$, $\SET W = s(\SET W) \in \Delta$. By axiom I7, (*): $\Box\SET W = s(\SET W) \in \Delta$.  


On the other hand, since $\Diamond \SET W = s(\SET W)\in \Gamma$, by axiom J15 $\Diamond (\SET W = s(\SET W)\with \Diamond \top)\in \Gamma$. Thus, $\Diamond \top \in \Delta$. Together with (*), this gives (by lemma \ref{lemma: basic properties of B}, 10.) $\Diamond\SET W = s(\SET W) \in \Delta$. Thus, by definition of $\T^\Delta$, $s\in (\T^\Delta)^-$.

If $t\neq s$, then  $\Diamond\SET W = t(\SET W) \in \Delta$, together with (*), would imply (lemma \ref{lemma: basic properties of B}, 10.) that $\Diamond (\SET W = s(\SET W) \with \SET W = t(\SET W))\in \Delta$. Together with I1 and J0, this yields $\Diamond \bot$, contradicting lemma \ref{lemma: properties of MCS}, 5. Thus, for each $t\neq s$, $\Diamond\SET W = t(\SET W) \notin \Delta$. So, by definition of $\T^\Delta$, $(\T^\Delta)^-$ is the singleton $\{s\}$.

Lastly, let us prove that $\Gamma R_\top \Delta$. We already know that $\Gamma,\Delta \supseteq \Bx_\sigma$. We need to verify the three clauses of the definition of $R_\top$.

\underline{Clause 1}. Since obviously $(\{s\},\F^\Gamma)\models \Diamond\top, \Box \SET W =  s(\SET W)$ and we have shown $(\{s\},\F^\Gamma) \equiv \T^\Delta$, we have $T^\Delta\models \Diamond\top, \Box \SET W = s(\SET W)$.  Since $\Diamond\top, \Box \SET W = s(\SET W)$ are $\Hal$ formulas, 
and furthermore $\T^{\Delta_0} = \T^\Delta$ is a singleton, by corollary \ref{cor: truth corollary} we obtain $\Diamond\top, \Box \SET W = s(\SET W) \in \Delta_0 \subseteq \Delta$, as needed.

\underline{Clause 2}. Let $Y\in \mathcal I$ (where $\sigma = (\mathcal E, \mathcal I,\ran)$), and suppose $\langle \SET W_Y = \SET w\rangle Y=y \in \Gamma$. Since 
we have $\Diamond \SET W = s(\SET W) \in \Delta$, 
by definition of $\Delta$, $\Diamond (\SET W =  s(\SET W) \with \SET W =  s(\SET W)) \in \Gamma$. By J0 and replacement, $\Diamond \SET W = s(\SET W) \in \Gamma$. On the other hand, by axiom J11, from the initial assumption we get $\Box\langle \SET W_Y = \SET w\rangle Y=y \in \Gamma$. So, by lemma \ref{lemma: basic properties of B}, 10., we obtain $\Diamond (\SET W = s(\SET W) \with \langle \SET W_Y = \SET w\rangle Y=y)\in \Gamma$. Thus, by definition of $\Delta$, $\langle \SET W_Y = \SET w\rangle Y=y \in \Delta$.

\underline{Clause 3}. Suppose $\chi\in\Delta$. By definition of $\Delta$, then, $\Diamond (\SET W = s(\SET W) \with \chi) \in\Gamma$. Thus, $\Diamond \chi \in \Gamma$ by lemma \ref{lemma: basic properties of B}, 11.
\end{proof}

\begin{lemma}\label{lemma: singleton subteams of intervened canonical teams}
    Let $\tau\in\hat\sigma$, and $\SET X$ be a tuple of variables in the domain of $\sigma$. Let $\Gamma\supseteq \Bx_\tau$ be a maximal $\Bx$-consistent set of $\Hal^+_\sigma$ formulas. If $s\in (\T^\Gamma_{\SET X = \SET x})^-$, then there is a a maximal consistent set $\Delta\supseteq \Bx_{\tau_{\SET X}}$ of $\Hal^+_\sigma$ formulas such that 1) $\Gamma R_{\SET X = \SET x} \Delta$ and 2) $\T^\Delta_{\tau_{\SET X}} \equiv (\{s\},\F^\Gamma_{\SET X = \SET x})$. 
\end{lemma}

\begin{proof}
In case $\SET X = \emptyset$, this is just lemma \ref{lemma: singleton subteams as canonical teams}. Let us assume henceforth that $\SET X \neq \emptyset$.

We let $\Delta := \{\varphi \in \Hal^+_\sigma  \mid  \langle \SET X = \SET x \rangle (\SET W = s(\SET W) \with \varphi) \in \Gamma\}$. Let us verify that $\Delta$ is maximally consistent and $\Bx_{\tau_{\SET X}} \subseteq \Delta$ (so that $\T^\Delta$ is well-defined), and that it satisfies 1) and 2).

Observe first that, for any formula $\varphi$, $(\{s\},\F^\Gamma_{\SET X = \SET x})\models \varphi$ iff $\T^\Gamma_{\tau} \models  \langle \SET X = \SET x \rangle(\SET W=s(\SET W) \with \varphi)$. Therefore, $(\{s\},\F^\Gamma_{\SET X = \SET x})$ is a model for $\Delta$, and it satisfies no $\Hal^+ _\sigma$ formulas outside $\Delta$. From this we obtain, exactly as in the proof of lemma \ref{lemma: singleton subteams as canonical teams}, that $\Delta$ is maximally consistent and (since $(\{s\},\F^\Gamma_{\SET X = \SET x})$ is a model of signature $\tau_{\SET X}$) $\Bx_{\tau_{\SET X}} \subseteq \Delta$. 

1) We need to show that clauses 1.-3. of the definition of $R_{\SET X = \SET x}$ are satisfied.

\underline{Clause 1}. Since $s\in (\T^\Gamma_{\SET X = \SET x})^-$, we have $\T^\Gamma \models \langle \SET X = \SET x \rangle \SET W = s(\SET W)$. But then, there is a $t\in (\T^\Gamma)^-$ such that $(\{t\}, \F^\Gamma) \models \langle \SET X = \SET x \rangle \SET W = s(\SET W)$. Now, by lemma \ref{lemma: singleton subteams as canonical teams}, $(\{t\}, \F^\Gamma)\equiv \T^\Lambda$ for the maximal consistent set $\Lambda:= \{\varphi\in\Hal^+_\sigma  \mid  \Diamond(\SET W=t(\SET W) \with \varphi)\in\Gamma \}$; thus, by theorem \ref{thm: equivalence implies HPLUS equivalence}, $\T^\Lambda \models \langle \SET X = \SET x \rangle \SET W = s(\SET W)$.  Since this is an $\Hal$ formula, and $\T^\Lambda$ is a singleton model, by corollary \ref{cor: truth corollary} we have $\langle \SET X = \SET x \rangle \SET W = s(\SET W) \in \Lambda$. Therefore, $\Diamond (\SET W =  t(\SET W) \with \langle \SET X = \SET x \rangle \SET W = s(\SET W) )  \in \Gamma$. By lemma \ref{lemma: basic properties of B}, 11., $\Diamond\langle \SET X = \SET x \rangle \SET W = s(\SET W)  \in \Gamma$. Thus, by axiom J14, $\langle \SET X = \SET x \rangle \SET W = s(\SET W)  \in \Gamma$. By axiom J15, $\langle \SET X = \SET x \rangle (\SET W = s(\SET W) \with \Diamond \top) \in \Gamma$. Thus, by definition of $\Delta$, we have $\Diamond\top\in \Delta$, as needed.

Note also that, from $\langle \SET X = \SET x \rangle (\SET W = s(\SET W)  ) \in \Gamma$, by J0 and replacement we obtain $\langle \SET X = \SET x \rangle (\SET W = s(\SET W) \with \SET W = s(\SET W) ) \in \Gamma$. Thus, by definition of $\Delta$, $\SET W = s(\SET W) \in \Delta$. Since we have already shown $\Bx_{\tau_{\SET X}} \subseteq \Delta$, and $\Delta$, being maximally consistent, is closed under $\vdash_{\Bx_{\tau_{\SET X}}}$ (lemma \ref{lemma: properties of MCS}, 1.) by axiom I7 we conclude $\Box\SET W = s(\SET W) \in \Delta$. If we can prove that $s(\SET X)= \SET x$, then, we are done. But that immediately follows from the assumption $s\in (\T^\Gamma_{\SET X = \SET x})^-$.

\underline{Clause 2}. Suppose $\langle \SET W_Y = \SET w\rangle Y=y \in \Gamma$ for an internal $Y\notin\SET X$. From this we obtain, by axiom J11, that (*): $[\SET X = \SET x]\langle \SET W_Y = \SET w\rangle Y=y \in \Gamma$. On the other hand, note that, since $\Diamond\top, \Box\SET W = s(\SET W)\in \Delta$, by lemma \ref{lemma: basic properties of B}, 10, we have $\Diamond\SET W = s(\SET W)\in\Delta$. By definition of $\Delta$, then,  $\langle \SET X = \SET x \rangle (\SET W = s(\SET W) \with \Diamond\SET W = s(\SET W)) \in \Gamma$. By lemma \ref{lemma: basic properties of B}, 11., $\langle \SET X = \SET x \rangle\Diamond\SET W = s(\SET W) \in \Gamma$. By axiom J13, then, $\langle \SET X = \SET x \rangle \SET W = s(\SET W) \in \Gamma$.  
This, together with (*), yields $\langle\SET X = \SET x\rangle (\SET W = s(\SET W) \with  \langle \SET W_Y = \SET w\rangle Y=y) \in \Gamma$ by lemma \ref{lemma: basic properties of B}, 10. Thus, by definition of $\Delta$, $\langle \SET W_Y = \SET w\rangle Y=y \in \Delta$.

\underline{Clause 3}. Suppose $\chi\in \Delta$. By definition of $\Delta$, $\langle\SET X = \SET x\rangle (\SET W = s(\SET W)\with \chi)\in \Gamma$. By lemma \ref{lemma: basic properties of B}, 11., then, $\langle\SET X = \SET x\rangle  \chi\in \Gamma$. 

2) 
Let $\Delta_0$ be the set of $\Hal$ formulas in $\Delta$. Since $\T^\Delta$ is a singleton, by corollary \ref{cor: truth corollary} we have $\T^\Delta\models \Delta_0$. Thus, $\T^\Delta$ and $(\{s\},\F^\Gamma_{\SET X = \SET x})$ satisfy the same $\Hal_\sigma$ formulas. By theorem \ref{thm: equivalence implies HPLUS equivalence}, then, $\T^\Delta \equiv (\{s\},\F^\Gamma_{\SET X = \SET x})$.
\end{proof}

\begin{lemma}[Representation lemma]\label{lemma: connection lemma}
    Let $\tau\in\hat\sigma$, and let $\Gamma \supseteq \Bx_\tau$ be a maximally consistent set of $\Hal^+_\sigma$ formulas, $\varphi$ an $\Hal^+_\sigma$ formula. Then, $\M,\Gamma \Vdash \varphi$ iff $\T^\Gamma \models \varphi$. 
\end{lemma}

\begin{proof}
    By induction on $\varphi$. The cases for $\with$ and $\cneg$ are straightforward.

    Atomic case: $\varphi$ is $Y=y$. Since $Y=y$ is atomic, corollary \ref{cor: truth corollary} 
    tells us that $\T^\Gamma\models Y=y$ iff $Y=y \in \Gamma$, iff (by definition of $\M$) $V(\Gamma,Y=y)=1$ iff (by definition of $\Vdash$) $M,\Gamma \Vdash Y=y$.

    Case $\varphi$ is $[\SET X=\SET x]\psi$. 

    $\Rightarrow$) Suppose $\M,\Gamma \Vdash [\SET X=\SET x]\psi$. 
     Since $\M,\Gamma \Vdash [\SET X=\SET x]\psi$, we have that, for all $\Delta$ such that $\Gamma R_{\SET X=\SET x}\Delta$, $\M, \Delta \Vdash \psi$. 
    By the i.h., $\T^\Delta \models \psi$. Therefore, if we show that 1) $\T^\Delta$ is a singleton model, 2) every singleton submodel of $\T^\Gamma_{\SET X=\SET x}$ is $\equiv\T^\Delta$ for some $\Delta$ with  $\Gamma R_{\SET X=\SET x}\Delta$, 
    we will have $\T^\Gamma\models [\SET X=\SET x]\psi$ by the semantic clauses for $\models$.

    2) This point is given by lemma \ref{lemma: singleton subteams of intervened canonical teams} (or lemma \ref{lemma: singleton subteams as canonical teams} in the special case for $\SET X = \emptyset$).

    1) Since $\Gamma R_{\SET X=\SET x}\Delta$, by definition, $\Diamond \top\in\Delta$ and $\Box \SET W = \SET w \in \Delta$ for some $\SET w$ such that $\SET w_{\upharpoonright \SET X} = \SET x$. 
    By lemma \ref{lemma: basic properties of B}, 9., $\Diamond \SET W = \SET w \in \Delta$; thus, $s\in (\T^\Delta)^-$, where $s$ is the assignment $s(\SET W) = \SET w$. Suppose, for the sake of contradiction, that $t\in (\T^\Delta)^-$, where $t(\SET W) = \SET w' \neq \SET w\in \Delta$. By definition of $\T^\Delta$, then, $\Diamond \SET W = \SET w'\in \Delta$. On the other hand, $\Box \SET W = \SET w \in \Delta$ together with axiom I1 yields $\Box \SET W \neq \SET w'$. Thus, by lemma \ref{lemma: basic properties of B}, 10., $\Diamond (\SET W = \SET w' \with \SET W \neq \SET w') \in \Delta$. By I0 and monotonicity, then, $\Diamond \bot\in\Delta$. Since $\Delta$ is consistent, we contradict lemma \ref{lemma: properties of MCS}, 5. 


    $\Leftarrow$) Assume $\T^\Gamma \models [\SET X=\SET x]\psi$. Then, for all $s\in (\T^\Gamma_{\SET X =\SET x})^-$, $(\{s\},\F^\Gamma_{\SET X =\SET x}) \models \psi$. Suppose we manage to prove the following, for all maximally 
    consistent sets of $\Hal_\sigma^+$ formulas $\Delta \supseteq \Bx_{\tau_{\SET X}}$:
    \begin{center}
    (*): if $\Gamma R_{\SET X=\SET x}\Delta$, then for some $s\in (\T^\Gamma_{\SET X =\SET x})^-$, $\Delta = \{\chi \in \Hal^+_\sigma  \mid  (\{s\},\F^\Gamma_{\SET X =\SET x})\models \chi  \}$.
    \end{center}
    If so, then by the above 
    $\psi \in \Delta$ if $\Gamma R_{\SET X=\SET x}\Delta$. But then, by the truth lemma for $\M$ (lemma \ref{lemma: truth lemma for M}), $\M,\Delta \Vdash \psi$ if $\Gamma R_{\SET X=\SET x}\Delta$. Thus, $\M, \Gamma \Vdash [\SET X=\SET x]\psi$, as needed. 
    
    Let us then prove (*). If  $\Gamma R_{\SET X=\SET x}\Delta$, then by definition of $R_{\SET X=\SET x}$ we have $\Diamond \top, \Box \SET W = \SET w^* \in \Delta$ for some $\SET w^*$; let $s$ be the assignment such that $s(\SET W)= \SET w^*$. Let $\Delta_0$ be the set of $\Hal$ formulas in $\Delta$. By lemma \ref{lemma: canonical singletons}, we have that $\Delta_0 = \{\chi \in \Hal_\sigma \mid (\{s\}, \F^\Delta) \models \chi \}$. By lemma \ref{lemma: post-intervention equivalence}$, \F^\Delta \equiv  \F^\Gamma_{\SET X = \SET x}$; but then, by theorem \ref{thm: equivalence implies HPLUS equivalence}, $(\{s\}, \F^\Delta)$ and $(\{s\}, \F^\Gamma_{\SET X = \SET x})$ satisfy the same $\Hal^+_\sigma$ formulas. Thus, $\Delta_0 = \{\chi \in \Hal_\sigma \mid (\{s\}, \F^\Gamma_{\SET X = \SET x}) \models \chi \}$. Since $\Delta_0$ has a singleton model of signature $\tau_\SET X$, by the soundness of $\Ax$ we have $\Ax_{\tau_{\SET X}} \subseteq \Delta_0$, and $\Delta_0$ is maximally $\Ax_{\tau_{\SET X}}$-consistent. Thus, by theorem \ref{thm: rigidity theorem}, $\Delta$ is the only maximal consistent set of $\Hal^+$ formulas that extends both $\Delta_0$ and $\Bx_{\tau_{\SET X}}$. Now, $\Delta':= \{\chi \in \Hal^+_\sigma \mid (\{s\}, \F^\Gamma_{\SET X = \SET x}) \models \chi \}$ also extends $\Delta_0$. Since $(\{s\}, \F^\Gamma_{\SET X = \SET x})$ is a model of $\Delta'$, we also have $\Bx_{\tau_\SET X}\subseteq \Delta'$. Thus, $\Delta=\Delta'$, as needed.
\end{proof}

\begin{theorem}[Strong completeness of $\Bx$]\label{thm: completeness of Bx}
For $\Gamma \cup \{\varphi\} \subseteq \mathcal H^+_\sigma$, 
\[
\Gamma \models_\sigma \varphi \iff \Gamma \vdash_{\Bx_\sigma} \varphi.
\]
\end{theorem}

\begin{proof}
  The right-to-left direction is given by the soundness theorem \ref{thm: soundness of B}. In the other direction, suppose $\Gamma \not\vdash_{\Bx_\sigma} \varphi$. By the usual reasoning, then, $\Gamma \cup \Bx_\sigma \cup \{\cneg \varphi\}$ is $\Bx_\sigma$-consistent. By the Lindenbaum lemma, there is a maximal 
  consistent set of $\Hal^+_\sigma$ formulas $\Delta \supseteq \Gamma \cup \Bx_\sigma \cup \{\cneg \varphi\}$. By the truth lemma for $\M$ (lemma \ref{lemma: truth lemma for M}), $\M, \Delta \Vdash \Gamma, \cneg\varphi$. But then, by the representation lemma (lemma \ref{lemma: connection lemma}), $\T^\Delta \models \Gamma, \cneg\varphi$. Since $\T^\Delta$ is a model (by lemma \ref{lemma: canonical team for B}), we conclude that $\Gamma \not \models \varphi$. 
\end{proof}



As a final remark, we note that now we may obtain an axiomatization of the $\Hal^+_\sigma$ languages over \emph{singleton} models as an extension of $\Bx_\sigma$. 
Consider the following axiom:

\vspace{5pt}

Sing. $\Diamond\top \with \bigsqcup_{\SET w \in \ran(\SET W)} \SET W = \SET w$. 

\begin{theorem}
For $\Gamma \cup \{\varphi\} \subseteq \mathcal H^+_\sigma$, we have $\Gamma \models_\sigma^1 \varphi \iff \Gamma, \operatorname{Sing} \vdash_{\Bx_\sigma} \varphi$.
\end{theorem}

\begin{proof}
    This is an immediate consequence of theorem \ref{thm: B as conservative extension of A} and theorem \ref{thm: completeness of Ax}.
\end{proof}

We are left instead with the open issue of axiomatizing language $\Hal_\sigma$ over the class of all models (of signature $\sigma$). We might conjecture that system $\Bx_\sigma$ could be conservative over the subsystem of axioms involving only unnested formulas (axioms J0, I1-I5, J6, I7-I9 and J16); the completeness for $\Hal_\sigma$ formulas of this smaller set of axioms would immediately follow. But nothing of what has been done here supports this conjecture.


\section{Parents, direct causes and dummy arguments}\label{sec: PA, direct causes, dummy arguments}

\corrf{The literature on causal inference defines a multitude of notions of causation in terms of interventionist counterfactuals \citep[see e.g.][]{Woo2003}. Among these, a key foundational role is played by the notion of \emph{direct cause}; roughly speaking, $X$ is a direct cause of $Y$ when there is some intervention on $X$ that alters the value of $Y$ \emph{while all other variables in the system are held fixed} (at \emph{some} tuple of values). In most discussions of causal models, saying that $X$ is a direct cause of $Y$ is equivalent to saying that $X$ is a parent of $Y$; furthermore, it is equivalent to saying that $X$ is a non-dummy argument of the causal law for $Y$. As hinted at in section \ref{sec: differences}, these three perspectives diverge in the indeterministic case. In the present section, we give precise definitions of these notions and prove some basic connections about them. In particular, we show that, for any model $T$,  
\[
PA_V \supseteq^* \{\text{direct causes of } V\} \supseteq \{\text{non-dummy arguments of } V\}
\]
and there are models that make these inclusions strict. We marked the first inclusion with a star because it holds only under the assumption that the model is \emph{total}.} 

\corrf{
We will also consider a fourth relationship, that of \emph{visible direct cause}, which, while not being properly causal in nature, will be useful for the purposes of axiomatization. We will prove the following important inclusion:
\[
\{\text{direct causes of } V\} \supseteq \{\text{visible direct causes of } V\}.  
\]
}

\subsection{Definitions}\label{subs: definitions of direct cause}

1. Whether $X$ is a parent of $Y$ in a model $T$ is arbitrarily decided by the model (this relation determines the \textbf{parenthood graph}). 
As we have seen, the effect of this stipulation is that, if $X$ is updated by an intervention, then also $Y$ is updated. If the parenthood graph of $T$ is acyclic, we say that $T$ is \textbf{strictly recursive}. As we already remarked, this class of models has many analogies with the recursive class of (the usual, deterministic) causal models. While it is no more true that the values for the external variables uniquely determine the values of the internal variables, it is still the case that the effects of interventions can be calculated by a straightforward recursive procedure. 

 Another analogy is that, if furthermore the causal laws are total, then intervening on a nonempty model always produces solutions:

 \begin{proposition}\label{prop: recursive total is solutionful}
   Let $T = (T^-,\F)$ be a nonempty, strictly recursive total model, $\SET X$ variables in its domain, $\SET x \in \ran(\SET X)$, \corrf{$s\in T^-$}. Then, \corrf{$s^\F_{\SET X = \SET x}$ is nonempty}. In particular, $T_{\SET X = \SET x}$ is nonempty.  
 \end{proposition}

 \begin{proof}
    As already mentioned in section \ref{sec: formal framework}, the acyclicity of the parenthood graph ensures that, if $Z\in PA_Y$, then $d(\SET X, Z) < d(\SET X, Y)$. Furthermore, acyclicity and finiteness ensure that each variable has a (finite) distance from $\SET X$.
     
     Fix an $s\in T^-$. We define a $t\in s^\F_{\SET X = \SET x}$
     by recursion on the distance of variables from $\SET X$. First, we let $t(\SET X):= \SET x$ and $t(\SET N_\SET X):=s(\SET N_\SET X)$. Each of the remaining variables, say  $Y$, is internal and not in $\SET X$. Since $d(\SET X, Z) < d(\SET X, Y)$ for each $Z\in PA_Y$, we can assume by i.h. that $t(Z)$ is already defined for each $Z\in PA_Y$. Then, we can let $t(Y):= y$ for an arbitrary $y\in \F_Y(t(PA_Y)) \neq \emptyset$. Since there is a maximum distance, this defining procedure ends.
 \end{proof}

2A. We define the notion of direct cause following the suggestion of \cite{Wys2023}: that $X$ should count as a direct cause of $Y$ if intervening on $X$ (while all the remaining variables are fixed at some values) will change the range of available values of $Y$. More precisely, let $T=(T^-,\F)$ be a model of signature $\sigma$. 
We will say that $X$ is a \textbf{direct cause} of $Y$ (in $T$) if, for some assignment $s$ of signature $\sigma$ compatible with $\F$, the following formula $X\rightsquigarrow Y$ 
is satisfied by $(\{s\},\F)$:
\[
X\rightsquigarrow Y: \hspace{10pt} \bigsqcup_{(\SET z,x,y) \in \ran(\SET ZXY)} \cneg (\langle\SET ZX = \SET zx\rangle Y=y  \leftrightarrow \langle\SET Z = \SET z\rangle Y=y ) 
\]
(where $\SET Z$ stands for $\dom \setminus \{XY\}$).\footnote{In a 2-variable model, the formula in the right-hand side of the equivalence would read $\Diamond Y=y$.} 
Note that this notion of direct cause, in agreement with previous literature on causation, only depends on the law component, and not the team component of the model.
If no variable is a direct cause of $Y$ we say that $Y$ is \textbf{exogenous}; otherwise, it is \textbf{endogenous}.

 We call \textbf{causal graph} the graph of the direct cause relation, and say that a model is \textbf{recursive} if its causal graph is acyclic.  
 In section \ref{sec: undefinability} we prove that direct cause and the related notions  are not definable in $\Hal$. We do not know if they may be definable in some stronger language such as $\Hal^+$. \\ 

2B. \corrf{We will say that $X$ is a \textbf{visible direct cause} of $Y$ (in $T$) if $T\models X\rightsquigarrow Y$.} 
This is not equivalent to $X$ being a direct cause of $Y$, and there are reasons to think that such a definition is not causal. In the example of two-coin Bob (\ref{example: dummy parent}) $B$ turns out to be both a direct cause of $O$ according to the definition in 2A and a visible direct cause. 
But imagine a variant $T'$ of the two-coin Bob model $T$ in which, instead of one assignment, we have two:
\begin{center}
\begin{tabular}{|c|c|}
\hline
 \multicolumn{2}{|l|}{ \ B \ \ \ \ \ \ \ \  \ O } \\
\hline
 \phantom{a}1\phantom{a} & \phantom{a}heads\phantom{a} \\
 \hline
\phantom{a}1\phantom{a} & \phantom{a}tails\phantom{a} \\ 
\hline
\end{tabular}
\end{center}
Intervening on $B$ does not change the available values of $O$ in $T'$, thus $B$ is not a visible direct cause of $O$ in $T'$. 
Then we have two models ($T$ and $T'$) that have the same set of causal laws, but  disagree on whether $B$ is a visible direct cause of $O$ or not. We believe that a causal notion should not depend on the team component (which is typically used to describe \emph{epistemic} uncertainty). Thus, ``visible direct cause'' does not fit the bill. 
On the other hand, clearly the definition of direct cause does not depend on the team component (in particular, $T$ and $T'$ agree about it).  

We consider the notion of visible direct cause because it is a useful tool for the sake of axiomatization. Differently from direct cause, it is easily definable in $\Hal$ (by the formula $X\rightsquigarrow Y$); we can then define related notions such as $Y$ being \textbf{visibly exogenous}, resp. \textbf{visibly endogenous}, by saying that the model satisfies the following formulas:
\columnratio{0.607}
\setlength\columnsep{0pt}
\footnotelayout{m}
\begin{paracol}{2}
\begin{leftcolumn}
\begin{itemize}
    \item $\varphi_{\vexo(Y)}: \bigwith_{X\in \dom \setminus \{Y\}} \cneg X\rightsquigarrow Y$
\end{itemize}
\end{leftcolumn}
\begin{rightcolumn}
    \begin{itemize}
        \item $\varphi_{\ven(Y)}: \cneg \varphi_{\vexo(Y)}$.
    \end{itemize}
\end{rightcolumn}
\end{paracol}
\corrf{We may call \textbf{visible causal graph} the graph of the relation $\rightsquigarrow$, and say that a model is \textbf{visibly recursive} if its visible causal graph is acyclic. }

\corrf{A key property of an external variable $Y$ is the fact that it is not affected when intervening on the set of all other variables in the system.\footnote{More generally, interventions on smaller sets of variables will not produce new values for $Y$ -- but some values may disappear, at least in the non-strictly recursive case.} For a visibly exogenous variable we do not get that much (we will soon show that such an intervention may make some new values for the visibly exogenous variable appear), but at least we can prove that such an intervention does not make values disappear. 
Remember that we write $T^-(Y)$ for the set of values that $Y$ takes in $T^-$, i.e., $T^-(Y):= \{y\in \ran(Y) \mid \text{there is an } s\in T^- \text{ such that } s(Y)=y \}$.}

\begin{proposition}\label{prop: visibly exogenous after total intervention}
    \corrf{Let $T = (T^-,\F)$ be a model such that $T\models \varphi_{\vexo(Y)}$; furthermore, let $y\in T^-(Y)$. Let $\SET w$ be a tuple of values for $\SET W_Y$ (a list of all variables except $Y$). Then, $y\in T_{\SET W_Y = \SET w}^-(Y)$.} 
\end{proposition}


\begin{proof}
    \corrf{
    Let $\SET w,\SET w'$ be two tuples of values for $\SET W_Y$. We first show that $y\in  T_{\SET W_Y = \SET w}^-(Y)$ iff $y\in  T_{\SET W_Y = \SET w'}^-(Y)$, by induction on the number $n$ of variables on which $\SET w,\SET w'$ disagree. If $n=1$, this follows immediately from $T\models \varphi_{\vexo(Y)}$. Suppose that it holds for $n$, and assume that $\SET w,\SET w'$ differ on $n+1$ variables $X_1,\dots, X_{n+1}$; say, these take values $x_1,\dots, x_{n+1}$ in $\SET w$ and $x_1',\dots, x_{n+1}'$ in $\SET w'$. Let also $\SET Z$ list the variables in $\SET W_Y\setminus \{X_1,\dots, X_{n+1}\}$ and $\SET z$ be the restriction of $\SET w$ to $\SET Z$. By the inductive hypothesis (for $n$) we have $y\in  T_{\SET W_Y = \SET w'}^-(Y)$ iff $y\in T_{\SET Z,X_1,\dots, X_{n+1} = \SET z,x_1,\dots,x_n,x'_{n+1}}^-$. By the base case ($n=1$), the latter is equivalent to $y\in T_{\SET W_Y = \SET w}^-(Y)$.
    }
    
    \corrf{
    Now let $s\in T^-$ such that $s(Y)=y$, and let $\SET w^* := s(\SET W_Y)$. We can show that $s\in  T_{\SET W_Y = \SET w^*}^-$. This is by definition in case $Y$ is external; if it is internal, from $s\in T^-$ we know that $y=s(Y)\in \F_Y(\SET w^*_{\upharpoonright PA_Y})$, thus again, by definition of intervention, $s\in  T_{\SET W_Y = \SET w^*}^-$. Thus, $y\in T_{\SET W_Y = \SET w^*}^-(Y)$. By the above, then, $y\in T_{\SET W_Y = \SET w}^-(Y)$ for any $\SET w\in\ran(\SET W_Y)$.
    }
\end{proof}

\noindent In other words, any model $T$ satisfies the formula $\varphi_{\vexo(Y)} \rightarrow ( \Diamond Y=y\rightarrow \langle \SET W_Y = \SET w\rangle Y=y)$. As we have seen in section \ref{sec: H language}, if $Y$ is an external variable, the stronger principle $\Diamond Y=y\leftrightarrow \langle \SET W_Y = \SET w\rangle Y=y$ holds. Let us see an example showing that the right-to-left implication may fail if $Y$ is visibly exogenous, but not external.

\begin{example}[Refereed two-coin Bob]
We have an external variable $A$ for whether Arbiter chooses coin $1$ or coin $2$; after Arbiter makes their choice,  Bob will toss the corresponding coin ($B=1$ or $2$); variable $C$ registers the outcome of the toss (heads or tails). Let $T$ be the model with $PA_B = {A}$, $PA_C ={B}$, deterministic law $\F_B(x) ={x}$ and indeterministic law $\F_C(0) = \F_C(1) = \{heads,tails\}$,  and containing only the assignment $s(A,B,C) =(1,1,heads)$ (the scenario in which Arbiter chose coin $1$, Bob tossed coin $1$ and it came $heads$). Now, it can be checked that in this model neither $A$ nor $B$ is a visible direct cause of $C$ ($A$ is not even a direct cause), the reason being that, once we intervene on $A$ (resp. $B$) all possible values of $C$  become available, and they will not disappear when intervening \emph{also} on the other variable. Thus, $C$ is visibly exogenous. On the other hand, intervening on $A$ alone makes the value $tails$ possible, even if that was not possible in the initial model. In other words, we have $T\models \langle A=1  \rangle C = tails$ but $T\not\models \Diamond C = tails$.    
\end{example}

3. Finally, 
we say $X(\neq Y)$ is a \textbf{non-dummy argument} of $Y$ if $X\in PA_Y$ and  there are $pa,pa'\in \ran(PA_Y)$ that coincide on all variables except $X$, such that $\F_Y(pa) \neq \F_Y(pa')$. If $X\in PA_Y$  but the other condition is not met, $X$ is a \textbf{dummy argument} $Y$. Note that, by definition, if a variable is not a parent of $Y$, then it is neither a dummy nor anon-dummy argument of $Y$. If the graph of the non-dummy argument relation is acyclic, we may say that the model is \textbf{weakly recursive}.

\subsection{Parents vs. direct causes}

\begin{theorem}\label{thm: parents vs direct causes}

    \begin{enumerate}
        \item Let $T$  be a model, with a variable $X$ that is either external or internal with a total causal law. If $X$ is a direct cause of $Y$ in $T$, 
        then $X\in PA_Y$.
        \item There are (total) models $T$ such that $X\in PA_Y$ but $X$ is not a direct cause of $Y$ in $T$. 
    \end{enumerate}
\end{theorem}

\begin{proof}
1) 
Let $T = (T^-,\F)$.  
 Assume $X\notin PA_Y$, and that $\F_X$ is a total multivalued function. We show that then, for any $\SET w\in\ran(\SET W_{XY}), x\in \ran(X), y\in\ran(Y)$, and any $s\in T^-$, $(\{s\},\F)\models \langle\SET W_{XY}X = \SET wx\rangle Y=y  \leftrightarrow \langle\SET W_{XY} = \SET w\rangle Y=y$ (thus, $X$ is not a direct cause of $Y$).

Suppose first that $(\{s\},\F)\models \langle\SET W_{XY} = \SET w\rangle Y=y$. \corrf{In case $Y$ is internal,} since we assumed $X\notin PA_Y$, by definition of intervention we have $y\in\F_Y(\SET w_{\upharpoonright PA_Y})$. Thus, 
$(\{s\},\F)_{\SET W_{XY}X = \SET wx}$ is nonempty (it contains the assignment $s( W_{XY}XY = \SET wxy)$); thus, 
we have again by definition of intervention that  $(\{s\},\F)\models \langle\SET W_{XY}X = \SET wx\rangle Y=y$. 
\corrf{In case $Y$ is external, the reasoning is similar (since $(\{s\},\F)\models \langle\SET W_{XY} = \SET w\rangle Y=y$ then entails $s(Y)=y$).}

Vice versa, suppose $(\{s\},\F)\models \langle\SET W_{XY}X = \SET wx\rangle Y=y$. Since we assumed $X\notin PA_Y$, we can observe that the parenthood graph $G_{\SET W_{XY}}$ is acyclic. Thus, since $y\in \F_Y(\SET wx_{\upharpoonright PA_Y})=\F_Y(\SET w_{\upharpoonright PA_Y})$ (so that, in particular, $\F_Y(\SET w_{\upharpoonright PA_Y})\neq\emptyset$), and by the totality assumption also $\F_X(\SET w_{\upharpoonright PA_X})\neq \emptyset$ in case $X$ is internal,  by definition of intervention we have then that $(\{s\},\F)_{\SET W_{XY} = \SET w}\neq \emptyset$. Thus, since $y\in \F_Y(\SET w_{\upharpoonright PA_Y})$, we conclude by definition of intervention that $(\{s\},\F)\models \langle\SET W_{XY} = \SET w\rangle Y=y$. 

2) Consider a model with internal variable $Y$ and external variables $Z,X$; $PA_Y = \{Z,X\}$; $\F_Y$ being the full ternary relation on $\{0,1\}$; and team component $\{s\} = \{ \{(Z,z^*),(X,x^*),(Y,y^*)\}  \}$, for any $z^*,x^*, y^* \in \{0,1\}$. It can be checked systematically that any intervention on $Z$ alone, and any intervention on both $Z$ and $X$ lead to a model in which both values $0$ and $1$ are available for $Y$. Thus, for any $z,x,y\in \{0,1\}$, $(\{s\},\F)\models \langle ZX = zx\rangle Y=y \leftrightarrow \langle Z = z\rangle Y=y$. Since this holds for all $s$ allowed by the signature, $X$ is not a direct cause of $Y$.
 \end{proof}

\noindent To see that the assumption that $X$ has a total causal law is needed for part 1., we may consider the following model
\begin{center}
\begin{tabular}{|c|c|c|}
\hline
 \multicolumn{3}{|l|}{  \ Z \  $\rightarrow$  \ X \  $\leftarrow$     Y } \\
\hline
 \phantom{a}1\phantom{a} & \phantom{a}1\phantom{a} & \phantom{a}1\phantom{a} \\
\hline
\end{tabular}
\end{center}
\corrf{where the causal law $\F_X(Z,Y)$ for $X$ is such that $\F_X(1,1)=\{1\}$ and $\F_X(2,1)=\emptyset$. Then, $T\models \langle Z = 2, X = 1 \rangle Y\neq 2$ (since $T_{Z = 2, X = 1}^-$ contains the assignment $s(Z,X,Y) =(2,1,1)$) while  $T\not\models \langle Z = 2\rangle Y\neq 2$ (since $T_{Z=2}$ is empty). Thus, $X$ is a direct cause of $Y$ even if $X$ is not a parent of $Y$.}

\subsection{Direct causes vs. non-dummy arguments}

\corrf{If $\SET z$ is a tuple of values for $\SET Z$, and $X\in \SET Z$, we write $\SET z_{-X}$ for the restriction of $\SET z$ to $\SET Z\setminus \{X\}$.}

\begin{theorem}
     \begin{enumerate}
        \item  If $X$ is a non-dummy argument of $Y$, then it is a direct cause of $Y$. 
        \item There are models $T$ such that $X$ is a direct cause of $Y$ but also a dummy argument of $Y$. 
    \end{enumerate}
\end{theorem}

\begin{proof}
    1) Let $T$ be a model, and suppose $X$ is a non-dummy argument of $Y$ in $T$. Then, there are $pa,pa'\in \ran(PA_Y)$ such that $pa_{-X}=pa'_{-X}$ but $\F_Y(pa) \neq \F_Y(pa')$. Let $x = pa_{\upharpoonright\{X\}}$, $x' = pa'_{\upharpoonright\{X\}}$  and pick a $\SET w \in \ran(W_{XY})$ such that $\SET w_{\upharpoonright PA_Y \setminus \{X\}} = pa_{-X}$.  Let $s\in T^-$ and $S = \{y\in \ran(Y) \mid (\{s\},\F) \models \langle \SET W_{XY} = \SET w\rangle Y=y\}$. $S$ cannot be equal to both $\F_Y(pa)$ and $\F_Y(pa')$; wlog suppose $S\neq \F_Y(pa)$. 

    We note first that there cannot be a $y\in S\setminus \F_Y(pa)$. Indeed, $(\{s\},\F) \models \langle \SET W_{XY} = \SET w\rangle Y=y$ would imply the existence of an assignment in $s^\F_{\SET W_{XY} = \SET w}$ which is not compatible with $\F_Y$.

    Thus, there is instead a $y \in \F_Y(pa) \setminus S$. Then $(\{s\},\F)\models \langle\SET W_{XY}X = \SET wx\rangle Y=y$, but $(\{s\},\F)\not\models \langle\SET W_{XY} = \SET w\rangle Y=y$. 
    So, $X$ is a direct cause of $Y$.


    2) In the two-coin Bob example, $B$ is a direct cause and a dummy argument of $O$.
\end{proof}

\subsection{Direct causes vs. visible direct causes}

\begin{theorem}\label{thm: visible cause is direct cause}
    \corrf{If $X$ is a visible direct cause of $Y$ in $T$, then it is a direct cause of $Y$ in $T$.}
\end{theorem}

\begin{proof}
    Suppose $T=(T^-,\F) \models X\rightsquigarrow Y$. Then, there are $\SET w,x,y \in \ran(\SET W_{XY}XY)$ such that either:
    \begin{itemize}
        \item $T\models \langle \SET W_{XY} = \SET w\rangle Y=y$ but $T\not\models \langle \SET W_{XY}X = \SET wx\rangle Y=y$. By the definition of intervention (plus the related observations in section \ref{subs: relational causal teams}), $T\models \langle \SET W_{XY} = \SET w\rangle Y=y$ entails that there is an $\hat s\in T^-$ and a $t\in \hat s^\F_{\SET W_{XY} = \SET w}$ such that $t(Y)=y$. But then $(\{\hat s\},\F) \models \langle \SET W_{XY} = \SET w\rangle Y=y$. On the other hand, by a similar argument $T\not\models \langle \SET W_{XY}X = \SET wx\rangle Y=y$ entails that, for all $s\in T^-$ (in particular $\hat s$), $(\{s\},\F) \not\models \langle \SET W_{XY}X = \SET wx\rangle Y=y$. Thus, $(\{\hat s\},\F) \models X\rightsquigarrow Y$. So, $X$ is a direct cause of $Y$.
        
        \item $T\models \langle\SET W_{XY}X = \SET wx\rangle Y=y$ but $T\not\models  \langle \SET W_{XY} = \SET w\rangle Y=y$. The proof for this case is analogous.
    \end{itemize}
\end{proof}

\section{Acyclicity of causal laws}\label{subs: causes and recursivity}

\subsection{Axiomatization}

Besides the full class of models, we want to characterize axiomatically the class of models in which the causal laws are acyclic. In the deterministic context, such models are called \emph{recursive}. 
It is not straightforward how to identify the analogue of this notion in the indeterministic context; 
 in the previous section, we have identified four possible candidates: the \emph{strictly recursive}, \emph{recursive}, \emph{visibly recursive} and \emph{weakly recursive} indeterministic models. We also identified a few logical connections between them, as illustrated in the following picture 
 
\begin{center}
\begin{tikzpicture}[->,thick, node distance=2.5cm]
  \node (strict) {strictly recursive};
  \node (rec) at (3,0) {recursive};
  \node (vis) [above right of=rec] {visibly recursive};
  \node (weak) [below right of=rec] {weakly recursive};

  \draw[double,->] (strict) -- (rec) node[midway, above]{\textasteriskcentered};
  \draw[double,->,sloped] (rec) -- (vis);
  \draw[double,->,sloped] (rec) -- (weak);
\end{tikzpicture}
\end{center}
where the asterisked entailment only holds for total models.
 
  We show here how to axiomatize the \emph{strictly} and the \emph{visibly} recursive classes. We do not know how to axiomatize the remaining two classes, but in the next section we will see that the class of recursive, \emph{total} models can be straightforwardly axiomatized.

In order to turn an axiomatization of the general class of models into one for the \corrf{strictly (resp. visibly)} recursive class, it will suffice to add a recursivity axiom of the form
\[
\text{R. }(X_1\rightsquigarrow X_2 \with \dots\with X_{n-1} \rightsquigarrow X_n) \rightarrow \cneg X_n\rightsquigarrow X_1. \hspace{30pt}  \text{[Generalized recursivity]} \\
\]
\corrf{Remember that $\rightsquigarrow$ does not capture the notion of direct cause but rather the stricter notion of visible direct cause. Thus, this axiom explicitly forbids cycles of visible direct causes, but might be insufficient by itself to forbid cycles in the causal graph.} 
Axioms of this form, using different notions of causal dependence in place of $X\rightsquigarrow Y$, are common in the literature; e.g., \citet{BarSchVelXie2021} uses the deterministic notion of direct cause, while \citet{Hal2000} uses a weaker dependence called ``causally affecting''.  



From the definition of canonical model we immediately obtain the following facts.

\begin{proposition}\label{prop: canonical coincidence of PA and VDC}
Let $\Gamma$ be a maximally consistent set of formulas. Then:
\begin{enumerate}
    \item $X$ is a parent of $Y$ in $\T^\Gamma$ iff $X$ is a visible direct cause of $Y$ in $\T^\Gamma$.  
    \item $\T^\Gamma$ is strictly recursive iff it is visibly recursive. 
\end{enumerate}
\end{proposition}







\begin{restatable}{lemma}{RCanonicity}\label{lemma: canonicity of R}
Suppose $\Gamma\supseteq \Ax \cup \{\operatorname{R}\}$ (or $\Gamma\supseteq \Bx  \cup \{\operatorname{R}\}$) is maximally consistent set of $\Hal_\sigma$ (resp. $\Hal^+_\sigma$) formulas. Then, $\T^\Gamma$ is \corrf{strictly recursive and visibly recursive}.
\end{restatable}

\begin{proof}
\corrf{By proposition \ref{prop: canonical coincidence of PA and VDC}, it suffices to prove that $\T^\Gamma$ is visibly recursive.} 
Suppose the visible causal graph of $\T^\Gamma$ has a cycle $X_1,\dots X_n$. Then, by the definition of $\T^\Gamma$, for all $i = 1, \dots, n$, 
 $X_i \rightsquigarrow X_{i+1} \in \Gamma$ and 
 $X_n \rightsquigarrow X_1 \in \Gamma$. 
But, since $\Gamma$ contains all the instances of axiom R, we also obtain $\cneg X_n \rightsquigarrow X_1 \in \Gamma$. This contradicts the consistency of $\Gamma$.
\end{proof}

Fix a signature $\sigma$. We write $\Gamma\models^{\operatorname {SR}}\varphi$ (resp. $\Gamma\models^{\operatorname {VR}}\varphi$) 
if every strictly recursive (resp. visibly recursive) model of signature $\sigma$ that satisfies $\Gamma$ also satisfies $\varphi$. We write $\Gamma\models^{\operatorname {SR}}_1\varphi$ ($\Gamma\models^{\operatorname {VR}}_1\varphi$) 
if every strictly recursive (resp. visibly recursive) \emph{singleton} model of signature $\sigma$ that satisfies $\Gamma$ also satisfies $\varphi$.

\begin{restatable}[Strong completeness for strictly/visibly recursive models]{theorem}{StrongCompletenessRecursive} \
\begin{enumerate}
    \item For $\Gamma \cup \{\varphi\} \subseteq \mathcal H_\sigma$, $\Gamma \models^{\operatorname{SR}}_1 \varphi  \iff 
    \Gamma \models^{\operatorname{VR}}_1 \varphi  \iff \Gamma, \operatorname{R} \vdash_\Ax \varphi$.
    \item For $\Gamma \cup \{\varphi\} \subseteq \mathcal H^+_\sigma$, $\Gamma \models^{\operatorname{SR}} \varphi  \iff 
    \Gamma \models^{\operatorname{VR}} \varphi  \iff \Gamma, \operatorname{R} \vdash_\Bx \varphi$.
\end{enumerate}
\end{restatable}

\begin{proof} 
 As for theorem \ref{thm: completeness of Ax} (resp. \ref{thm: completeness of Bx}), using the fact that $\T^\Delta$ is strictly recursive if $\Ax^R\subseteq \Delta$  (lemma \ref{lemma: canonicity of R}).
\end{proof}


\subsection{Failure of composition}\label{subs: failure of composition}


There are further significant differences between the deterministic and indeterministic recursive cases that are not evident from the proposed axiomatization. In the deterministic case, intervening on a single state of affairs produces again a single state of affairs (the unique solution of a certain system of equations). It then follows that counterfactuals
    $[\SET X = \SET x]   \psi$ 
and \emph{might}-counterfactuals
    $\langle\SET X = \SET x\rangle   \psi$  
are equivalent, and so the latter are redundant. 
If indeterministic laws are involved, instead, interventions  on a single state of affairs may produce multiple possible states of affairs even if there are no cyclic causal laws. The operator $\langle\SET X = \SET x\rangle$, then, albeit definable as $\cneg[\SET X = \SET x]\cneg$, seems to be vital for expressing properties of the solution sets in a natural way.

Another important difference is the failure of one of Galles and Pearl's principles for recursive models, the law of \emph{Composition}. This can be expressed as
\[
([\SET X = \SET x] W=w \with [\SET X = \SET x] Y=y) \rightarrow [\SET X = \SET x, W=w] Y=y 
\]
\corrf{which can sometimes\footnote{E.g., in the systems considered in \cite{BarYan2022}, the two are provably equivalent. Nested conditionals seem to be needed in the equivalence proof.} be replaced, in axiomatizations for recursive models, by the more intuitive \emph{Conjunction conditionalization}} 
\[
(\SET X = \SET x \with \eta) \rightarrow [\SET X = \SET x]\eta 
\]
where, crucially, $\eta$ is a formula without counterfactuals. 
 The example given in section \ref{subs:  uncertainty} shows that both laws fail for indeterministic \corrf{(even weakly)} recursive models. Indeed, in it $A=1$ and $C=heads$ hold, but $[A=1]C=heads$ does not.

As we have seen, it turns out that the weakened form of \emph{Composition},
\[
\langle\SET X = \SET x\rangle  (W =w \with  \SET Y = \SET y) \rightarrow \langle\SET X = \SET x,  W =w \rangle \SET Y = \SET y
\]
which was proposed in \citet{Hal2000} for the axiomatization of the (possibly) \emph{cyclic} case, is also sound for indeterministic models (both in the recursive and nonrecursive case); it is just axiom I3. This law illustrates the importance of \emph{might}-counterfactuals in the indeterministic context.



\subsection{Strong reversibility}

Besides \emph{Recursivity} and \emph{Composition}, the third principle proposed by \citet{GalPea1998} for recursive models is the (strong) \emph{Reversibility} axiom:
\[
([\SET X = \SET x, W = w]Y=y \with [\SET X = \SET x, Y = y]W=w) \rightarrow [\SET X = \SET x]Y=y \hspace{30pt} \corrf{(\text{when } Y\neq W)}.
\]
In \citet{GalPea1998}, this axiom was thought of as a characterization of the \emph{unique solution} property\footnote{A structural equation model has this property if 1) its system of equations is satisfied by a unique assignment once values for the exogenous variables are fixed, and 2) the same holds after any intervention.}, which (in the deterministic case) is more general than recursivity. 
The axiom is indeed part of Halpern's axiomatization of unique solution causal models in \citet{Hal2000}; it does not feature in the axiomatization of recursive causal models because it is derivable from the other axioms including \emph{Recursivity}. 
\corrf{The insights from \citet{ZhaLamDec2013} and \cite{FanZha2023}, however, clearly show that this hypothesized connection between \emph{Reversibility} and unicity of solutions is incorrect.}

It turns out that strong \emph{Reversibility} is valid also on indeterministic \emph{strictly} recursive models. Since multiplicity of solutions is the norm when the laws are indeterministic, looking at indeterministic, strictly recursive models one may see even more clearly that \emph{Reversibility} has nothing to do with the unicity of solutions. 

\begin{restatable}[Strong reversibility]{theorem}{StrongReversibility}
    Let $T = (T^-,\F)$ be a 
    \corrf{strictly} recursive model. Then, $T\models ([\SET X = \SET x, W = w]Y=y \with [\SET X = \SET x, Y = y]W=w) \rightarrow [\SET X = \SET x]Y=y$.
\end{restatable}

\begin{proof}
Suppose $T\models [\SET X = \SET x, W = w]Y=y$ and $T\models [\SET X = \SET x, Y = y]W=w$. 
    Since $T$ is \corrf{strictly} recursive, \corrf{in the parenthood graph} either $W$ is not an ancestor of $Y$ or $Y$ is not an ancestor of $W$.

    Case 1: $W$ is not an ancestor of $Y$. 
    %
    Then, by \corrf{definition of intervention}, 
    intervening on $W$ does not affect $Y$ \corrf{(at most, some assignments may disappear).} 
    Thus, from the assumption $T\models [\SET X = \SET x, W = w]Y=y$ we immediately obtain $T\models [\SET X = \SET x]Y=y$. 

    Case 2: $Y$ is not an ancestor of $W$. By the assumption that $T\models [\SET X = \SET x, Y = y]W=w$, we get that $T_{\SET X = \SET x, Y = y} \models W=w$. Since $Y$ is not an ancestor of $W$, we get $T_{\SET X = \SET x} \models W=w$. But then the intervention $do(W=w)$ does not modify the team component of $T_{\SET X = \SET x}$, i.e. $T_{\SET X = \SET x}^- = T_{\SET X = \SET x, W=w}^-$. Now,  for each $s\in T_{\SET X = \SET x, W=w}^-$ we have $s(Y)=y$ by the assumption that $T\models [\SET X = \SET x, W = w]Y=y$; so, $T\models [\SET X = \SET x]Y=y$.   
\end{proof}




\section{Further axiomatizations}\label{sec: further axiomatizations}

\subsection{The deterministic case}\label{sec: determinism}

We now consider the issues of definability and axiomatizability for the subclass of deterministic models. Remember that $T = (T^-,\F)$ is deterministic if, for each internal variable $Y$, $\F_Y$ assigns at most one value to each tuple of values for the parents of $Y$ ($|\F_Y(pa)|\leq 1$ for each such tuple $pa$). We characterize this property by the following axiom:

\begin{center}
Det. $\langle \SET W_Y = \SET w\rangle Y=y \rightarrow \cneg \langle\SET W_Y = \SET w\rangle Y=y'$ \hspace{15pt} (when $y\neq y'$ and $Y\in \mathcal I$)
\end{center}

\noindent We remark that this axiom scheme (like axiom schemes I8, J10, J11 J16 and J17) depends, for its formulation, on having the internal/external distinction in the signature; the formulas in this scheme are \emph{not} sound in deterministic models when $Y$ is an external variable. We will show in section \ref{sec: undefinability} that this dependency is ineliminable.

\begin{lemma}\label{lemma: semantic characterization of determinism}
  Let $T$ be a model. Then, $T\models \mathrm{Det}$ iff 
  $T$ is deterministic.
\end{lemma}

\begin{proof}
$\Rightarrow$) Suppose, for the sake of contradiction, that $T$ is not deterministic. Then, there is a variable $Y\in \inte(T)$, a value $pa \in \ran(PA_Y)$ and two values $y \neq y' \in \ran(Y)$ such that $y,y' \in \F_Y(pa)$. 
%
Pick a $\SET w \in \ran(\SET W_Y)$ such that $\SET w_{\upharpoonright PA_Y} = pa$. 
By the definition of intervention, $T\models \langle\SET W_Y = \SET w\rangle Y=y$ and $T\models \langle\SET W_Y = \SET w\rangle Y=y'$. From the latter we obtain $T\not\models \cneg\langle\SET W_Y = \SET w\rangle Y=y'$. But then, $T$ falsifies an instance of Det.

$\Leftarrow$) Suppose $T$ is deterministic, and assume that $T\models  \langle \SET W_Y = \SET w\rangle Y=y$ for some $Y,y,\SET w$, \corrf{with  $Y\in \inte(T)$}. 

Since $T\models \langle \SET W_Y = \SET w\rangle Y=y$ \corrf{and $Y$ is internal}, then, by the semantic clause and the assumption of determinism, \corrf{$T\not\models \langle \SET W_Y = \SET w\rangle Y=y'$ when $y'\neq y$. Thus, $T\models  \langle \SET W_Y = \SET w\rangle Y=y \rightarrow \cneg \langle\SET W_Y = \SET w\rangle Y=y'$ whenever $y\neq y'$ and $Y$ is internal, i.e., $T\models \operatorname{Det}$.} 
\end{proof}


\begin{lemma}
    Let $\Gamma \supseteq \Ax \cup \{\mathrm{Det}\}$ (resp. $\Gamma \supseteq \Bx \cup \{\mathrm{Det}\}$) be a maximal consistent set of $\Hal_\sigma$ (resp. $\Hal^+_\sigma$) formulas. 
    Then, $\T^\Gamma$ is deterministic.
\end{lemma}

\begin{proof}
\corrf{Suppose for the sake of contradiction that there is an internal variable $Y$, values $y\neq y'\in \ran(Y)$ and a tuple $pa \in \ran(PA_Y)$ such that $\{y,y'\}\subseteq \F^\Gamma_Y(pa)$. Then, by the definition of canonical model, there is a $\SET w \in \ran(\SET W_Y)$ such that $\SET w_{\upharpoonright PA_Y} = pa$ and $\langle \SET W_Y = \SET w\rangle Y=y \in \Gamma, \langle \SET W_Y = \SET w\rangle Y=y'\in\Gamma$. However,  $\{\operatorname{Det}\}\subseteq \Gamma$, so in particular $\cneg \langle\SET W_Y = \SET w\rangle Y=y'\in \Gamma$. Thus $\Gamma$ is inconsistent, contradicting the assumption.}
\end{proof}

\noindent By the usual argument, then, we obtain the following completeness result. Write $\models^{\operatorname D}$ for semantic entailment in the class of deterministic models, and $\models^{\operatorname D}_1$ for entailment in the class deterministic singleton models.

\begin{theorem}
    $\Gamma \models^{\operatorname D}_1 \varphi$ iff $\Gamma, \mathrm{Det} \vdash_\Ax \varphi$, and $\Gamma \models^{\operatorname D} \varphi$ iff $\Gamma, \mathrm{Det} \vdash_\Bx \varphi$.
\end{theorem}

\subsection{The total and recursive cases}\label{subs: total and recursive cases}

We consider here the axiomatization of total causal teams (of a given fixed signature). 
Remember that a causal team $T = (T^-,\F)$ is total if, for each of its internal variables $Y$, $\F_Y$ is a total multivalued function; i.e, for each tuple $pa \in PA_Y$, $\F_Y(pa) \neq \emptyset$. We claim that a complete axiomatization for the class of 
total models is obtained by adding to $\Bx$ the following axiom scheme:\footnote{This axiom is analogous to one of the conjuncts of axiom D9 in \cite{Hal2000}.}

\vspace{5pt}

Tot. $\langle \SET W_Y = \SET w\rangle \top$ \hspace{10pt} (when $Y\in \mathcal I$)

\vspace{5pt}

\noindent where $\SET W_Y$ denotes a list of all variables in the domain, except $Y$. 
In the following two lemmas we see that axiom Tot captures the notion of totality from a semantic point of view. 

\begin{proposition}\label{prop: characterizing nonempty total models}
    Let $T$ be a model. $T\models \mathrm{Tot}$ iff $T$ is total. 
\end{proposition}

\begin{proof}
$\Leftarrow$) 
%
%
Let $T = (T^-,\F)$ be total. Since $Y$ is internal, then, we may pick a $y\in \F_Y(\SET w_{\upharpoonright PA_Y})$. We note then that the assignment $s(\SET W_Y) = \SET w$, $s(Y) = y$ is in $T_{\SET W_Y = \SET w}^-$. Thus, $T \models \langle \SET W_Y = \SET w\rangle \top$.


$\Rightarrow$) Suppose $T$ satisfies all instances of Tot. Let $Y\in \inte(T)$, $pa\in \ran(PA_Y)$ and $\SET w\in \ran(\SET W_Y)$ \corrf{such that $\SET w_{\upharpoonright PA_Y} = pa$}. 
Then, $T\models \langle \SET W_Y = \SET w\rangle \top$. Thus, there is an $s\in T_{\SET W_Y = \SET w}^-$, and $s(PA_Y)=pa$. Then, by the definition of intervention, $(pa, s(Y))\in \F_Y$. Since the choice of $pa\in \ran(PA_Y)$ was arbitrary, $\F_Y$ is a total function. Since the choice of $Y$ among internal variables was arbitrary, $T$ is total.
%
%
\end{proof}




\begin{lemma}\label{lemma: nonempty totality is canonical}
  Let $\Gamma$ be a maximal consistent set of $\Hal$ (resp. $\Hal^+$) formulas containing all instances of $\Ax$ (resp. $\Bx$) 
  and \emph{Tot}. Then, $\T^\Gamma$ is 
  total.  
\end{lemma}

\begin{proof}
In the $\Ax$ case, since axiom Tot is an $\Hal$ formula, by the truth lemma \ref{lemma: truth lemma}, 
$\T^\Gamma\models \operatorname{Tot}$. Thus, $\T^\Gamma$ is total by proposition \ref{prop: characterizing nonempty total models}. 

In the $\Bx$ case we use lemmas \ref{lemma: truth lemma for M} and \ref{lemma: connection lemma}  instead of lemma \ref{lemma: truth lemma}.
%
%
\end{proof}

Let us fix a signature $\sigma$. Write $\Gamma\models^{\operatorname T}\varphi$ if, for all \emph{total} causal teams $T$ of signature $\sigma$, $T\models \Gamma$ implies $T\models \varphi$.
Similarly, we write $\Gamma\models^{\operatorname T}_1\varphi$ if this holds for all \emph{singleton total} causal teams of signature $\sigma$.

\begin{theorem} \
\begin{enumerate}
\item $\Gamma\models^{\operatorname T}_1\varphi$ iff $\Gamma, \mathrm{Tot} \vdash_\Ax \varphi$.
\item $\Gamma\models^{\operatorname T}\varphi$ iff $\Gamma, \mathrm{Tot} \vdash_\Bx \varphi$.
\end{enumerate}
\end{theorem}

\begin{proof}
We prove the left-to-right directions.
Suppose $\Gamma, \mathrm{Tot} \not\vdash \varphi$. Then, by the usual argument, $\Gamma \cup\mathrm{Tot} \cup \{\cneg\varphi\}\cup \Ax$ (resp. $\Gamma \cup\mathrm{Tot} \cup \{\cneg\varphi\}\cup \Bx$) is consistent, and by the Lindenbaum lemma it is included in a maximally consistent set $\Delta$. By lemma \ref{lemma: nonempty totality is canonical}, $\T^\Delta$ is total. Furthermore, $\T^\Delta\models\Delta$ (by the truth lemma \ref{lemma: truth lemma}, in the $\Ax$ case; by lemmas  \ref{lemma: truth lemma for M} and \ref{lemma: connection lemma} in the $\Bx$ case). In particular, $\T^\Delta\models \Gamma$ but $\T^\Delta\not\models \varphi$, and thus $\Gamma \not\models^{\operatorname{T}}_{(1)}\varphi$.  
\end{proof}

Secondly, we can see that the class of 
total \emph{recursive} models can be axiomatized by further adding axiom R (it turns out that, within the class of total models, the logic of recursivity coincides with that of strict/visible recursivity).

\begin{lemma}
    \corrf{Let $\Gamma \supseteq \Ax \cup \{ \text{Tot, R}\}$ be a maximally consistent set of $\Hal_\sigma$ formulas. Then, $\T^\Gamma$ is total and recursive.}
\end{lemma}

\begin{proof}
    Totality is given by lemma \ref{lemma: nonempty totality is canonical}. Let us prove recursivity. Suppose for the sake of contradiction that there is a cycle $X_1,\dots,X_n$ in the causal graph of $\T^\Gamma$. Since $\T^\Gamma$ is total, then, we may apply theorem \ref{thm: parents vs direct causes} and conclude that $X_1,\dots,X_n$ is also a cycle in the parenthood graph. By proposition \ref{prop: canonical coincidence of PA and VDC}, it is a cycle in the visible causal graph. The proof then proceeds as for lemma \ref{lemma: canonicity of R}. 
\end{proof}

\noindent Then, the usual line of reasoning tells us that $\Bx \cup \{ \text{Tot, R}\}$ is a sound and complete axiomatization for the class of total recursive models, while $\Ax \cup \{ \text{Tot, R}\}$ is a sound and complete axiomatization for the class of total recursive singleton models.

Finally, a moment of thought shows that the singleton causal teams that are total, deterministic and recursive essentially describe the same semantics as the recursive causal models of \citet{GalPea1998} and \citep{Hal2000,Hal2016}.\footnote{The reader should refer to \cite{Hal2016} for the clearest description of the semantics.} Indeed, first of all, while Halpern's causal settings only include an assignment $u$ over the \emph{exogenous} variables, the recursivity and determinism constraints make so that $u$ determines a unique assignment $s$ over all variables that complies with the causal laws. Secondly, even though our definition of intervention in the general case differs from Halpern's, it was shown in \citet{BarGal2022} that in the (deterministic and) recursive case the two notions of interventions coincide. On the side of the syntax, we note that our axioms are sound in Halpern's models if we follow him in interpreting the formulas $\psi$ without modal operators as abbreviations for $\Box\psi$. 
We can thus conclude that:

\begin{itemize}
    \item The axiom systems $\Ax$ + Rec + Tot + Det and $\Bx$ + Rec + Tot + Det + Sing are  alternative axiomatizations for the class of Halpern's recursive causal models. 
\end{itemize}
One might be tempted to surmise that the axiom systems $\Ax$ + Tot + Det and $\Bx$ + Tot + Det + Sing are then  axiomatizations of Halpern's general causal models, but that is incorrect. One reason, already hinted above, is that without the restriction of recursivity, our notion of intervention produces distinct results from  both the \cite{Hal2000} and \cite{Hal2016} versions of Halpern's definition. For example, Halpern's empty interventions can produce new assignments, so that axiom I7 ($\SET Y = \SET y \leftrightarrow \Box \SET Y = \SET y$) can be falsified. Note furthermore that, in our framework, if we apply the same intervention to two assignments $s,s'$ that agree on the values of the \corrf{external} variables, we might still get two distinct \corrf{sets of} post-intervention assignments, in case $s$ and $s'$ disagree on the values of some variables that are not descendants of the intervened variables. Thus, it is not even clear, in the nonrecursive case, to what singleton causal team should a causal setting $(M,u)$ of \cite{Hal2000}'s semantics correspond to.\footnote{This issue does not arise in the later \cite{Hal2016}, where formulas are evaluated on assignment over the set of all variables.} Most likely, such a causal setting should correspond to the \emph{maximal} causal team containing \emph{all} the assignments that assign $u$ to the exogenous variables and respect the constraint of compatibility with the causal laws (this shows that, implicitly, causal team semantics was already used in \cite{Hal2000}). Interestingly, it can be shown that this class of model satisfies axiom I6 (\emph{Weak reversibility}), and thus plausibly it can be axiomatized by some appropriate extension of system $\Ax \setminus \{\textrm{I}10\}$. We will not pursue this line of thought any further in this paper, as in any case the language from \cite{Hal2000} requires further complications in the semantics.\footnote{We are referring to the fact that the formulas used in \cite{Hal2000} may involve multiple contexts -- for example, the formula $Y(u)=y \with Y(u')=y'$ involves two different assignments $u,u'$ over the exogenous variables. Such formulas require a \emph{family} of maximal causal teams for their evaluation.}


\section{Simplified signatures}\label{sec: simplified signatures}

\corrf{We have been working with signatures that explicitly encode the distinction between internal and external variables. An alternative, closer to the tradition in the philosophical literature, is to take signatures to be pairs of the form
$\mathtt{s}=(\dom,\ran)$ (where $\dom$ and $\ran$ are as before); we will call these \textbf{simplified signatures}. Models of simplified signature $\mathtt{s}$ and interventions on them can be redefined with minimal changes to our earlier definitions; each model $T$ has its own set $\inte(T)$ of internal variables, which can be an arbitrary subset of the variable domain given by the signature. The result is a bit more elegant, as intervening on a model of simplified signature  $\mathtt{s}$ produces again a model of simplified signature  $\mathtt{s}$. In the same vein, we can define languages $\Hal_{\mathtt s}$.
A number of completeness results have been proved in the literature \citep{Bri2012, BarYan2022,BarVir2023b} relative to  simplified signatures; such results are more general than those concerning the other type of signatures, as they cover larger classes of models. The main reason why we do not adopt simplified signatures in our official framework is that the internal/external distinction is essential in order to formulate axiom I8 of our  axiomatizations (and, for system $\Bx$, axioms J10, J11, J16 and J17); we will see that the internal/external distinction is \emph{undefinable} in $\Hal_\mathtt{s}$. It may be, of course, that a different, still reasonable axiomatization exists for classes of models with simplified signature, but this needs remain for now an open problem. A second reason for favouring the ordinary signatures was the discovery that the deterministic class of models is undefinable if one uses simplified signatures. In this section we first explore these kinds of undefinability results. After that, we see three strategies for recovering the definability of determinism while keeping simplified signatures: by restricting attention to singleton models, by extending the language $\Hal$ with right-nested counterfactuals, or by extending it with a learning operator.}

\subsection{Undefinability results}\label{sec: undefinability}


In this section we see some limitations in the expressivity of language $\Hal$ in the context of simplified signatures; a few of them extend to the language with full signatures. We first identify a crucial closure condition for the classes of models definable in $\Hal_{\mathtt s}$. The following definition describes a type of internal variables that may be particularly difficult to distinguish from external variables.

\begin{definition}\label{df: quasiexternal and generic}
Let $T =(T^-,\F)$ be a model and $V\in \inte(T)$. We say that:
\begin{enumerate}
    \item $V$ is \textbf{quasi-external} for $T$ if 
    \corrf{$\F_V(pa)=\F_V(pa')$ for all $pa,pa'\in \ran(PA_V)$}.

    \item     $T$ is \textbf{generic} for $V$ if, for 
    each value $v\in \bigcup_{pa\in\ran(PA_V)}\F_V(pa)$, there is an $s\in T^-$ such that 
    $s(V)=v$. 
\end{enumerate}
 \end{definition}

\noindent If $\Gamma$ is a set of formulas of (simplified) signature $\mathtt s$, we denote as $\K_\Gamma^\mathtt{s}$ the set of models of signature $\mathtt s$ which satisfy all formulas of $\Gamma$.  We are now ready to formulate the closure condition.


\begin{theorem}\label{thm: closure condition for H}
    Let $T =(T^-,\F)$ be a model of signature $\mathtt s$ and $V\in \inte(T)$, and suppose furthermore that $V$ is quasi-external for $T$ and $T$ is generic for $V$. Consider the model $T^*=(T^-,\G)$, where $\G=\F_{\upharpoonright\inte(T)\setminus\{V\}}$ (i.e., the only difference between $T$ and $T^*$ is that $V$ is internal in $T$ and external in $T^*$). For all $\Gamma\subseteq \mathcal H_{\mathtt s}$,
    \[
    T\in \K_\Gamma^\mathtt{s} \iff T^* \in \K_\Gamma^\mathtt{s}.
    \]
\end{theorem}

\noindent For the proof, it may be useful to review here some earlier notational conventions. Remember that, if $G$ is the parenthood graph of a model $T$, we write  $G_{\SET X}$ for the graph obtained from $G$ by omitting all arrows that point to variables in $\SET X$; we write $G_V$ for $G_{\{V\}}$. If $G$ is the parenthood graph of $T$, we will write $\SET N^G_{\SET X}$ for the set of nondescendants of $\SET X$ in $G$, and $PA^G_V$ for the set of parents of $V$ in $G$.  And remember that, if $T^-$ is a team, $T^-(\SET X)$ denotes the set $\{s(\SET X) \mid s\in T^-\}$.

\begin{proof}
    Let $\varphi\in\Gamma$. It suffices to prove that $T\models \varphi$ iff $T^*\models \varphi$; we proceed by induction on $\varphi$.

    For the base case, we just observe that, since $T$ and $T^*$ have the same team component, $T\models Y=y$ iff $T^*\models Y=y$. The inductive steps for $\varphi$ of the form $\psi\with\chi$ or $\cneg\psi$ are straightforward. In case $\varphi$ is $\Box \psi$, we observe that, since $\psi$ contains no modal operators, we obviously have $(\{s\},\F)\models \psi$ iff $(\{s\},\G)\models \psi$ for each $s\in T^-$; thus, $T\models \Box\psi$ iff $T^*\models \Box\psi$.

    Lastly, suppose $\varphi$ is $[\SET X =\SET x]\psi$, with $\SET X\neq \emptyset$. Since $\psi$ has no occurrences of modal operators, the desired conclusion follows immediately if we prove that $(T_{\SET X = \SET x})^- = (T^*_{\SET X = \SET x})^-$. This is obvious if $V\in \SET X$. 
    Let us then assume $V\notin\SET X$. 

    $\subseteq$) Let $t\in (T_{\SET X = \SET x})^-$. From this it follows that: 1) $t(\SET X)=\SET x$, 2) $t$ is compatible with $\F_{\SET X = \SET x}$, 3) $t(\SET N^G_{\SET X})\in T^-(\SET N^G_{\SET X})$ (where $G$ is the parenthood graph of $T$). From 2), it follows a fortiori that $t$ is compatible with $\G_{\SET X = \SET x}$. It remains then to check that $t(V)\in T^-(V)$ (as $\SET N^{G_V}_{\SET X} = \SET N^G_{\SET X}\cup \{V\}$). If $V\in \SET N^G_{\SET X}$, this is immediate. Suppose then that $V\notin \SET N^G_{\SET X}$. Notice that, since $t\in (T_{\SET X = \SET x})^-$ and $V \notin \SET X$, $t(V)\in \F_V(t(PA_V))$; thus, since $T$ is generic for $V$,  there is an $s \in T^-$ such that $s(V)=t(V)$. Thus, $t(V)\in T^-(V)$, and we may conclude that $t\in (T^*_{\SET X = \SET x})^-$.

    $\supseteq$) Let $t\in (T^*_{\SET X = \SET x})^-$. Then, we have that 1) $t(\SET X)=\SET x$, 2) $t$ is compatible with $\G_{\SET X = \SET x}$, 3)  $t(\SET N^{G_V}_{\SET X})\in T^-(\SET N^{G_V}_{\SET X})$. Since $\SET N^G_{\SET X} \subseteq \SET N^{G_V}_{\SET X}$, from 3) we obtain $t(\SET N^G_{\SET X})\in T^-(\SET N^G_{\SET X})$. We only need to show, then, that $t$ is compatible with $(\F_{\SET X = \SET x})_V = \F_V$, i.e., $t(V) \in \F_V(t(PA^G_V))$. To this purpose note that, since $V$ is external in $T^*$, by definition of intervention $t(V)\in T^-(V)$. Thus, there is an $s\in T^-$ such that $s(V)=t(V)$. By the compatibility constraints of $T$, $t(V)\in \F_V(s(PA_V^G))= 
    \F_V(t(PA_V^G))$, where the equality holds because $V$ is quasi-external in $T$. Thus, $t$ is compatible with $\F_{\SET X = \SET x}$, and we conclude that $t\in (T_{\SET X = \SET x})^-$.    
\end{proof}

\begin{corollary}\label{cor: determinism is undefinable}
    Suppose the simplified signature $\mathtt s$ has at least two variables, each of them with at least two distinct possible values.\footnote{The same argument can be easily adapted also to the case that one of the two variables is single-valued, but we prefer to illustrate it in a less pathological context.} Then:
    \begin{enumerate}
        \item The set of deterministic models of signature $\mathtt s$ is not definable in $\mathcal H_\mathtt s$ (not even using an infinite  set of formulas). Neither is its complement.
        \item The same holds for its intersections with the classes of total models, recursive models, \corrf{nonempty models}, or any of their intersections.
    \end{enumerate}
\end{corollary}

\begin{proof}
    Let $\mathtt{s}=(\dom,\ran)$, where $\dom=\{B,C\}$, $\ran(B) =\ran(C)=\{1,2\}$, and let $\Gamma$ be an arbitrary set of $\mathcal H_\mathtt{s}$ formulas.
    We consider a small modification of the model given in example \ref{example: dummy parent}.  Let $T=(T^-,\F)$, where $T^- =\{\{(B,1),(C,1)\}, \{(B,1),(C,2)\}\}$, $\inte(T) = \{C\}$, $PA_C=\{B\}$ and $\F_C(1) = \F_C(2) = \{1,2\}$. This last condition says that $C$ is quasi-external in $T$; the fact that $T^-(C) = \F_C(1) = \bigcup_{b=1,2}\F_C(b)$ says that $T$ is generic for $C$. Thus, by theorem \ref{thm: closure condition for H}, $T\models \Gamma$ iff $T^* = (T^-, \F_{\upharpoonright\inte(T)\setminus \{C\}})= (T^-,\emptyset) \models \Gamma$. But $T^*$ is deterministic, while $T$ is not. Thus, $\K^\mathtt{s}_\Gamma$ is not the set of deterministic models of signature $\mathtt{s}$ (nor is the set of nondeterministic ones). This proves point 1. To prove 2., it suffices to note that both $T$ and $T^*$ are nonempty, total and (strictly) recursive models. 
\end{proof}

\noindent The undefinability of the deterministic class of models does not necessarily entail the impossibility of a reasonable axiomatization -- it might simply be that this class shares the same set of validities with a larger, definable class of models, and that such class admits a recursive axiomatization. We leave it as an open problem whether such class exists and whether it has a defining set of formulas that is canonical for the deterministic class.

A possible reaction to this kind of result is to think that the class of deterministic models, as described in here, is an unnatural one. After all, if we allow the external variables to take multiple values, aren't these models just forbidding indeterminism for the internal variables -- while allowing it for the external ones? If we forbid indeterminism at both levels, we obtain the singleton deterministic models; so, this should be the correct notion of deterministic model. We find this attitude to be untenable, for a multitude of reasons. First, we find it doubtful that the multiplicity of values for an external variable $U$ in a team may always be interpreted as the existence of an indeterministic causal law producing $U$ in the same sense as indeterminacy for the causal laws. 
Secondly, even the 1-assignment models exhibit forms of uncertainty that are not due to indeterminism in the laws -- the uncertainty that is triggered by the multiplicity of solutions in cyclic models. Should we then also forbid multiple-solution models, in order to properly characterize determinism? But the final blow for this attitude comes from the fact that the class of deterministic models (in our sense) \emph{is} definable when we use the ordinary signatures (see section \ref{sec: determinism}). Furthermore, even with simplified signatures, determinism becomes definable within very natural extensions of language $\mathcal H_{\textsf{s}}$, as we shall see in a moment. Thus, we take it that the problem lies not in the definition of the deterministic class, but rather in the limited expressivity of language $\mathcal H_{\textsf{s}}$.

As a second application of the closure condition outlined in theorem \ref{thm: closure condition for H}, we show that the distinction between internal and external variables cannot be captured in $\Hal_{\mathtt{s}}$ (nor can the notion of strict recursivity). If we had a formula $\varphi_{\ext(Y)}$ characterizing the fact that $Y$ is external, then we could reformulate axiom I8 in the form $\varphi_{\ext(Y)} \rightarrow (\langle \SET W_Y = \SET w\rangle Y=y \leftrightarrow \Diamond Y=y)$ and obtain a complete system; the undefinability result shows that this simple trick is unavailable. A quick argument to show this undefinability consists in observing that, if we had formulas $\varphi_{\ext(Y)}$, then we would be able to characterize the deterministic models of signature $\mathtt s$ by the formula $\varphi_{\ext(Y)} \rightarrow \operatorname{Det}$ (cp. section \ref{sec: determinism}), contrary to corollary \ref{cor: determinism is undefinable}. Our alternative argument will also show that the notion of direct cause and related concepts like exogeneity/endogeneity and recursivity are undefinable in $\mathcal H$; the same goes for the notion of a dummy argument. 



\begin{corollary}
    Let $\mathtt{s} = (\dom,\ran)$ be a simplified signature.
    \begin{enumerate}
    \item \corrf{The property of being internal (resp. external) is not definable in $\Hal_{\mathtt s}$, i.e., for each $Y$ and for all models of signature $\mathtt s$, there is no $\Gamma\subseteq\Hal_{\mathtt s}$ such that $T\models \Gamma \iff$ $Y$ is internal (resp. external) in $T$.}
    
    \item The notion of direct cause is not definable in $\Hal_{\mathtt s}$, i.e., there is no $\Gamma\subseteq\Hal_{\mathtt s}$ such that, for all models of signature $\mathtt s$, $T\models\Gamma \iff$ $X$ is a direct cause of $Y$ in $T$. The same for not being a direct cause.

     \item  The notion of recursivity is not definable in $\Hal_{\mathtt s}$, i.e., there is no $\Gamma\subseteq\Hal_{\mathtt s}$ such that, for all models of signature $\mathtt s$, $T\models\Gamma \iff$ $T$ is recursive. The same for its negation.

    \item The notion of endogeneity is not definable in $\Hal_{\mathtt s}$, i.e.,  there is no $\Gamma\subseteq\Hal_{\mathtt s}$ such that, for all models of signature $\mathtt s$, $T\models\Gamma \iff$ $X$ is endogenous in $T$. The same for exogeneity.

    \item \corrf{The notion of being a dummy argument is not definable in $\Hal_{\mathtt s}$, i.e., for all variables $X,Y\in \dom$, there is no $\Gamma\subseteq\Hal_{\mathtt s}$ such that, for all models of signature $\mathtt s$, $T\models\Gamma \iff X$ is a dummy argument for $Y$.} 
    \end{enumerate}
\end{corollary}

\begin{proof}
  Consider the simplified signature $\mathtt s = (\dom,\ran)$ such that $\dom=\{X,Y\}$, $\ran(X) = \ran(Y)=\{1,2\}$. Let $T^-$ be the full team for this signature:
  \begin{center}
\begin{tabular}{|c|c|}
\hline
 \multicolumn{2}{|l|}{ \ X \ \ \ \ \ \ Y } \\
\hline
 \phantom{a}1\phantom{a} & \phantom{a}1\phantom{a} \\
 \hline
\phantom{a}1\phantom{a} & \phantom{a}2\phantom{a} \\ 
\hline
\phantom{a}2\phantom{a} & \phantom{a}1\phantom{a} \\
 \hline
\phantom{a}2\phantom{a} & \phantom{a}2\phantom{a} \\ 
\hline
\end{tabular}
\end{center}
2) This team is compatible with the following causal laws: $PA_X=\{Y\}$, $PA_Y=\{X\}$,  $\F_X(1)=\F_X(2)=\{1,2\}$, $\F_Y(1)=\F_Y(2)=\{1,2\}$. Thus, in particular, $X$ is a direct cause of $Y$ in $T = (T^-,\F)$ (just look at the effect of any intervention on a single assignment). Instead, in $T^* = (T^-,\F_{\upharpoonright \{X\}})$, $X\notin PA_Y$, so by theorem \ref{thm: parents vs direct causes} (whose proof is unchanged if we use simplified signatures), $X$ is not a direct cause of $Y$ in $T^*$. But, on the other hand, $Y$ is quasi-external in $T$, and $T$ is generic for $Y$; thus, by theorem \ref{thm: closure condition for H}, $T$ and $T^*$ satisfy the same formulas.

1) In the example above, $Y$ is internal in $T$ but external in $T^*$. But, by theorem \ref{thm: closure condition for H}, $T$ and $T^*$ satisfy the same $\Hal$ formulas.

3)  Just observe that $T^*$ is recursive while $T$ is not.

    4)  Observe that $Y$ is endogenous in $T$ but exogenous in $T^*$.

    5) Observe that $X$ is a dummy argument for $Y$ in $T$ but not in $T^*$ (as, in $T^*$, $Y$ is external). 
\end{proof}

\noindent We remark that the notion of direct cause and those defined in terms of it do not really really depend on having the distinction between external and internal variables in the signature. Thus, an $\Hal_\sigma$ formula defining, say, "$X$ is a direct cause of $Y$" (where $\sigma = (\mathcal E, \mathcal I, \ran)$) does so also over models of signature $\mathtt s$ (where $\mathtt s = (\mathcal E \cup \mathcal I,\ran)$), so it would be an (impossible) $\Hal_{\mathtt s}$ definition of "$X$ is a direct cause of $Y$". Thus, some of the definability results also extend to full signatures.

\begin{corollary}
    The notions of direct cause, recursivity, endogeneity and their negations are not definable in $\Hal_\sigma$.
\end{corollary}

\noindent The previous shower of undefinability results suggests that, if we use simplified signatures, we should either modify the notion of model or consider a stronger language. We now turn to these two possibilities.

\subsection{Capturing determinism in the singleton case}\label{subs: determinism in singletons}

We have seen that the class of deterministic models is undefinable if we adopt the simplified signatures. We show here that, however, determinism can be characterized if we restrict attention to the class of singleton models. Consider the following axiom:

\vspace{5pt}

Det$_\mathtt{s}$. $\langle \SET W_Y = \SET w\rangle Y=y \rightarrow \cneg \langle\SET W_Y = \SET w\rangle Y=y'$. 

\vspace{5pt}

\noindent This is just axiom Det, without $Y$ being restricted to internal variables (we cannot make this distinction with simplified signatures). As already observed, such an axiom is not sound over the class of deterministic models (an external $Y$ may violate it); but it is sound for \emph{singleton} deterministic models, and characterizes them.





\begin{lemma}\label{lemma: semantic characterizartion of 1-determinism}
  Let $T= (\{s\},\F)$ be a singleton relational causal team with simplified signature. Then, $T\models \mathrm{Det}_\mathtt{s}$ iff 
  $T$ is deterministic.
\end{lemma}

\begin{proof}
$\Rightarrow$) 
This direction of the proof is identical to the left-to-right implication in the proof of lemma \ref{lemma: semantic characterization of determinism}.

$\Leftarrow$) Suppose $T$ is a deterministic singleton model, and assume that $T\models  \langle \SET W_Y = \SET w\rangle Y=y$ for some $Y,y,\SET w$. 
In case $Y$ is not in $\inte(T)$, and thus $Y$ is not a descendant of any other variable, by definition of intervention, and the fact that $T$ is a singleton model, $T^-(Y) = T_{ \SET W_Y = \SET w}^-(Y) =  \{y\}$. Thus, 
$T\not \models \langle\SET W_Y = \SET w\rangle Y=y'$ if $y'\neq y$, from which we immediately obtain $T \models \cneg \langle\SET W_Y = \SET w\rangle Y=y'$. 

If instead that $Y\in\inte(T)$, the proof proceeds as in lemma \ref{lemma: semantic characterization of determinism}.
%
%
%
\end{proof}










We also note that generic\footnote{In the sense of definition \ref{df: quasiexternal and generic}.} singleton models do not exist except for trivial signatures (which allow single-valued variables). Thus, for singleton models, we cannot prove undefinability results via theorem \ref{thm: closure condition for H}. Nonetheless, we are not aware of sets of formulas that may be used to define the relativizations to singleton models of notions like being internal/external, being a direct cause and so on.


 \subsection{Capturing determinism via nested counterfactuals}\label{sec: nested conditionals}

A natural way of extending language $\mathcal H$ is to allow for right-nested counterfactuals -- e.g., formulas of the form $[\SET X= \SET x][\SET Y=\SET y]\psi$. We have already introduced such languages; when discussing a simplified signature $\mathtt s$, the corresponding language with right-nested counterfactuals will be denoted as 
$\Hal^+_{\mathtt s}$. 
We immediately see that $\mathcal H^+_{\mathtt s}$ is more expressive than $\mathcal H_{\mathtt s}$, precisely because it captures the deterministic class.

\begin{theorem}\label{thm: semantic characterizartion of determinism}
    Let $T$ be a model with simplified signature. Then, $T$ is deterministic iff it satisfies the following axiom scheme: 
    \begin{center}
\emph{Det}$^\Box$. \hspace{10pt} $\Box(\langle \SET W_Y = \SET w\rangle Y=y \rightarrow \cneg \langle\SET W_Y = \SET w\rangle Y=y') \with$ 

$(\cneg\Diamond\top \rightarrow (\langle \SET W_Y = \SET w\rangle Y=y \rightarrow \cneg \langle\SET W_Y = \SET w\rangle Y=y') )$ \hspace{15pt} (when $y\neq y'$).
\end{center}
\end{theorem}

\begin{proof}
 $\Rightarrow$) Suppose $T = (T^-,\F)$ is deterministic. Then, for all $s\in T^-$, $(\{s\},\F)$ is deterministic. Thus, by lemma \ref{lemma: semantic characterizartion of 1-determinism}, $(\{s\},\F)\models \langle \SET W_Y = \SET w\rangle Y=y \rightarrow \cneg \langle\SET W_Y = \SET w\rangle Y=y'$ (for any choice of $Y,\SET w, y\neq y'$). Since this holds for all $s\in T^-$, $T\models \Box(\langle \SET W_Y = \SET w\rangle Y=y \rightarrow \cneg \langle\SET W_Y = \SET w\rangle Y=y')$. 

 Now suppose $T\models \cneg\Diamond\top$. Then, $T$ is empty. If $Y$ is external, then by definition of intervention $T\not\models \langle \SET W_Y = \SET w\rangle Y=y$, thus $T\models \langle \SET W_Y = \SET w\rangle Y=y \rightarrow \langle \SET W_Y = \SET w\rangle Y=y'$. If $Y$ is internal, then, since $T$ is deterministic, $T_{\SET W_Y = \SET w}$ has at most one assignment. If it is empty, then $T\not\models \langle \SET W_Y = \SET w\rangle Y=y$, and we conclude as before. Suppose instead $T_{\SET W_Y = \SET w}^- = \{s\}$, with $s(Y)=y^*$. If $y^*\neq y$, then $T\not\models \langle \SET W_Y = \SET w\rangle Y=y$ and we conclude as before. In case $y^*=y$, then $T\not\models \langle \SET W_Y = \SET w\rangle Y=y'$, so $T\models \langle \SET W_Y = \SET w\rangle Y=y \rightarrow \cneg \langle\SET W_Y = \SET w\rangle Y=y'$. 

 $\Leftarrow$) Suppose $T\models \mathrm{Det}^\Box$. Then, for any choice of $Y,\SET w, y\neq y'$, and any $s\in T^-$, we have $(\{s\},\F)\models \langle \SET W_Y = \SET w\rangle Y=y \rightarrow \cneg \langle\SET W_Y = \SET w\rangle Y=y'$. Thus, by lemma \ref{lemma: semantic characterizartion of 1-determinism}, $(\{s\},\F)$ is deterministic. Suppose $T$ is nonempty, i.e., there is such an $s$; then, since $T$ and $(\{s\},\F)$ have the same causal laws, also $T$ is deterministic. In case $T$ is empty, by the second conjunct of axiom Det we obtain $T\models \langle \SET W_Y = \SET w\rangle Y=y \rightarrow \cneg \langle\SET W_Y = \SET w\rangle Y=y')$ (for each pair of distinct $y,y'$). Thus, if $Y$ is internal in $T$, by definition of intervention $|\F_Y(\SET w_{\upharpoonright PA_Y})| \leq 1$.
\end{proof}

\noindent Thus, the axiom system $\Bx+\operatorname{Det}^\Box$ is sound and complete for $\Hal^+$ over the class of deterministic models.

\subsection{Capturing determinism via learning}\label{subs: determinism via learning}

A second natural extension of language $\mathcal H$ is obtained by adding to it a second conditional operator $\supset$; the intended interpretation of $\psi\supset \chi$ is that after \emph{learning} $\psi$ (e.g., after observing a feature of the world or the outcome of an experiment) one can conclude that $\chi$ holds. This operator is introduced in \cite{BarSan2020} under the name of \emph{selective implication} and is a sort of qualitative counterpart of probabilistic conditionalization \citep[see][]{BarSan2024}.

\[
\text{Language } \mathcal {HO}_{\mathtt{s}}: \hspace{20pt} 
X=x  \mid  \cneg\psi   \mid   \psi\with \chi    \mid \alpha\supset \chi \mid  [\SET X=\SET x] \eta    
\] 
where $\eta$ has no occurrences of modal operators, and $\alpha$ is restricted to formulas whose occurrences of $\cneg$ (if any) are exclusively inside the scope of modal operators. We will say that such an $\alpha$ \emph{has the $\cneg$-restriction}. The semantic clause for the additional operator is:
\begin{itemize}
\item $T \models \alpha \supset \chi$ iff $T^\alpha \models \chi$, where $T^\alpha = \{s \text{ of signature }\mathtt s \mid (\{s\},\F) \models \alpha \}$. 
\end{itemize}

\noindent The restriction on $\alpha$ guarantees that $\alpha$ is \emph{flat}, i.e., it holds of a causal team iff it holds of all its singleton subteams (this will be proved below); and, consequently, that the conditional can really be interpreted as saying that ``upon learning $\alpha$, $\chi$ holds''. Without this restriction,  unintended meanings easily arise. For example, this operator should satisfy the classical principle $A \supset A$. But using $\cneg$ we can define $\sqcup$; and a formula of the form $(X=0 \sqcup X=1) \supset (X=0 \sqcup X=1)$ is not satisfied by teams which feature both values $0$ and $1$ for $X$.

We may also write $\alpha \subset \supset \beta$ as an abbreviation for $(\alpha\supset\beta)\with(\beta \supset \alpha)$. Since both $\alpha$ and $\beta$ need to be flat,  $T=(T^-,\F)\models \alpha \subset \supset \beta$ means that, for all $s\in T^-$, $(\{s\},\F)\models \alpha$ iff $(\{s\},\F)\models \beta$; or equivalently, that $T^\alpha = T^\beta$. 

Also $\HO$ captures determinacy. 
Consider the following axiom scheme:
\begin{center}
Det$^\mathcal O$. 
$\SET W = \SET w\supset [\langle \SET W_Y = \SET w'\rangle Y=y \rightarrow \cneg \langle\SET W_Y = \SET w'\rangle Y=y']$ 
\hspace{20pt} (when $y\neq y'$)
\end{center}

\begin{theorem}
 Let $T$ be a 
 model with simplified signature. $T\models \mathrm{Det}^\mathcal O$ iff $T$ is deterministic.
\end{theorem}

\begin{proof}
$\Rightarrow$) Suppose $T\models \mathrm{Det}^\mathcal O$. We need to prove that, if $Y$ is internal, then for all $pa \in \ran(PA_Y)$, $\F_Y(pa)$ is contained in a singleton set. Pick an arbitrary $\SET w \in \ran(\SET W)$ 
and let $\SET w'\in \ran(\SET W_Y)$ be such that $\SET w'_{\upharpoonright PA_Y} = pa$. Since $T\models \mathrm{Det}^\mathcal O$, we have  $T^{\SET W = \SET w}\models \langle \SET W_Y = \SET w'\rangle Y=y \rightarrow \cneg \langle\SET W_Y = \SET w'\rangle Y=y'$ whenever $y\neq y'$; thus,  $(T^{\SET W = \SET w})^-_{\SET W_Y = \SET w'}$ contains at most one assignment. If it is empty, then by definition of intervention $\F_Y(pa)=\emptyset$, and if it has an assignment $s$, 
then $\F_Y(pa)=\{s(Y)\}$, as needed. 



$\Leftarrow$) Suppose $T = (T^-,\F)$ is deterministic. 
\corrf{Suppose first that $Y$ is internal, and assume that $T^{\SET W = \SET w}\models \langle\SET W_Y = \SET w'\rangle Y=y$. Let $y'\in\ran(Y)$ be such that $y'\neq y$. Since $\F_Y$ is deterministic, $T^{\SET W = \SET w}\models \cneg \langle\SET W_Y = \SET w'\rangle Y=y'$, as needed.}

\corrf{Suppose instead $Y$ is external. If $T^{\SET W = \SET w}$ is a singleton model, then, by definition of intervention and determinism also $(T^{\SET W = \SET w})_{\SET W_Y = \SET w'}$ is a singleton model, say with team $\{t\}$. If $t(Y)\neq y$, then $T^{\SET W = \SET w}\models \langle \SET W_Y = \SET w'\rangle Y=y \rightarrow \cneg \langle\SET W_Y = \SET w'\rangle Y=y'$ because the antecedent is false. If $t(Y) = y$, then $T^{\SET W = \SET w}\models \cneg \langle\SET W_Y = \SET w'\rangle Y=y'$, and so again $T^{\SET W = \SET w}\models \langle \SET W_Y = \SET w'\rangle Y=y \rightarrow \cneg \langle\SET W_Y = \SET w'\rangle Y=y'$.}

If instead $T^{\SET W = \SET w}$ is empty, since $Y$ is external, by definition of intervention also $(T^{\SET W = \SET w})_{\SET W_Y = \SET w'}$ is empty; so, $T^{\SET W = \SET w}\not\models \langle \SET W_Y = \SET w'\rangle Y=y$. Thus, $T\models \langle \SET W_Y = \SET w'\rangle Y=y \rightarrow \cneg \langle\SET W_Y = \SET w'\rangle Y=y'$.
\end{proof}

Again, we lack a complete axiomatization for $\HO$, either on ordinary or simplified signatures; and we remark that the Henkin method via canonical models seems ill-suited for treating the operator $\supset$.\footnote{Actually, an axiomatization via canonical models could be obtained for a probabilistic language featuring $\supset$ in \cite{BarVir2023b}, but the proof strategy does not apply to our case.} We think that, in any case, the additional operator is perfectly natural in the context of causal inference; thus, we use the rest of the section to understand language $\HO$ better and suggest a partial axiomatization.

 The following definitions are (analogues of) standard ones from the literature on team semantics; we assume a fixed simplified signature $\mathtt{s}$. $S =(S^-,\F)$ is a \textbf{causal subteam} of $T=(T^-,\G)$ ($S\leq T$) if $S^- \subseteq T^-$ and $\F=\G$.  
\begin{itemize}
    \item $\varphi$ is \textbf{flat} if, for all causal teams $T =(T^-,\F)$, $T\models \varphi$ iff for every $t\in T^-$, $(\{t\},\F)\models \varphi$. 
    \item $\varphi$ has the \textbf{empty team property} if it is satisfied by all empty causal teams of signature $\mathtt{s}$.
    \item $\varphi$ is \textbf{downward closed} if, whenever $T\models\varphi$ and $S\leq T$, then $S\models \varphi$.
    \item $\varphi$ is \textbf{upwards closed} if, whenever $S\models\varphi$ and $S\leq T$, then $T\models \varphi$.
\end{itemize}
Note in particular that flatness implies downward closure and the empty team property.

\begin{lemma}\label{lemma: closure properties of HO}
    \begin{enumerate}
        \item If $\alpha\in\HO$ has the $\cneg$-restriction, then it is flat (and therefore is downward closed and has the empty team property).
        \item $\HO$ formulas of the form $\langle\SET X = \SET x\rangle \psi$ are upwards closed.
        \item If $\psi,\chi$ are downward closed, then $\psi\sqcup\chi$ is downward closed.
    \end{enumerate}
\end{lemma}

\begin{proof}
    1) By induction on $\alpha$, simultaneously for all models. The cases for $\with$ and $Y=y$ are straightforward.

    \begin{itemize}
        \item Case $\varphi$ is $\beta\supset\gamma$. Suppose first that $T=(T^-,\F)\models\beta\supset\gamma$. Then, $T^\beta\models \gamma$. Since $\gamma$ has no occurrence of $\cneg$ except in the scope of modal operators,  by inductive hypothesis $(\{s\},\F)\models \gamma$ for each $s\in (T^\beta)^-$. Similarly, applying the i.h. for $\beta$ we obtain that $(\{s\},\F)\models \beta$ for each $s\in (T^\beta)^-$ Thus, if $s\in T^-$ is such that $(\{s\},\F)\models \beta$, then we also have $(\{s\},\F)^\beta = (\{s\},\F)\models \gamma$, and so $(\{s\},\F)\models \beta\supset\gamma$. If instead $s\in T^-$ is such that $(\{s\},\F)\not \models \beta$, then $(\{s\},\F)^\beta=(\emptyset,\F)$; since $\gamma$ is flat by the i.h., it has the empty team property, and so $(\{s\},\F)^\beta\models\gamma$. Thus, again, $(\{s\},\F)\models \beta\supset \gamma$.

        Vice versa, suppose $(\{s\},\F)\models \beta\supset \gamma$ for all $s\in T^-$. Then, for all such $s$, $(\{s\},\F)^\beta\models\gamma$. By the i.h., $T^\beta\models \gamma$. Thus, $T\models \beta\supset\gamma$.

        \item Case $\varphi$ is $[\SET X = \SET x]\beta$. Suppose $T\models [\SET X = \SET x]\beta$. Then, for all $t\in T_{\SET X = \SET x}^-$, $(\{t\},\F_{\SET X = \SET x}) \models \beta$. Note that in case $T$ is empty, flatness just amounts to the assumption that $T\models [\SET X = \SET x]\beta$; thus, we may henceforth assume that $T$ is nonempty. We then have $T_{\SET X = \SET x}^- = \bigcup_{s\in T^-} s^\F_{\SET X = \SET x}$, and so, for all $s\in T^-$ and all $t\in s^\F_{\SET X = \SET x}$, $(\{t\},\F_{\SET X = \SET x})\models \beta$. Thus, for all $s\in T^-$, $(\{s\},\F)\models [\SET X = \SET x]\beta$. All the steps in this argument are reversible.

    \end{itemize}

    2) Let $S=(S^-,\F)\leq T=(T^-,\F)$ and $S\models \langle\SET X = \SET x\rangle\psi$. Then, there is a $t\in S_{\SET X = \SET x}^-$ such that $(\{t\},\F_{\SET X = \SET x})\models\psi$. Since $S_{\SET X = \SET x} \leq T_{\SET X = \SET x}$, such $t$ is in $T_{\SET X = \SET x}^-$. Thus, $T\models \langle\SET X = \SET x\rangle\psi$.

    3) This is a well-known result.
\end{proof}

We remark that $\supset$ does not coincide with the previously defined material implication $\rightarrow$. The subtle relationship between the two operators is described in the following proposition.

\begin{proposition}
\begin{enumerate}
    \item $\psi\rightarrow \chi \not\models \psi\supset\chi$, even when $\psi,\chi$ are atomic formulas. 
    \item If $\alpha$ has the $\cneg$-restriction, then $\alpha\supset\chi \models \alpha\rightarrow\chi$. So, this holds for $\HO$ formulas.
    \item  If $\alpha$ has the $\cneg$-restriction, then $\models \alpha\rightarrow \chi$ iff $\models \alpha\supset \chi$.
\end{enumerate}
\end{proposition}

\begin{proof}
1) Let $T$ be a causal team of domain $\{X\}$, and which features at least the two values $0,1$ for $X$. Then, $T\models X= 0\rightarrow X=1$ because the antecedent is false; but $T^{X=0}\not\models X=1$, so $T\not\models X=0\supset X=1$. 

2) Suppose $T\models \alpha\supset \chi$ and $T\models \alpha$. The latter (by the flatness of $\alpha$) gives $T^\alpha = T$, and so the former gives $T\models \chi$. Thus, $T\models \alpha\rightarrow \chi$.
%

3) The right-to-left direction immediately follows from point 2.

For the left-to right direction, assume that $\models \alpha\rightarrow \chi$, and let $T$ be a causal team. Since $\alpha$ is flat, $T^\alpha\models\alpha$. But then $T^\alpha \models \chi$. So, $T\models\alpha\supset \chi$. Thus, $\models\alpha\supset \chi$.
\end{proof}

\noindent In particular, $\psi\supset\chi$ as a subformula cannot be replaced with $\psi\rightarrow\chi$, or vice versa (we will see that these kinds of  substitutions are allowed within the scope of a modal operator). Point 3. can be taken as a pair of inference rules, which may be applied only to valid formulas.

  To see that the $\cneg$-restriction is needed in 2., and in the right-to-left direction of 3.,
  consider a signature with a Boolean variable $X$; 
  any causal team of such signature which features both values for $X$ satisfies $\cneg X=0 \supset X=1$; but such a causal team does not satisfy 
  $\cneg X=0 \rightarrow X=1$ (the consequent is false). 
   For point 3, a counterexample to the left-to-right implication is given by the formulas $(X=0 \sqcup X=1) \rightarrow (X=0 \sqcup X=1)$ and $(X=0 \sqcup X=1) \supset (X=0 \sqcup X=1)$ .

The following are a number of formal properties of $\supset$, which might serve as a starting point towards axiomatization.

\begin{proposition}
If $\alpha,\beta,\psi,\chi$ are $\HO$ formulas, and $\alpha, \beta$ satisfy the $\cneg$-restriction, then:
\begin{enumerate}
    \item $\alpha \supset (\psi\with \chi) \equiv (\alpha \supset \psi) \with (\alpha \supset \chi)$.
    \item $\alpha \supset \cneg \chi  \equiv  \cneg (\alpha \supset \chi)$.
    \item $\alpha \supset (\beta \supset\psi)  \equiv  (\alpha\with\beta) \supset \psi$.
    \item $\langle\SET X = \SET x\rangle\alpha \rightarrow ([\SET X = \SET x](\alpha \supset \beta) \equiv ([\SET X = \SET x]\alpha) \supset ([\SET X = \SET x]\beta))$.
    \item $(\Diamond \top \with \langle\SET X = \SET x\rangle\alpha) \rightarrow ([\SET X = \SET x](\alpha \supset \chi) \equiv ([\SET X = \SET x]\alpha) \supset ([\SET X = \SET x]\chi))$.
    \item $\alpha \supset (\psi\sqcup \chi) \equiv (\alpha \supset \psi) \sqcup (\alpha \supset \chi)$.
    \item $\alpha \supset (\psi\rightarrow \chi) \equiv (\alpha \supset \psi) \rightarrow (\alpha \supset \chi)$.
    \item 
    $\models(\alpha\subset\supset\beta)\rightarrow [(\alpha\supset \chi) \rightarrow (\beta \supset \chi)]$. 
    \item 
    If $\models \alpha \leftrightarrow \beta$, then $\models(\alpha\supset \chi) \rightarrow (\beta \supset \chi)$.
    \item If $\models\psi\rightarrow\chi$, then $\models (\alpha\supset \psi) \rightarrow (\alpha \supset \chi)$.
     \item $\langle \SET X = \SET x\rangle \varphi \leftrightarrow \langle \SET X = \SET x\rangle \varphi[\alpha \rightarrow\chi / \alpha\supset\chi]$.
    \item $[\SET X = \SET x] \varphi \leftrightarrow [\SET X = \SET x] \varphi[\alpha \rightarrow\chi / \alpha\supset\chi]$.
    \item $\langle \SET X = \SET x\rangle \varphi \leftrightarrow \langle \SET X = \SET x\rangle \varphi[\cneg(\alpha \with \cneg\chi)  / \alpha\supset\chi]$.
    \item $[\SET X = \SET x] \varphi \leftrightarrow [\SET X = \SET x] \varphi[\cneg(\alpha \with \cneg\chi) / \alpha\supset\chi]$.
\end{enumerate}
\end{proposition}

\begin{proof}
1) Straightforward.

2) $T\models \alpha \supset \cneg \chi$ iff $T^\alpha\models \cneg \chi$ iff $T^\alpha \not\models \chi$ iff $T\not\models \alpha\supset\chi$ iff $T\models \cneg(\alpha\supset\chi)$.

3) $T\models\alpha \supset (\beta \supset\psi)$ iff $(T^\alpha)^\beta \models \psi$. But $(T^\alpha)^\beta = T^{\alpha\with\beta}$, so the former holds iff $T^{\alpha\with\beta}\models \psi$, iff $T\models (\alpha\with\beta) \supset \psi$.

4) Let $T=(T^-,\F)$. If $T$ is empty, note that, since $\alpha$ and $\beta$ have the $\cneg$-restriction, also the two formulas in the equivalence have it. Thus, by lemma \ref{lemma: closure properties of HO}, 1. they both have the empty team property; so, both formulas are satisfied by $T$. Let us then assume that $T$ is nonempty. Note that $T\models [\SET X = \SET x](\alpha \supset \beta)$ iff $(T_{\SET X = \SET x})^\alpha\models \beta$, while $T\models ([\SET X = \SET x]\alpha) \supset ([\SET X = \SET x]\beta)$ iff $(T^{[\SET X = \SET x]\alpha})_{\SET X = \SET x}\models\beta$. Then, it suffices to prove that $((T_{\SET X = \SET x})^\alpha)^- = ((T^{[\SET X = \SET x]\alpha})_{\SET X = \SET x})^-$. 

If $s\in ((T_{\SET X = \SET x})^\alpha)^-$, then $(\{s\},\F_{\SET X = \SET x})\models\alpha$ and (since $T$ is nonempty) there is a $t\in T^-$ such that $s\in t^\F_{\SET X = \SET x}$. Thus, $t\in (T^{[\SET X = \SET x]\alpha})^-$. But then $s\in ((T^{[\SET X = \SET x]\alpha})_{\SET X = \SET x})^-$.

Vice versa, assume $s\in ((T^{[\SET X = \SET x]\alpha})_{\SET X = \SET x})^-$. We want to prove, first, that $s\in (T_{\SET X = \SET x})^-$. Since $T\models \langle\SET X = \SET x\rangle\alpha$, $T^{[\SET X = \SET x]\alpha}$ is nonempty; but then, $s\in (T^{[\SET X = \SET x]\alpha})_{\SET X = \SET x}^- = \bigcup_{t\in (T^{[\SET X = \SET x]\alpha})^-} t^\F_{\SET X = \SET x} \subseteq \bigcup_{t\in T^-} t^\F_{\SET X = \SET x} = T_{\SET X = \SET x}^-$.\footnote{This step could also be proved, by a longer argument, without the assumption of nonemptyness of $T^{[\SET X = \SET x]\alpha}$.}

 Since $s\in (T_{\SET X = \SET x})^-$ and $T$ is nonempty, there is a $t\in T^-$ such that $s\in t^\F_{\SET X = \SET x}$. Since furthermore $s\in (T^{[\SET X = \SET x]\alpha})_{\SET X = \SET x})^-$ and $T^{[\SET X = \SET x]\alpha}$ is nonempty, $t$ can be chosen so that $(\{t\},\F)\models [\SET X = \SET x]\alpha$. Thus, by the clause for the modal operator, $(\{s\},\F)\models \alpha$. Thus, $s\in ((T_{\SET X = \SET x})^\alpha)^-$.

5) The proof proceeds as for 4., using the fact that $T\models \Diamond\top$ to exclude the case that $T$ is empty.




6) Straightforward.

7) We observe that
\begin{align*}
    T\models\alpha \supset (\psi\rightarrow \chi) & \iff  T^\alpha\models\psi\rightarrow\chi \\ 
    & \iff T^\alpha\not\models \psi \text{ or } T^\alpha\models \chi \\
    & \iff T\not\models \alpha \supset \psi \text{ or } T\models \alpha\supset\chi \\
    & \iff T\models (\alpha \supset \psi) \rightarrow (\alpha \supset \chi).\\
\end{align*}

8) Suppose $T\models \alpha\subset\supset\beta$ and $T\models \alpha\supset \chi$. From the latter, $T^\alpha\models \chi$. From the former, we obtain $T^\alpha = T^\beta$. Thus, $T^\beta\models \chi$, i.e., $T\models \beta \supset \chi$.

9) Analogous to the proof of 7, using the flatness of $\alpha,\beta$.

10) Straightforward.

11) Fix a $T = (T^-,\F)$. For any given assignment $s\in T^-$, we will use the notation $\eta \equiv_s\theta$ to say that $(s,\F)\models \eta$ iff $(s,\F)\models \theta$. We also write $\eta \equiv_\emptyset\theta$ in case $(\emptyset,\F)\models \eta$ iff $(\emptyset,\F)\models \theta$.

The statement is equivalent to saying that there is an assignment $s\in T_{\SET X =\SET x}^-$ such that $\varphi \equiv_s\varphi[\alpha \rightarrow\chi / \alpha\supset\chi]$. We prove this simultaneously with the statement that  $\varphi \equiv_\emptyset\varphi[\alpha \rightarrow\chi / \alpha\supset\chi]$, by induction on $\varphi$.

If $\varphi$ has no occurrence of $\alpha\supset\chi$ then the statement trivially holds; this covers the atomic case. The inductive steps are easy to prove, except for the case in which the most external operator of $\varphi$ is $\supset$. We have two subcases, depending on whether $\varphi$ is the occurrence $\alpha\supset \chi$ to be substituted, or another formula $\beta\supset \psi$.

In the former case, showing $\varphi \equiv_s \varphi[\alpha \rightarrow\chi / \alpha\supset\chi]$ is straightforward. We show the statement for $\equiv_\emptyset$. Note that, if $(\emptyset,\F)\models \alpha\supset \chi$, then $(\emptyset,\F)^\alpha\models \chi$. But $(\emptyset,\F)^\alpha = (\emptyset,\F)$. Thus, $(\emptyset,\F)\models \chi$. Thus, $(\emptyset,\F)\models \alpha\rightarrow \chi$. In the other direction, if $(\emptyset,\F)\models \alpha\rightarrow \chi$ we have either $(\emptyset,\F)\not\models \alpha$ or $(\emptyset,\F)\models \chi$. The former is excluded by the $\cneg$-restriction on $\alpha$ (since that makes $\alpha$ flat, and thus satisfy the empty team property). So, $(\emptyset,\F) = (\emptyset,\F)^\alpha\models \chi$, i.e., $(\emptyset,\F)\models \alpha\supset \chi$. 

If $\varphi$ is of the form $\beta\supset \psi$, first note that $(\emptyset,\F)\models \beta\supset \psi$ iff $(\emptyset,\F)^\beta =(\emptyset,\F)\models \psi$ iff (by the i.h.) $(\emptyset,\F)=(\emptyset,\F)^{\beta[\alpha \rightarrow\chi / \alpha\supset\chi]}\models \psi[\alpha \rightarrow\chi / \alpha\supset\chi]$ iff $(\emptyset,\F)\models \beta[\alpha \rightarrow\chi / \alpha\supset\chi] \supset \psi[\alpha \rightarrow\chi / \alpha\supset\chi]$. On the other hand, $(\{s\},\F)\models \beta\supset \psi$ iff $(\{s\},\F)^\beta\models \psi$. We can show that the latter is equivalent to $(\{s\},\F)^\beta\models \psi[\alpha \rightarrow\chi / \alpha\supset\chi]$. Indeed, if $(\{s\},\F)^\beta$ is nonempty, then the equivalence is given by the i.h. $\psi\equiv_s \psi[\alpha \rightarrow\chi / \alpha\supset\chi]$; if it is empty, the equivalence is given by the i.h. $\psi\equiv_\emptyset \psi[\alpha \rightarrow\chi / \alpha\supset\chi]$. Finally, $(\{s\},\F)^\beta\models \psi[\alpha \rightarrow\chi / \alpha\supset\chi]$ is equivalent to $(\{s\},\F)^{\beta[\alpha \rightarrow\chi / \alpha\supset\chi]}\models \psi[\alpha \rightarrow\chi / \alpha\supset\chi]$ by the inductive hypothesis on $\beta$. Thus, it is equivalent to  $(\{s\},\F)\models \beta[\alpha \rightarrow\chi / \alpha\supset\chi] \supset \psi[\alpha \rightarrow\chi / \alpha\supset\chi]$.

The proofs of 11., 12. and 13. are analogous.
\end{proof}

\section{Comparisons with related work}\label{sec: comparisons}

As mentioned in the introduction, the idea of extending the Galles-Pearl framework of causal models with \emph{one-to-many} causal laws is already suggested as a final remark in Halpern's original axiomatization work \citep{Hal2000}; on the other hand, we are not aware of any further development in this direction in the following two decades. 

In 2021, Peters and Halpern \citep{PetHal2021,HalPet2022} developed a vast generalization of causal models (the \emph{generalized structural equation models}, GSEM). Their generalizations go into two directions: on one hand, they remove the restriction to finite variable domains and finite value ranges, seemingly with the aim of modeling physical phenomena; and on the other hand, they allow for highly arbitrary interventions, which do not require causal laws for their computation. Such interventions associate a \emph{set} of possible outcomes for the endogenous variables to each setting for the exogenous variables, and the only constraint that the interventions are enforced to abide to is that, say, an intervention $do(\SET X = \SET x)$ should only produce sets of assignments that set the variables $\SET X$ to $\SET x$. It is clear that a causal model with indeterministic causal laws can be seen as a rather special case of a GSEM, although this is not remarked upon in these works; no talk of laws at all seems to be made in these papers. Given the large increase in generality -- well beyond the simple addition of indeterministic laws -- the GSEM framework seems quite unsuitable for obtaining a clear comparison between the usual causal models and their indeterministic generalization. And indeed, even when restricting attention to finite signatures, some of the usual axioms like \emph{Weak composition} (our I3) and \emph{Weak reversibility} (our I6) fail.

In 2023, \cite{Wys2023} focused explicitly on indeterministic causal laws (there called ``underdeterministic'') as an addition to the usual recursive causal models (although, not having to deal with the problem of axiomatization, he does not bother restricting value ranges to he finite case). The paper is mainly a philosophical defense of qualitative indeterministic laws as opposed to probabilistic laws, and provides e.g. several examples where indeterministic laws may be available (either as a representation of real indeterminism or of lack of exact knowledge) while a probabilistic law is unavailable due to pragmatic or mathematical constraints. Its most interesting technical contributions are, arguably, 1) the idea of redefining direct causation in terms of variation in the range of possible values of $Y$ when intervening on $X$ having fixed all other variables (which we tried to formalize in section \ref{subs: definitions of direct cause}); and 2) the observation that the $d$-separation criterion is correct also for qualitative independence statements. For the rest, the presentation and some details of the framework - e.g. the definition of intervention, the semantics, and the syntax - are unusual and a bit vague, which makes it rather difficult to say to what extent Wysocki's framework agrees or differs with that which is considered in the present paper. It would look like formulas of the form $\Diamond\psi$ or $\Box\psi$ are evaluated over teams that are \emph{maximal}, in the sense we mentioned in section \ref{subs: total and recursive cases}, but allowing, rather than a single tuple of values for the exogenous variables, a cartesian product of value ranges for each exogenous variable (so as to ensure a qualitative form of independence among the exogenous variables). We have not considered this class of causal teams in the present paper.  

In 2024, the proceedings version of the present paper \citep{Bar2024} appeared in print. Our work on the subject of indeterministic causal models began as early as 2020, and was influenced only in the later stages by the publication of \cite{Wys2023} (from which we essentially only borrowed the generalized idea of direct causation that is discussed here in section \ref{subs: definitions of direct cause}). The paper introduced a formal definition of relational causal team (with \emph{simplified} signatures) and sketched a description of the formalization obstacles that are presented more sharply here (sections \ref{sec: differences} and \ref{subs: causes and recursivity}). The paper also made some claims about axiomatization; regrettably, it was later discovered that the system therein introduced (as a purported axiomatization of the general class of models) is unsound. Indeed, the system in question includes axiom I6 (\emph{Weak reversibility}) which, as we have seen, can be falsified by teams with more than one assignment. Secondly, its version of axiom I8 (which, differently from ours, does not rely on having the external/internal distinction in the signature) relies on the definability, in $\Hal$, of what we call here \emph{visible direct cause} in order to enforce syntactically the assumption that this axiom only applies to external variables; unfortunately, doing so amounts to evaluating externality by looking at the incorrect graph (the visible causal graph instead of the parenthood graph). Lastly, there is an axiom $\Diamond\top \leftrightarrow\langle \SET W = \SET w\rangle\top$. Its right-to-left direction is in general false, since, as we have seen, the semantics allows for the possibility that intervening on an empty model may produce a nonempty team (this is actually \emph{always} the case when intervening on the set $\SET W$ of all variables). Our system $\Ax$ replaces this incorrect principle with axioms I9 and I10.
We also remark that, while I6 is at least sound on singleton models, the other two axioms we discussed here are not.

In \cite{Bec2025}, Beckers contributed his own theory of indeterministic counterfactuals \citep[which is then used in][to formulate a new definition of actual causality]{Bec2025b}. The indeterministic causal models used by Beckers (at least, in what he calls \emph{swc semantics}) are essentially identical to our total recursive singleton models \citep[the paper is directly inspired by our][]{Bar2024}, but Beckers proposes a different definition of counterfactual. Roughly, the truth value of his counterfactuals is assessed after a phase of observation (of values actually taken by the variables) followed by an intervention in the sense of Halpern. The observation phase is used (to the best of our understanding) to exclude interventions that would modify the actual value of a variable $Y$ even though they do not change the values of the parents of $Y$ (thus in practice enforcing the validity of the \emph{Strong composition} axiom we discussed in section \ref{subs: failure of composition}). So, for example, in a coin-tossing scenario where the coin has been tossed and it came heads, the counterfactual ``had the coin been tossed, it would have come heads'' is taken to be true (while it is false in our framework). Our impression is that Beckers' counterfactuals might be closer to Lewisian counterfactuals; their consequents are evaluated at (some of the) worlds that differ from the actual one only for making the antecedent true. The counterfactuals considered in our paper seem to be closer to the spirit of \emph{interventionist} counterfactuals; their consequents are evaluated in worlds where not only the antecedent is true, but it is also enforced by external intervention. Note also that it is only our paper's version of the counterfactual that preserves the \emph{predictive} power of causal statements. Suppose for example that the Federal Council of Switzerland decides to declare war to Luxembourg upon the outcome of a coin toss (war shall be declared if a \emph{tails} is tossed). \emph{Heads} is tossed, thus preventing a suicidal war declaration. If we made the mistake of using Beckers' notion of counterfactual for policy-making, we would reach the obviously mistaken conclusion that it will be a good policy to use coin-tossing as a device to decide on the future diplomatic relations between Switzerland and Luxembourg. Using the conterfactuals from our paper, one will instead conclude that further coin-tossing \emph{might} lead to a counterproductive war.\footnote{One might retort that it is only a lucky accident that interventionist counterfactuals turn out to have predictive power in the deterministic case; after all, counterfactuals concern past, not future events. We thank S. Beckers for this observation.} It would seem, then, that we have a distinction betwen a philosophical and a practical notion of counterfactual; differences between the two approaches are certainly worth future investigation. For what concerns axiomatization, Beckers provides deduction systems for two rather special classes of models. First, he axiomatizes total recursive singleton models (what he calls \emph{swc semantics}); his system differs from our $\Ax+\operatorname{Tot}+\operatorname{R}$ (see section \ref{subs: causes and recursivity}) in that it features both \emph{Strong composition} (which is unsound in our context) and \emph{Weak composition} (our I3), it lacks \emph{Nonemptyness} (I10) and has axiom $\Diamond \varphi \rightarrow \Box \varphi$  in place of our I12b. Furthermore, axioms I7 and I8 are absent due to differences in the syntax.\footnote{Becker's syntax only uses endogenous variables -- thus our I8 cannot even be formulated; and identifies modal-free formulas $\varphi$ with $\Box\varphi$, making axiom I7 useless.} The recursivity axiom is formulated in terms of an indeterministic version of ``causally affecting'' instead of our choice of direct cause. Secondly, Beckers provides an axiomatization for what he calls \emph{scc semantics}, i.e., essentially for the (total, recursive) \emph{maximal} causal teams that we briefly mentioned in section \ref{subs: total and recursive cases}.

There is a different line of inquiry connecting counterfactuals and (qualitative) indeterminism, and that is the study of (using Bohm's terminology) \emph{many-to-one} causation, i.e. cases where ``many different kinds of causes can produce essentially the same effect'' \citep{Boh1957}. We are aware of only three papers attempting to develop the technicalities behind this idea. The first one is \cite{BarYan2022}, in whose conclusion section an idea is described of \emph{indeterminate interventions}, i.e. interventions that are in some ways underspecified (e.g. ``pressing \emph{any} button on the remote'', ``picking one direction'', ``switching to the third or fifth gear''). Such interventions admit a natural definition in the \emph{generalized causal teams} studied in that paper\footnote{While a causal team encodes uncertainty about the state of the variables under a fixed hypothesis about the causal laws, a generalized causal team encodes uncertainty about both the state of the variables and the causal laws.}. These interventions are introduced with the purpose of giving an intervention-based semantics for interventionist counterfactuals with disjunctive antecedents; while several semantic frameworks for such counterfactuals had been proposed elsewhere \citep[e.g.][]{Bri2012,CiaZhaCha2018,Sant2019}, all these earlier proposals seemed rather \emph{ad hoc} and ungrounded in interventionist ideas. In 2023, \cite{Wys2023b} makes a similar proposal under the curious name of \emph{semaphore interventions}\footnote{The idea is also described, less explicitly, in \cite{Wys2023}.}; since he does not cite \cite{BarYan2022}, we can only assume that the idea was developed independently. This paper makes the interesting observation that for a simple enough language (say, $\Hal$ extended with the possibility of having disjunctive counterfactual antecedents) the semantics returns the same judgements of truth and falsity as Briggs's theory. Thus, the indeterministic interventions provide a clear interventionist motivation for Briggs's seemingly \emph{ad hoc} truth-maker semantics.

\section{Conclusions}

We have shown that extending causal modeling to the case of indeterministic causal laws, as suggested in \citet{Hal2000}, is doable but not as straightforward as Halpern suggested. We have seen that the notions of direct cause and causal parenthood are more complex than in the deterministic case, and even more complicated than what suggested in \citet{Wys2023}, since it turns out that even the dummy arguments of causal laws can be direct causes. These insights lead us to the definition of appropriate models (relational causal teams).

We then produced two types of axiomatizations; one, exemplified by system $\Ax$ and its extensions, axiomatizes the Halpern-style language $\Hal$ but is only applicable to singleton models (since the \emph{Weak reversibility} axiom I6 is unsound over multiple-assignment causal teams). This semantics is the closest to the original causal models, and we used it to compare the deterministic and indeterministic cases. The second type of proof system, given by system $\Bx$ and its extensions, axiomatizes the language $\Hal^+$ with \emph{right-nested} counterfactuals and applies to the class of \emph{all} models (i.e., all relational causal teams). The axiomatization is somewhat unsatisfactory, as one of the axioms (J17) is rather \emph{ad hoc} and heavily relies on the semantics; what remains unsettled is finding an elegant family of ``rules'' to transform $\Hal^+$ formulas into unnested, $\Hal$ formulas. We believe that the result has in any case some value, in that it is (to the best of our knowledge) the first known axiomatization of right-nested counterfactuals outside of the recursive deterministic case.\footnote{The recursive deterministic case is easily dealth with, as in that case the diamond modalities are interchangeable with box modalities, and the sequence of modalities $[\SET X = \SET x][\SET Y = \SET y]$ can always be contracted into a single modality. See \cite{Bri2012} and \cite{BarSan2020}.} Its proof also provides  a detailed exploration (perhaps the first) of the connections between interventionist counterfactuals and (dynamic) modal logic. On the other hand, while system $\Bx$ does provide a recursive enumeration of the validities of language $\Hal$, the direct axiomation of $\Hal$ over the class of all models remains as an open question.

We  also saw how to specialize systems  $\Ax$ and $\Bx$ (by adding new axioms) to a number of significant subclasses of models: 
the strictly/visibly recursive, deterministic, total and total recursive 
subclasses (the corresponding additional axioms can also be combined modularly). The logic of the recursive class differs from its deterministic counterpart in a few respects, among which we may remark the failure of the (strong) \emph{Composition} law. 
We also observed that the \emph{Reversibility} law holds in the strictly recursive case, showing that its traditional connection to the property of uniqueness of solutions breaks in the recursive indeterministic context. 

Our axiomatization results rely to some extent on using signatures that fix the distinction between internal and external variables. We have seen that, abandoning this constraint, some important notions, such as determinism and the external/internal distinction, become undefinable in the usual language $\Hal$. 
We have seen that the severity of these problems decreases when restricting attention to singleton models (determinism becomes again definable), so these technical obstacles might be seen as a counterindication to using teams rather than singleton models. However, they may instead be seen as evidence that the usual causal language $\mathcal H$ is insufficiently expressive in the indeterministic context; we have seen that determinism is definable when using the ordinary signatures, and becomes definable also with simplified signatures as soon as one allows for (right-)nested counterfactuals. 
\corrf{We have also seen that something similar happens if $\Hal$ is enriched  with the conditional $\supset$ expressing the results of learning and observations that was proposed in \citet{BarSan2020}. We conjecture that these two extended languages, $\Hal^+$ and $\HO$, might be expressively complete (further work will be needed just to make this statement precise).} For the moment, an important open question is whether using such languages, or by restricting attention to singleton models, one might be able to define the important notion of direct cause; we have seen that $\Hal$ is insufficient for this purpose, with both types of signatures. Other language extensions might be worth considering in the future; for example, in \citet{BarYan2022} we suggested a strategy for allowing complex, disjunctive antecedents, by assigning them a semantics by means of \emph{indeterminate interventions}. This extension requires a further generalization of the models and is not explored in the present paper, although plausibly the generalization is straightforward. 

A number of natural directions of investigation opens ahead, among which is the axiomatixation problem for classes of models with simplified signatures; the axiomatization of 
$\HO$; a systematical comparison with the logic of Stalnaker-Lewis counterfactuals \citep{Lew1973} in the spirit of \citet{Hal2013,Zha2013,FanZha2023}; and the analysis of the computational and descriptive complexity of the formalism \citep[cf., for the deterministic case,][Chapter 5]{MosIbeIca2022,Hal2016}.
A rather different direction to explore is the development of languages that can express data dependencies along with interventionist counterfactuals. This was done in the deterministic, recursive case in \cite{BarSan2020} and \cite{BarYan2022}. We note that the definition of counterfactual that was given in the present paper is clearly inadequate for this task: the consequent of a counterfactual is evaluated assignment by assignment, while data dependencies are global properties of a whole team. The natural move is then to use the following definition of interventionist counterfactual \cite[from][]{BarSan2020}:
\begin{itemize}
    \item $\SET X = \SET x \cf \psi$ iff $T_{\SET X = \SET x}\models \psi$.
\end{itemize}
One might also need a setup more similar to that of common logics with team semantics, featuring connectives such as the \emph{dual negation} and the \emph{tensor disjunction} instead of those used here. In this case, the \emph{might} counterfactual would need to be introduced as a distinct operator, as follows: 
\begin{itemize}
    \item $\SET X = \SET x \mcf \psi$ iff there is a nonempty $S\leq T_{\SET X = \SET x}\models \psi$ such that $S\models \psi$.
\end{itemize}
We may note that $\mcf$ is an operator that preserves interesting closure properties of formulas -- such as union closure and convexity -- but it does not preserve downward closure. Thus, the purely counterfactual part of such a language -- without the addition of data dependencies -- is already quite more complex than in the deterministic recursive case, where it happened to be just a flat, ``classical'' language. The properties and proof theory of such languages are still largely unexplored, and plausibly depart considerably from what may be familiar to the expert of causal reasoning; for these reasons, in this paper we chose to stick to languages closer to the Galles-Pearl-Halpern tradition. 
\\




\noindent \textbf{Acknowledgements.} The author wishes to thank Christopher Hitchcock, Sanders Beckers and the late Joseph Halpern for intense discussions on the topic and contents of the paper. 

The author's research was supported by the Research Council of Finland's grant n. 349803.

\vspace{20pt}

\noindent \textbf{Author contributions:} Single-author paper. 

\noindent \textbf{Conflict of interest:} The author declares none.

\noindent \textbf{Data availability:} No data was produced.

\vspace{10pt}


\bibliographystyle{abbrvnat}   
\bibliography{iilogics}

\pagebreak

\appendix

\vspace{15pt}

\begin{center}
    \Large
    APPENDIX
    \normalsize
\end{center}

We include here all the omitted proofs.

\section{Basic properties of the axiom system}

These are just results in normal modal logic; we include the proofs for reference. 

\BasicProperties*


\begin{proof}
1) By induction on the length of a proof of $\Gamma,\psi \vdash \chi$.

Suppose $\chi$ is an axiom. Then $\vdash \chi$. By I0, $\vdash \chi \rightarrow (\psi \rightarrow \chi)$. Thus, by MP,  $\Gamma \vdash \psi\rightarrow \chi$.

Suppose $\chi$ is in $\Gamma$. Then, since $\vdash \chi \rightarrow (\psi \rightarrow \chi)$ by I0, we have $\Gamma \vdash \psi \rightarrow \chi$ by MP. 

Suppose $\chi$ is $\psi$. Then $\Gamma\vdash \psi\rightarrow\psi$ by I0.

Suppose the last rule used is MP; then $\chi$ follows from assumptions $\theta$ and $\theta \rightarrow \chi$, such that $\Gamma\cup\{\psi\} \vdash \theta$ and $\Gamma\cup\{\psi\} \vdash \theta \rightarrow \chi$ (via subderivations). By i.h., $\Gamma \vdash \psi \rightarrow \theta$ and $\Gamma\vdash \psi \rightarrow ( \theta \rightarrow \chi)$. Since, by I0, $\Gamma \vdash (\psi \rightarrow \theta)\rightarrow [(\psi \rightarrow ( \theta \rightarrow \chi))\rightarrow (\psi\rightarrow \chi)]$, we get $\Gamma \vdash \psi\rightarrow \chi$ by two applications of MP.


Suppose the last rule used is NEC; then $\chi$ is of the form $[\SET X = \SET x]\varphi$, and $\vdash \varphi$. Using NEC, we obtain $\vdash [\SET X = \SET x]\varphi$. By I0, $\vdash [\SET X = \SET x]\varphi \rightarrow (\psi\rightarrow[\SET X = \SET x]\varphi)$, so, by MP, $\vdash\psi\rightarrow[\SET X = \SET x]\varphi$. \emph{A fortiori}, then,  $\Gamma\vdash\psi\rightarrow[\SET X = \SET x]\varphi$.



2) If $[\SET X = \SET x]\psi, [\SET X = \SET x]\psi'\in\Hal$, then $\psi,\psi'$ are modal-free. Thus, from  $\vdash \psi \rightarrow \psi'$, by NEC, we get $\vdash [\SET X = \SET x](\psi \rightarrow \psi')$. By I5 and MP,  $\vdash [\SET X = \SET x]\psi \rightarrow [\SET X = \SET x]\psi'$. Thus, by MP, from the assumption that $\Gamma\vdash[\SET X = \SET x]\psi$ we obtain $\Gamma\vdash[\SET X = \SET x]\psi'$. 

3) Remember that $\langle \SET X = \SET x\rangle$ abbreviates $\neg [\SET X = \SET x]\neg$. From  $\vdash \psi \rightarrow \psi'$ we obtain  $\vdash \cneg\psi' \rightarrow \cneg\psi$ by I0 and MP. As in the proof of point 2., we can show that $\cneg\psi' \rightarrow \cneg\psi$ is modal-free. Thus, by NEC and I5, $\vdash [\SET X = \SET x]\cneg\psi' \rightarrow [\SET X = \SET x]\cneg\psi$. Again by I0 and NEC, we obtain $\vdash \cneg [\SET X = \SET x]\cneg\psi \rightarrow \cneg [\SET X = \SET x]\cneg\psi'$. Thus, from $\Gamma \vdash \cneg [\SET X = \SET x]\cneg\psi$ we obtain $\Gamma \vdash \cneg [\SET X = \SET x]\cneg\psi'$ by MP.


4) Assume $[\SET X = \SET x]\psi \with  [\SET X = \SET x]\chi$. We obtain $[\SET X = \SET x]\psi$ and $[\SET X = \SET x]\chi$ by I0 and MP. Since $\psi$ and $\chi$ are modal-free, by I0 and NEC, we have $[\SET X = \SET x](\psi\rightarrow(\chi\rightarrow (\psi\with \chi)))$. Applying I5 and MP, we obtain  $[\SET X = \SET x](\chi\rightarrow (\psi\with \chi))$. Again by  I5 and MP, we obtain  $[\SET X = \SET x](\psi\with \chi)$. Then $([\SET X = \SET x]\psi \with  [\SET X = \SET x]\chi) \rightarrow [\SET X = \SET x](\psi \with \chi)$ by the deduction theorem.  

In the opposite direction, assume $[\SET X = \SET x](\psi\with \chi)$. By I0 and NEC we have $[\SET X = \SET x]((\psi\with \chi)\rightarrow \psi)$. By I5, $[\SET X = \SET x](\psi\with \chi)\rightarrow [\SET X = \SET x]\psi$. We then obtain $[\SET X = \SET x]\psi$ by MP. $[\SET X = \SET x]\chi$ is obtained analogously. 


5) Induction on $\varphi$.

If $\varphi$ is an atom, then either $\theta = \varphi$ or $\theta$ is not a subformula of $\varphi$. Both cases are easy.

Case $\varphi$ is $\psi\with\chi$. We prove this case by the deduction theorem. Assume $\Gamma$ and $\varphi$; then, we get $\psi$ and $\chi$ by I0+MP. By the i.h., $\Gamma\vdash \psi\leftrightarrow\psi[\theta'/\theta]$ and $\Gamma\vdash \chi\leftrightarrow\chi[\theta'/\theta]$. Thus, by MP, we get $\Gamma,\varphi\vdash\psi[\theta'/\theta]$ and $\Gamma,\varphi\vdash\chi[\theta'/\theta]$. By I0+MP, we conclude $\Gamma,\varphi\vdash\varphi[\theta'/\theta]$. So, by the deduction theorem, $\Gamma\vdash\varphi\rightarrow \varphi[\theta'/\theta]$. The converse direction is analogous.

Case $\varphi$ is $\cneg\psi$. We have $\Gamma\vdash \theta \leftrightarrow \theta'$; by the i.h., $\Gamma\vdash \psi \leftrightarrow \psi[\theta'/\theta]$. Since, by I0, $\Gamma\vdash (\psi \leftrightarrow \psi[\theta'/\theta])\leftrightarrow (\cneg\psi \leftrightarrow \cneg\psi[\theta'/\theta])$, we conclude $\Gamma\vdash \cneg\psi \leftrightarrow \cneg\psi[\theta'/\theta]$.

Case $\varphi$ is $[\SET X = \SET x]\psi$. By the i.h., $\Gamma\vdash \psi \leftrightarrow \psi[\theta'/\theta]$. By NEC, $\Gamma\vdash [\SET X = \SET x](\psi \leftrightarrow \psi[\theta'/\theta]$). By point 4.,  plus I0+MP, we obtain $\Gamma\vdash [\SET X = \SET x](\psi \rightarrow \psi[\theta'/\theta])$ and $\Gamma\vdash [\SET X = \SET x]( \psi[\theta'/\theta]\rightarrow \psi)$. By axiom I5 and I0+MP, we obtain the desired conclusion. 


6) We use once more the deduction theorem. Assume first $\cneg[\SET X = \SET x]\psi$. We have $\vdash\psi\leftrightarrow \cneg\cneg \psi$ by I0, so by replacement and MP we obtain $\cneg[\SET X = \SET x]\cneg\cneg\psi$. This is $\langle\SET X = \SET x\rangle\cneg\psi$ by definition. 

Vice versa, assume $\langle\SET X = \SET x\rangle \cneg\psi$, i.e.$\cneg[\SET X = \SET x]\cneg\cneg\psi$. By I0, replacement and MP, we obtain $\cneg[\SET X = \SET x]\psi$. 

7) $\cneg\langle\SET X = \SET x\rangle\psi$ abbreviates $\cneg\cneg [\SET X = \SET x] \cneg\psi$. By I0 and MP, it is equivalent to $[\SET X = \SET x]\cneg\psi$.

8) Assume  $\langle\SET X = \SET x\rangle \psi \sqcup \langle\SET X = \SET x\rangle \chi$;  it abbreviates $\cneg(\cneg\langle\SET X = \SET x\rangle \psi \with  \cneg \langle\SET X = \SET x\rangle \chi)$. By part 7. and replacement (twice) we obtain $\cneg([\SET X = \SET x]\cneg\psi \with  [\SET X = \SET x]\cneg\chi)$. Then, by 
part 4.  and MP, we obtain $\cneg[\SET X = \SET x](\cneg\psi \with  \cneg\chi)$.  By part 6., then, $\langle\SET X = \SET x\rangle\cneg(\cneg\psi \with  \cneg\chi)$, i.e. $\langle\SET X = \SET x\rangle (\psi \sqcup \chi)$. The converse is similar.


10)  From the negation of the conclusion we obtain $[\SET X = \SET x]\cneg(\psi \with \chi)$ by point 7. Together with the assumption $[\SET X = \SET x]\psi$, by point 4.,  we obtain $[\SET X = \SET x](\cneg(\psi \with \chi)\with \psi)$, and thus by I0 and replacement $[\SET X = \SET x]\cneg\chi$. Finally, by point 7. again, $\cneg\langle \SET X = \SET x \rangle\chi$.

9) By point 10. we have $\vdash([\SET X = \SET x]\psi \with \langle \SET X = \SET x \rangle \top) \rightarrow \langle \SET X = \SET x \rangle (\psi \with \top)$. By I0 and replacement, $\langle \SET X = \SET x \rangle (\psi \with \top) \rightarrow\langle \SET X = \SET x \rangle \psi$. Thus, by I0 and point 2., $\vdash([\SET X = \SET x]\psi \with \langle \SET X = \SET x \rangle \top) \rightarrow \langle \SET X = \SET x \rangle \psi$.

11) Assume $\langle\SET X = \SET x\rangle(\psi \with \chi)$. Since $\vdash (\psi \with \chi) \rightarrow \psi$, by monotonicity (point 3.) we obtain $\Diamond \psi$. We obtain analogously $\Diamond \chi$. Thus, $\Diamond \psi \with \Diamond \chi$ by I0.  
\end{proof}

\section{Maximally consistent sets of formulas}

The following proofs 
are routine, and included for reference.

 \begin{restatable}{lemma}{Contradiction}\label{lemma: contradiction}
  Let $\Gamma \cup \{\chi\}\subseteq \mathcal H$. If $\Gamma\vdash \chi$ and $\Gamma\vdash \cneg\chi$, then $\Gamma\vdash \bot$.
  \end{restatable}


\begin{proof}
Remembering that $\bot$ abbreviates $X=x \with \cneg X=x$, 
We observe that $\chi\rightarrow (\cneg\chi \rightarrow\bot)$ is an instance of a tautology, so we obtain $\Gamma\vdash \bot$ by I0 and two applications of MP.
\end{proof}

\begin{restatable}{lemma}{Consistency}\label{lemma: consistency}
Suppose $\Gamma \not\vdash \chi$. Then, $\Gamma \cup \{\cneg\chi\}$ is consistent.
\end{restatable}


\begin{proof}
Suppose it is not; then $\Gamma \cup \{\cneg\chi\}\vdash\psi,\cneg\psi$ for some formula $\psi$. By I0,  $\vdash\psi\rightarrow (\cneg\psi \rightarrow \psi \with \cneg \psi$), so by MP, $\Gamma \cup \{\cneg\chi\}\vdash \psi \with \cneg \psi$. By the deduction theorem, $\Gamma \vdash \cneg\chi \rightarrow (\psi \with \cneg \psi)$.
By I0 and MP, we obtain $\Gamma \vdash \chi$, contrarily to the assumption.
\end{proof}



\MCSProperties*

\begin{proof}
1) Suppose $\Gamma\vdash\psi$. Then, $\Gamma$ and $\Gamma\cup \{\psi\}$ have the same set of consequences; thus, by the maximality of $\Gamma$, $\psi\in\Gamma$.

2) Suppose first that $\Gamma\vdash\psi$. Then, $\psi\in\Gamma$ by point 1.

If instead $\Gamma\not\vdash\psi$, then, by lemma \ref{lemma: consistency}, $\Gamma\cup \{\cneg\psi\}$ is consistent, and $\Gamma\cup \{\cneg\psi\}\supseteq \Gamma$; thus, by
the maximality of $\Gamma$, $\cneg \psi \in \Gamma$.






3) 
By the assumptions, $\Gamma\vdash\psi,\chi$. By I0, $\Gamma\vdash \psi\rightarrow(\chi\rightarrow (\psi \with  \chi))$. Thus, by two applications of MP, $\Gamma\vdash\psi \with  \chi$. Thus,  by 1., $\psi \with  \chi \in \Gamma$.

4) Suppose $\psi\notin\Gamma$ and $\chi\notin\Gamma$. By 2., then, $\cneg\psi\in\Gamma$ and $\cneg\chi\in\Gamma$. By 3., $\cneg\psi\with\cneg\chi\in\Gamma$. By I0 and part 2., $\cneg\cneg(\cneg\psi\with\cneg\chi)\in\Gamma$, i.e. $\cneg (\psi \sqcup \chi)\in\Gamma$. By the consistency of $\Gamma$, then, $\psi \sqcup \chi\notin\Gamma$.
%

5) Note that $\Diamond \bot$ abbreviates $\cneg \Box \cneg \bot$, i.e., $\cneg \Box \top$. On the other hand, $\Box \top\in \Gamma$ by I0 and NEC. Thus, by the consistency of $\Gamma$, $\cneg \Box \top \notin \Gamma$.
\end{proof}

\end{document}